\documentclass[12pt,onecolumn]{aa}

\usepackage{lscape}
\usepackage{graphicx,amssymb,amsmath,times}
\usepackage{longtable}
\usepackage{bm} 
\usepackage{url,hyperref}
\usepackage[varg]{txfonts}
\usepackage[T1]{fontenc}
\usepackage{natbib}
\bibpunct{(}{)}{;}{a}{}{,} 
\usepackage[normalem]{ulem}

\def\simlt{\lower.5ex\hbox{\ltsima}}
\def\simgt{\lower.5ex\hbox{\gtsima}}

\def\wcen{$\omega$Cen}

\def\gtsim{\;\lower.6ex\hbox{$\sim$}\kern-6.7pt\raise.4ex\hbox{$>$}\;}
\def\ltsim{\;\lower.6ex\hbox{$\sim$}\kern-6.9pt\raise.4ex\hbox{$<$}\;}

\def\sec{${}^{\prime\prime}$}
\def\hst{{\it HST \/}}
\def\gaia{{\it Gaia \/}}
\def\bmv{\hbox{\it B--V\/}}

\titlerunning{NIR light-curve templates for RR Lyrae variables}
\authorrunning{Braga et al.}

\begin{document}

\title{New Near-Infrared $JHK_s$ light-curve templates for RR Lyrae variables\footnote{
The coefficients of the templates will be provided only once the paper is published.}}

\author{
V.~F.~Braga \inst{1,2}
\and P.~B.~Stetson \inst{3}
\and G.~Bono \inst{4,5}
\and M.~Dall'Ora \inst{6}
\and I.~Ferraro \inst{5}
\and G.~Fiorentino \inst{7}
\and G.~Iannicola \inst{5}
\and L.~Inno \inst{8}
\and M.~Marengo \inst{9}
\and J.~Neeley \inst{10}
\and R.~L.~Beaton \inst{11}
\and R.~Buonanno \inst{12}
\and A.~Calamida \inst{13}
\and R.~Contreras Ramos \inst{14}
\and B.~Chaboyer \inst{15}
\and M.~Fabrizio \inst{16}
\and W.~L.~Freedman \inst{17}
\and C.~K.~Gilligan \inst{15}
\and K.~V.~Johnston\inst{18}
\and B.~F.~Madore \inst{11}
\and D.~Magurno \inst{4}
\and M.~Marconi \inst{6}
\and S.~Marinoni \inst{16}
\and P.~Marrese \inst{16}
\and M.~Mateo\inst{19}
\and N.~Matsunaga \inst{20}
\and D.~Minniti \inst{1,2,21}
\and A.~J.~Monson \inst{11}
\and M.~Monelli \inst{22}
\and M.~Nonino \inst{23}
\and S.~E.~Persson \inst{11}
\and A.~Pietrinferni \inst{12}
\and C.~Sneden\inst{24}
\and J.~Storm \inst{25}
\and A.~R.~Walker \inst{26}
\and E.~Valenti \inst{27}
\and M.~Zoccali \inst{14}
}
\institute{
Instituto Milenio de Astrof{\'i}sica, Santiago, Chile 
\and Departamento de F{\'i}sica, Facultad de Ciencias Exactas, Universidad Andr{\'e}s Bello, Fern{\'a}ndez Concha 700, Las Condes, Santiago, Chile 
\and Herzberg Astronomy and Astrophysics, National Research Council, 5071 West Saanich Road, Victoria, British Columbia V9E 2E7, Canada 
\and Department of Physics, Universit\`a di Roma Tor Vergata, via della Ricerca Scientifica 1, 00133 Roma, Italy 
\and INAF-Osservatorio Astronomico di Roma, via Frascati 33, 00040 Monte Porzio Catone, Italy 
\and INAF-Osservatorio Astronomico di Capodimonte, Salita Moiariello 16, 80131 Napoli, Italy 
\and INAF-Osservatorio Astronomico di Bologna, Via Ranzani 1, 40127 Bologna, Italy 
\and Max Planck Institute for Astronomy K\"onigstuhl 17 D-69117, Heidelberg, Germany 
\and Department of Physics and Astronomy, Iowa State University, Ames, IA 50011, USA 
\and Department of Physics, Florida Atlantic University, 777 Glades Rd, Boca Raton, FL 33431 USA 
\and The Observatories of the Carnegie Institution for Science, 813 Santa Barbara St., Pasadena, CA 91101, USA 
\and INAF-Osservatorio Astronomico d'Abruzzo, Via Mentore Maggini snc, Loc. Collurania, 64100 Teramo, Italy 
\and Space Telescope Science Institute, 3700 San Martin Drive, Baltimore, MD 21218, USA
\and Pontificia Universidad Catolica de Chile, Instituto de Astrofisica, Av. Vicu\~na Mackenna 4860, Santiago, Chile 
\and Department of Physics and Astronomy, Dartmouth College, Hanover, NH 03755, USA
\and Space Science Data Center, via del Politecnico snc, 00133 Roma, Italy
\and Department of Astronomy \& Astrophysics, University of Chicago, 5640 South Ellis Avenue, Chicago, IL 60637, USA 
\and Department of Astronomy, Columbia University, 550 W 120th st., New York, NY 10027, USA
\and Department of Astronomy, University of Michigan, Ann Arbor, MI, USA
\and Kiso Observatory, Institute of Astronomy, School of Science, The University of Tokyo, 10762-30, Mitake, Kiso-machi, Kiso-gun, 3 Nagano 97-0101, Japan 
\and Vatican Observatory, V00120 Vatican City State, Italy 
\and Instituto de Astrof\'isica de Canarias, Calle Via Lactea s/n, E38205 La Laguna, Tenerife, Spain 
\and INAF, Osservatorio Astronoico di Trieste, Via G.~B. Tiepolo 11, 34143 Trieste, Italy 
\and Department of Astronomy and McDonald Observatory, The University of Texas, Austin, TX 78712, USA
\and Leibniz-Institut f\"ur Astrophysik Potsdam, An der Sternwarte 16, 14482, Potsdam, Germany
\and Cerro Tololo Inter-American Observatory, National Optical Astronomy Observatory, Casilla 603, La Serena, Chile 
\and European Southern Observatory, Karl-Schwarzschild-Str. 2, 85748 Garching bei Munchen, Germany 
}

\date{\centering Submitted \today\ / Received / Accepted }

\abstract{
We provide homogeneous optical ($UBVRI$) and near-infrared ($JHK$) 
time series photometry for 254 cluster ($\omega$ Cen, M4) 
and field RR Lyrae (RRL) variables. We ended up with more than 551,000 
measurements and only a minor fraction (9\%) were collected in the 
literature. For 94 fundamental (RRab) and 51 first overtones (RRc) 
we provide a complete optical/NIR characterization (mean magnitudes, 
luminosity amplitudes, epoch of the anchor point).    
The NIR light curves of these variables were adopted to provide new 
and accurate light-curve templates for both RRc (single period bin) 
and RRab (three period bins) variables.  The templates for the $J$ 
and the $H$ band are newly introduced, together with the use of the 
pulsation period to discriminate among the different RRab templates.    
To overcome subtle uncertainties in the fit of secondary features of 
the light curves (dips, bumps) we provide two independent sets of 
analytical functions (Fourier series, Periodic Gaussian functions). 
The new templates were validated by using 26 $\omega$~Cen and 
Bulge RRLs covering the four period bins. We found that the 
difference between the measured mean magnitude along the light 
curve and the mean magnitude estimated by using the template 
on a single randomly extracted phase point is better than 0.01 mag 
($\sigma$=0.04 mag). We also validated the template on variables 
for which at least three phase points were available, but without 
information on the phase of the anchor point. We found that the 
accuracy of the mean magnitudes is also $\sim$0.01 mag 
($\sigma$=0.04 mag).  
The new templates were applied to the Large Magellanic Cloud (LMC) 
globular Reticulum and by using literature data and predicted PLZ 
relations we found true distance moduli of 18.47$\pm$0.10$\pm$0.03 mag 
($J$) and 18.49$\pm$0.09$\pm$0.05 mag ($K$). We also used 
literature optical and mid-infrared data and we found a mean true 
distance modulus of 18.47$\pm$0.02$\pm$0.06 mag, suggesting that 
Reticulum is $\sim$1 kpc closer than the LMC.}

\keywords{Stars: variables: RR Lyrae; Methods: data analysis; 
Globular clusters: individual: M4; Globular clusters: individual: $\omega$ Centauri;
Globular clusters: individual: Reticulum}
\maketitle

\section{Introduction}\label{intro}

RR Lyrae (RRLs), are very accurate distance indicators and 
solid tracers of old (age > 10 Gyr) stellar populations. The 
near-infrared (NIR) Period-Luminosity (PL) relations of RRLs will be
the first calibrator of the extragalactic distance scale based
on population II stars \citep{beaton2016}, which will provide 
an independent estimate of $H_0$. The NIR bands, when compared 
with the optical bands, present several advantages.
These become even more compelling for variable stars, 
like RRLs. It is therefore mandatory to
fully exploit the advantages that the NIR bands bring up. They 
are the following.



{\it i)} The NIR bands are less prone to uncertainties in reddening corrections 
and are less affected by the occurrence of differential reddening.
Indeed, the $K$ band is one order of magnitude less affected than the visual band.   
For this reason, the highly reddened regions of the Galactic center and of 
the inner bulge can only be investigated effectively in NIR bands. At these
Galactic latitudes, the absorption in the $K$ band becomes of the 
order of 2.5-3.0 mag \citep{gonzalez2012}, meaning 
$\sim$25-30 mag in the $V$ band. This is well beyond the capabilities of 
current and near-future optical observing facilities.

{\it ii)} The luminosity variation in the optical bands is dominated by variations 
in effective temperature, while in the NIR bands it is dominated by variations 
in stellar radius \citep{madore13}. This means that the NIR light curves are minimally affected by nonlinear 
phenomena like shock formation and propagation, which cause the appearance of either bumps 
and/or dips along the light curves. Moreover, the luminosity amplitudes steadily 
decrease when moving from the optical to the NIR bands and approach an almost constant 
value for wavelengths equal or longer than 2.2 $\mu$m \citep{madore13}. 
Indeed, the ratio in luminosity amplitudes $A_{Ks}/A_V$ 
and $A_{[3.6]}/A_V$ attain values ranging from 0.22 
to 0.41 \citep[RRc and RRab, respectively, ][]{braga2018} 
and from 0.18 to 0.22 \citep[RRc and RRab, respectively, ][]{neeley15}.

{\it iii)} The typical sawtooth shape of the light curves of RRab in the optical is 
less sharp in the NIR, where light curves become more symmetrical. This means that,
even with a modest number of phase points, the light curve can be well fitted.

The NIR bands, together with these ``intrinsic advantages'' also bring up several 
``extrinsic advantages'' concerning the RRL distance scale. 

{\it i)} Solid theoretical \citep{bono01,marconi15} 
and empirical \citep{longmore86,bono03c} evidence indicates that RRLs obey well-defined 
Period-Luminosity-Metallicity (PLZ) relations in the NIR bands. The slope of the 
relation becomes steeper and its standard deviation decreases when 
moving toward longer wavelengths. 
The RRLs in the optical bands also obey mean Magnitude-Metallicity 
\citep[MZ, ][]{Sandage81a,Sandage81b} relations, but 
these are affected by non-linearity and evolutionary effects, and are
less precise than the PLZ relations in the NIR bands \citep{caputo00}.   

{\it ii)} In the case that both optical and NIR bands are available, one can 
adopt the newly developed algorithm REDIME \citep{bono2018}.
REDIME is capable of providing homogeneous and simultaneous estimates of metal content, 
distance and reddening.

However, the NIR bands also bring up some cons. 

{\it i)} The identification and the characterization 
of RRLs is more difficult in the NIR bands, due to the decrease 
in luminosity amplitude and the less characteristic shape of the 
light curve, when moving from shorter- to longer-period variables.

{\it ii)} Accurate and deep NIR photometry is never trivial, in particular in 
crowded stellar fields. NIR observations are quite demanding of telescope time, 
since specific observing strategies must be devised to properly subtract the sky background.  
This means that NIR observations typically have shallower limiting magnitudes and longer observing runs
when compared with optical bands. 
A practical example of this disadvantage is the comparison
between the OGLE and VVV surveys in the Bulge. While the first covers a larger sky area and 
provides time series with thousands of phase points, the second achieved $\sim$100 phase point 
per time series in a smaller area, although being much more capable of piercing the dust 
in the Galactic plane.
A very interesting and promising approach to overcome 
several of the limitations affecting the NIR bands is to use observing facilities that are 
assisted by an adaptive optics system. However, these complex detectors have a quite limited 
field of view, typically of the order of one arcminute or even smaller. This means that they 
can hardly be adopted for a photometric survey and/or for an efficient detection and 
characterization of variable stars. 

{\it iii)} To overcome possible nonlinear effects in cameras 
and/or the saturation of bright stars, and to 
improve sky subtraction, the NIR images are collected 
as series of short-exposure images, arranged in  
specific dithering pattern. Several approaches have been suggested in the literature to 
perform PSF photometry of NIR images and all of them present pros and cons.    

These limitations of NIR photometry are at the base of the development of NIR light 
curve template. More than twenty years ago, 
\citet[][henceforth, J96]{jones1996} provided, in a seminal 
investigation, the first NIR light-curve templates for RRLs. What really matters 
in this context is that---once the period of an RRL is already known,
preferentially from optical data, together with its 
optical amplitude and its epoch of maximum light---a template provides the opportunity to 
estimate its mean $K$-band magnitude on the basis of a single NIR measurement.  
However, the J96 templates were provided only for 
the $K$-band. Furthermore, owing to the limited number of 
NIR measurements available at that time it was only based on 17 RRab and 4 RRc. 
J96 divided RRab variables into four subgroups and kept the RRc variables within 
a single group, therefore obtaining four and one light-curves template, respectively.
However, the bins in luminosity amplitude adopted to split the 
fundamental pulsators into different sub-groups 
did not overlap one another (see Fig.~\ref{fig:bailey}). 
It is also worth mentioning that the use of the 
luminosity amplitude to discriminate RRLs with different shapes of the light curve might 
also be affected by degeneracy. Indeed, the Bailey diagram (luminosity amplitude versus 
period) shows that the trend of both RRc and RRab luminosity amplitudes is not linear over 
their typical period range \citep{cacciari05,kunder13}. This means 
that two variables that have the same amplitude might have different periods. 

To overcome some of these intrinsic limitations of the J96 NIR light-curve 
templates, new approaches have been recently proposed in the literature. 
It has been suggested by \citet{freedman2010} that accurate optical bands 
for Classical Cepheids can be transformed into the NIR bands using 
only a few measurements. The same approach was also applied to RRLs by 
\citet{beaton2016}, the experiment was limited to a single RRL as a 
preliminary result of their ongoing investigation based on $HST$ data. 
The key advantage of this method is that it does not require knowledge of the epoch 
of maximum light to phase the NIR measurements.
More recently, \citep{hajdu2018} suggested an interesting new
method to use a well-sampled $K_s$-band light curve to estimate 
the $J$- and the $H$-band mean magnitude of an RRL from single-epoch 
measurements. They used data from the VISTA Variables in the 
V{\'i}a L{\'a}ctea (VVV) survey and decomposed the $K_s$-band light 
curves of 101 RRab variables into orthogonal Principal Components. 
Their method also provides estimates of photometric metallicities.

Light-curve templates of RRLs have also been developed in the visual 
bands. \citet{layden1998} obtained six $V$-band light-curve
templates, but they were limited to RRab variables. The adopted sample of 
103 field RRLs was divided according to the shape of the light curve 
(Bailey types $a$ and $b$, plus the phase range of the rising branch). 
They were used to estimate simultaneously mean magnitude and luminosity 
amplitude. More recently, optical ($ugriz$) light-curve templates of 
RRLs were derived by \citet{sesar2010} from SDSS photometry of 
379 RRab and 104 RRc. They provided 22 RRab templates and two RRc templates 
for the five $ugriz$ SDSS bands. They found evidence that the shape of the 
RRab light curves steadily changes when moving from the blue to the red edge 
of the instability strip, while RRc light curves are dichotomous. They claim 
that this evidence might suggest the possible occurrence of 
second-overtone RRLs. However, theoretical models and spectroscopic measurements 
indicate that shorter period RRc variables are, on average, more metal-rich 
than the bulk of field RRc variables \citep{bono97b,sneden2017}, providing
an alternative explanation to the hypothesis of second overtone RRLs. 
In passing, it is worth mentioning that the light-curve templates by \citet{sesar2010}  
were mainly developed for RRL identification---especially within the upcoming
LSST survey---rather than to determine their mean magnitudes.

The main aim of this investigation is to provide new NIR light-curve templates for RRLs 
based on a detailed optical and NIR data set that our group collected for RRLs 
in the Galactic Globular Clusters (GGC) $\omega$ Cen and M4, supplemented by 
literature photoelectric photometry of Milky Way RRLs. 

The structure of the paper is as follows. In Section~\ref{sect_data},
we describe the optical and the NIR photometric data sets adopted 
for the current analysis. In \S˜3 we deal with the NIR light-curve templates 
and, in particular, with the criteria adopted to select the period bins and 
the normalization of the light curves. The analytical form of the light-curve
templates are discussed in Section~\ref{sect_template} together with 
a detailed discussion of the adopted anchor point to phase NIR measurements. 
Section~\ref{validation} is dedicated to the validation of the templates. 
The validation is based on $\omega$ Cen data and 
OGLE+VVV \citep{udalski92,minniti2010} data and it was performed for 
single and triple phase points. 
In Section~\ref{sect_reticulum} we apply the new NIR templates to the 
$J$ and $K_s$ light curves of RRLs in the extragalactic GC 
Reticulum and provide a new true distance modulus determination. 
We summarize our results in Section~\ref{sect_conclusions} and briefly outline 
future developments of the current project.

\section{Optical and near-infrared data sets}\label{sect_data}

We use proprietary---still unpublished until this 
work---optical and NIR PSF-reduced photometry of RRLs in M4 \citep{stetson14a}
and in $\omega$ Cen \citep{braga16,braga2018}. Optical data 
are in the \citep{landolt83,landolt92} system, and NIR 
data are in the 2MASS photometric system \citep{skrutskie2006}. 
Note that the NIR data were binned by epoch, therefore each phase point 
is actually an average of three to five phase points belonging to the 
same dithering sequence. The binning process is 
described in detail in \citet{braga2018}. More insights on the 
data (telescopes, cameras, and reduction) can be found 
in \citet{stetson14a}, \citet{braga16} and \citet{braga2018}.

These data were supplemented with 
{\it i)} relatively old optical and NIR 
photoelectric photometry of 26 Milky Way (MW) field RRLs, mostly collected to
perform Baade-Wesselink (BW) analysis 
\citep{carney84,cacciari87,jones87a,barnes88,jones88a,jones88b,liujanes89,skillen89,
clementini90,fernley90b,barnes92,cacciari92,jones92,skillen93b,skillen93a}, which
we call the ``BW'' sample, and {\it ii)} optical data from long-term photometric surveys 
(ASAS: \citealt{pojmanski1997}; NSVS: \citealt{wozniak04}). 
Note that the photoelectric data were not available in machine-readable format, 
therefore we have digitized the tables available in the original papers. 
Moreover, to deal with a homogeneous data set, we transformed all the photoelectric 
NIR data to the 2MASS system. We have used the transformations 
by \citet{carpenter2001} to convert the magnitudes from the CIT system 
\citep{jones87a,liujanes89,jones88a,jones88b,barnes92,jones92}, UKIRT 
system \citep{skillen89}, SAAO system \citep{fernley90b,skillen93b}
and ESO system \citep{cacciari92} to the 2MASS system. 
Note that more recent transformations between
the SAAO and 2MASS system are available \citep{koen2007}. However, 
these would require measurements in the $H$-band which
are not available for the \citet{fernley90b} data.
On the other hand, the optical photoelectric data are all in the 
Johnson system. However, we only use these optical data to derive the 
epoch of the mean magnitude on the rising branch ($t_{ris}$), independently
for each variable. Therefore, they were not homogenized with the CCD data 
in the Landolt system.

The key advantage of $\omega$ Cen RRLs is that this stellar system contains
almost 200 RRLs and they cover a range in metal content of at least one 
dex \citep{rey2000,sollima06a}.  Moreover, it is the only cluster---with 
the exception of the peculiar metal-rich clusters NGC~6388 \citep{pritzl2002a} 
and NGC~6441 \citep{pritzl2001}---hosting a sizable sample of long-period 
(P$>$0.7 days) RRLs.


To make a homogeneous Optical and NIR 
data set available to the entire astronomical community, Table~\ref{tab:allcv_opt} and Table~\ref{tab:allcv} give, respectively, the 
$UBVRI$ and $JHK_s$ light curves of 233 RRLs in $\omega$ Cen and M4;
Table~\ref{tab:allcv}, also provides $JHK_s$ light curves for 21 RRLs 
in the BW sample. Table~\ref{tab:allcv_opt} is based on the optical data
collected during $\sim$20-year-long campaigns \citep{stetson14a,braga16} and 
are calibrated to the Landolt ($UBV$) and Kron-Cousins 
($RI$) photometric system. Note that Table~\ref{tab:allcv_opt} contains
also literature data (\citealt{sturch1978,kaluzny1997,kaluzny04} plus 
the CATALINA \citealt{catalina} and ASAS-SN surveys \citealt{shappee2014,kochanek2017}) 
which we used to supplement our photometry 
\citep[more details in Section 3.1 of][]{braga16}.
Table~\ref{tab:allcv} includes 
objects for which either we collected NIR time series data during 
10-year-long observation campaigns \citep{stetson14a,braga2018} 
or NIR photometry was available in the literature. NIR measurmenets listed in 
Table~\ref{tab:allcv} are in the 2MASS photometric system. Note that the 
fraction of objects adopted for the NIR light-curve templates is 57\% of the 
total number of objects listed in Table~\ref{tab:allcv}.


\begin{table*}[!htbp]
 \footnotesize
 \caption{$UBVRI$ light curves of $\omega$ Cen and M4 RRLs.}
  \centering
 \begin{tabular}{l c c c c c c}
 \hline\hline  
Name\tablefoottext{a}  & band\tablefoottext{b} & HJD--2,400,000 & mag & err & dataset\tablefoottext{c} \\ 
 & & days & mag & mag & \\ 
\hline 
    \wcen-V3 & 1 & 54705.4699 &  14.950 & 0.005 & B19 \\
    \wcen-V3 & 1 & 50601.5756 &  14.767 & 0.022 & B19 \\
    \wcen-V3 & 1 & 50601.5802 &  14.775 & 0.022 & B19 \\
    \wcen-V3 & 1 & 50920.7741 &  15.335 & 0.017 & B19 \\
    \wcen-V3 & 1 & 53795.8733 &  14.428 & 0.039 & B19 \\
    \wcen-V3 & 1 & 53795.8858 &  14.487 & 0.065 & B19 \\
    \wcen-V3 & 1 & 51368.4851 &  15.320 & 0.003 & B19 \\
    \wcen-V3 & 1 & 51369.4884 &  15.339 & 0.003 & B19 \\
    \wcen-V3 & 1 & 52443.4731 &  15.245 & 0.011 & B19 \\
    \wcen-V3 & 1 & 52443.4773 &  15.223 & 0.011 & B19 \\
  \hline
 \end{tabular}
 \tablefoot{Only the first ten entries are displayed. 
 The full table is provided in electronic form.
 \tablefoottext{a}{An asterisk close to the name, in the 
 first column, indicates that the star was not used to derive the templates.}
 \tablefoottext{b}{Photometric flag:``1'' indicates data in the $U$ band, 
 ``2'' data in the $B$, ``3'' data in the $V$ band, ``4'' data in the $R$ band
 and ``5'' data in the $I$ band.}
  \tablefoottext{c}{Literature data flag:
  ``B19'' indicates data from our own photometry;
  ``asa'' indicates data from the ASAS-SN; 
  ``kal'' indicates data from \citet{kaluzny1997,kaluzny04}; 
  ``lub'' indicates unpublished data by J.H. Lub; 
  ``kal'' indicates data from \cite{sturch1978}.}}
\label{tab:allcv_opt}
\end{table*}

\begin{table*}[!htbp]
 \footnotesize
 \caption{$JHK_s$ light curves of $\omega$ Cen, M4 and BW RRLs.}
  \centering
 \begin{tabular}{l c c c c c c}
 \hline\hline  
Name & Flag\tablefoottext{a}  & band\tablefoottext{b} & HJD--2,400,000 & mag & err \\ 
 & & & days & mag & mag \\ 
\hline 
  \wcen-V3 & & 1 & 55341.5764 &  13.137 & 0.006 \\
  \wcen-V3 & & 1 & 55341.5878 &  13.147 & 0.007 \\
  \wcen-V3 & & 1 & 55341.6006 &  13.175 & 0.004 \\
  \wcen-V3 & & 1 & 55341.6074 &  13.179 & 0.013 \\
  \wcen-V3 & & 1 & 51946.8638 &  13.053 & 0.013 \\
  \wcen-V3 & & 1 & 51948.7440 &  13.154 & 0.012 \\
  \wcen-V3 & & 1 & 51948.8136 &  13.170 & 0.008 \\
  \wcen-V3 & & 1 & 52308.7494 &  13.075 & 0.013 \\
  \wcen-V3 & & 1 & 52308.8255 &  13.117 & 0.016 \\
  \wcen-V3 & & 1 & 52308.8674 &  13.145 & 0.011 \\
  \hline
 \end{tabular}
 \tablefoot{Only the first ten entries are displayed. 
 The full table is provided in electronic form.
 \tablefoottext{a}{An asterisk close to the name, in the 
 first column, indicates that the star was not used to derive the templates.}
 \tablefoottext{b}{Photometric flag:``1'' indicates data in the $J$ band, 
 ``2'' data in the $H$ and ``3'' data in the $K_s$.}}
\label{tab:allcv}
\end{table*}

\subsection{Sample selection}

To derive accurate and precise NIR light-curve templates we selected from the 
initial RRL sample the variables satisfying the following criteria.

{\it 1)} --- At least 10 phase points in $J$, $H$ or $K_s$.

{\it 2)} --- An accurate estimate of $t_{ris}$ (see Appendix~\ref{sec:tris} for the 
calculation of $t_{ris}$), i.e., the epoch to which the template is anchored.

{\it 3)} --- A small dispersion ($\sigma \lesssim$0.1) of the phase points 
along the normalized light curve. To derive the light-curve templates, all the 
light curves were divided by their amplitude (see Section~\ref{templatefit}). 
The variables with limited photometric accuracy are more likely to increase 
the dispersion of the normalized light curve, and in turn of the light curve 
template. Our approach was conservative: we only included variables with a 
``clean'' trend in the normalized light-curve fit. 

{\it 4)} --- Special care was taken to include variables that trace 
the shape of the light curve of both RRab and RRc when moving from 
shorter to longer period RRLs. This means the occurrence of either dips 
just before the phase of maximum light and/or bumps just before the phase 
of minimum light.

Once we apply these selection criteria we are left with a subsample 
of 94 RRab and 51 RRc variables. 
The excluded variables are marked with an 
asterisk in Table~\ref{tab:allcv}. In the following, the selected  
objects, belonging to $\omega$ Cen, to  M4, or to the BW sample
are called ``Template Data Sample'' (TDS);their pulsation properties are listed in  
Table~\ref{tab:sample}. The reader interested 
in a more detailed discussion of the approach adopted to derive 
periods, mean magnitudes, amplitudes and their uncertainties is referred
to \citet{stetson14a}, \citet{braga16} and \citet{braga2018}. The 
photometric properties of field RRLs were derived using the PLOESS 
polynomial fit \citep{braga2018}.
Note that the number of variables with accurate light curves in all 
three filters is limited. More specifically, the light-curve templates 
rely on a number of variables ranging from 142 for the $J$ band to 101 for the 
$H$ band and 112 for the $K_s$ band. The difference among the three bands 
is mainly caused by the paucity of $H$-band data for field and M4 RRLs. 
Moreover, the $H$- and $K_s$-band light curves have luminosity amplitudes 
that are half the $J$-band amplitudes. This means that the photometric 
scatter in the normalized light curves appears larger.



\longtab{
\begin{landscape}
\begin{longtable}{l c c c c c c c c c}
\caption{Properties of the RRLs in the Template Data Sample.}\\
\hline\hline
ID &  Period & $\langle J \rangle$ & $\langle H \rangle$ & $\langle K_s \rangle$ & $AJ$ & $AH$ & $AK_s$ & flag\tablefootmark{a} & $t_{ris}$\tablefootmark{b} \\
 &  days & mag & mag & mag & mag & mag & mag & & days \\
 \hline
\endfirsthead
\caption{continued.}\\
\hline\hline
ID &  Period & $\langle J \rangle$ & $\langle H \rangle$ & $\langle K_s \rangle$ & $AJ$ & $AH$ & $AK_s$ & flag\tablefootmark{a} & $t_{ris}$\tablefootmark{b} \\
 &  days & mag & mag & mag & mag & mag & mag & & days \\
\hline
\endhead
\hline
\endfoot
\multicolumn{10}{c}{---RRc; total used for template: 51 ($J$), 41 ($H$), 38 ($K_s$)---} \\
\wcen-V98     & 0.2805656 & 13.916$\pm$0.008 & 13.716$\pm$0.012 & 13.725$\pm$0.011 &  0.196$\pm$0.021 &  0.120$\pm$0.021 &  0.094$\pm$0.011 & 111 & 55715.6665  \\
\wcen-V19     & 0.2995517 & 13.864$\pm$0.004 & 13.661$\pm$0.004 & 13.653$\pm$0.005 &  0.169$\pm$0.017 &  0.088$\pm$0.010 &  0.111$\pm$0.009 & 111 & 49869.6627  \\
\wcen-V184    & 0.3033717 & 13.748$\pm$0.003 & 13.554$\pm$0.004 & 13.553$\pm$0.005 &  0.080$\pm$0.007 &  0.062$\pm$0.008 &  0.045$\pm$0.005 & 111 & 57049.7917  \\
\wcen-V121    & 0.3041817 & 13.689$\pm$0.004 & 13.484$\pm$0.004 & 13.483$\pm$0.005 &  0.123$\pm$0.012 &  0.064$\pm$0.008 &  0.051$\pm$0.007 & 111 & 57050.0006  \\
\wcen-V127    & 0.3052727 & 13.745$\pm$0.003 & 13.553$\pm$0.003 & 13.559$\pm$0.006 &  0.104$\pm$0.010 &  0.074$\pm$0.007 &  0.073$\pm$0.010 & 110 & 57049.6869  \\
\wcen-V119    & 0.3058754 & 13.743$\pm$0.005 & 13.543$\pm$0.007 & 13.529$\pm$0.007 &  0.161$\pm$0.017 &  0.083$\pm$0.011 &  0.058$\pm$0.010 & 110 & 57049.6907  \\
\wcen-V276    & 0.3078034 & 13.714$\pm$0.004 & 13.510$\pm$0.005 & 13.518$\pm$0.006 &  0.066$\pm$0.006 &  0.031$\pm$0.007 &  0.032$\pm$0.007 & 110 & 57049.8504  \\
TV Boo        & 0.3126494 & 10.290$\pm$0.010 &       \ldots     & 10.175$\pm$0.010 &  0.259$\pm$0.030 &       \ldots     &  0.124$\pm$0.015 & 100 & 47228.1993  \\
\wcen-V163    & 0.3132315 & 13.723$\pm$0.004 & 13.530$\pm$0.005 & 13.542$\pm$0.005 &  0.100$\pm$0.009 &  0.041$\pm$0.008 &  0.031$\pm$0.007 & 110 & 57049.8922  \\
\wcen-V169    & 0.3191135 & 13.744$\pm$0.005 & 13.550$\pm$0.006 & 13.533$\pm$0.006 &  0.102$\pm$0.008 &  0.062$\pm$0.009 &  0.062$\pm$0.009 & 111 & 52743.8624  \\
M4-V6         & 0.3205151 & 11.692$\pm$0.018 & 11.452$\pm$0.200 & 11.313$\pm$0.056 &  0.180$\pm$0.023 &       \ldots     &       \ldots     & 100 & 55412.8765  \\
\wcen-V168    & 0.3212974 & 14.175$\pm$0.006 & 13.965$\pm$0.005 & 13.959$\pm$0.005 &  0.195$\pm$0.017 &  0.113$\pm$0.012 &  0.111$\pm$0.010 & 111 & 57049.8723  \\
\wcen-V264    & 0.3213933 & 13.832$\pm$0.010 & 13.590$\pm$0.009 & 13.569$\pm$0.011 &  0.168$\pm$0.015 &  0.095$\pm$0.014 &  0.113$\pm$0.015 & 110 & 57048.7618  \\
T Sex         & 0.3247132 &  9.325$\pm$0.010 &       \ldots     &  9.152$\pm$0.010 &  0.202$\pm$0.017 &       \ldots     &  0.085$\pm$0.011 & 101 & 47228.0473  \\
\wcen-NV346   & 0.3276184 & 13.645$\pm$0.014 & 13.436$\pm$0.008 & 13.426$\pm$0.010 &  0.203$\pm$0.023 &  0.124$\pm$0.022 &  0.123$\pm$0.012 & 111 & 57049.7612  \\
\wcen-V285    & 0.3290152 & 13.700$\pm$0.003 & 13.514$\pm$0.004 & 13.509$\pm$0.005 &  0.066$\pm$0.006 &  0.038$\pm$0.007 &  0.046$\pm$0.008 & 101 & 57049.7154  \\
\wcen-V16     & 0.3301961 & 13.688$\pm$0.003 & 13.468$\pm$0.005 & 13.481$\pm$0.004 &  0.182$\pm$0.015 &  0.072$\pm$0.009 &  0.103$\pm$0.010 & 101 & 57049.5969  \\
\wcen-V124    & 0.3318616 & 13.670$\pm$0.004 & 13.465$\pm$0.004 & 13.462$\pm$0.005 &  0.176$\pm$0.022 &  0.097$\pm$0.011 &  0.118$\pm$0.014 & 111 & 57049.7567  \\
\wcen-V110    & 0.3321024 & 13.694$\pm$0.008 & 13.510$\pm$0.009 & 13.478$\pm$0.009 &  0.164$\pm$0.017 &  0.086f$\pm$0.016 &  0.086$\pm$0.014 & 110 & 57049.8486  \\
\wcen-V105    & 0.3353309 & 13.756$\pm$0.003 & 13.530$\pm$0.004 & 13.534$\pm$0.004 &  0.168$\pm$0.016 &  0.122$\pm$0.011 &  0.107$\pm$0.010 & 111 & 57049.9780  \\
\wcen-V82     & 0.3357655 & 13.591$\pm$0.004 & 13.381$\pm$0.005 & 13.386$\pm$0.007 &  0.165$\pm$0.015 &  0.106$\pm$0.012 &  0.092$\pm$0.013 & 110 & 57049.7786  \\
\wcen-V101    & 0.3409995 & 13.655$\pm$0.006 & 13.434$\pm$0.006 & 13.443$\pm$0.006 &  0.120$\pm$0.017 &  0.038$\pm$0.008 &  0.046$\pm$0.011 & 101 & 57049.9680  \\
\wcen-V126    & 0.3417339 & 13.677$\pm$0.006 & 13.464$\pm$0.005 & 13.443$\pm$0.006 &  0.222$\pm$0.020 &  0.089$\pm$0.009 &  0.099$\pm$0.010 & 111 & 57049.6277  \\
AU Vir        & 0.3432070 & 10.968$\pm$0.020 & 10.853$\pm$0.020 & 10.787$\pm$0.020 &  0.187$\pm$0.023 &  0.102$\pm$0.021 &  0.097$\pm$0.022 & 111 & 46556.1941  \\
\wcen-V64     & 0.3444460 & 13.634$\pm$0.005 & 13.409$\pm$0.005 & 13.413$\pm$0.006 &  0.195$\pm$0.018 &  0.133$\pm$0.013 &  0.106$\pm$0.012 & 111 & 58190.9600  \\
\wcen-V156    & 0.3592530 & 13.585$\pm$0.004 & 13.358$\pm$0.010 & 13.345$\pm$0.006 &  0.152$\pm$0.014 &  0.123$\pm$0.021 &  0.074$\pm$0.009 & 111 & 57050.1801  \\
\wcen-V89     & 0.3740372 & 13.576$\pm$0.008 & 13.316$\pm$0.008 & 13.276$\pm$0.012 &  0.177$\pm$0.017 &  0.116$\pm$0.014 &  0.123$\pm$0.019 & 111 & 57050.1179  \\
\wcen-V145    & 0.3741985 & 13.599$\pm$0.007 & 13.390$\pm$0.010 & 13.334$\pm$0.009 &  0.156$\pm$0.013 &  0.142$\pm$0.017 &  0.053$\pm$0.012 & 110 & 57049.5704  \\
\wcen-V10     & 0.3747561 & 13.551$\pm$0.006 & 13.316$\pm$0.017 & 13.313$\pm$0.009 &  0.153$\pm$0.013 &  0.092$\pm$0.013 &  0.080$\pm$0.010 & 110 & 57049.9700  \\
\wcen-V14     & 0.3771026 & 13.575$\pm$0.003 & 13.352$\pm$0.005 & 13.347$\pm$0.005 &  0.205$\pm$0.018 &  0.113$\pm$0.011 &  0.110$\pm$0.009 & 111 & 57049.8382  \\
\wcen-NV350   & 0.3791090 & 13.530$\pm$0.012 & 13.324$\pm$0.011 & 13.284$\pm$0.011 &  0.154$\pm$0.020 &  0.044$\pm$0.014 &  0.098$\pm$0.017 & 101 & 51276.6883  \\
\wcen-V36     & 0.3799093 & 13.529$\pm$0.004 & 13.305$\pm$0.008 & 13.295$\pm$0.004 &  0.188$\pm$0.015 &  0.113$\pm$0.011 &  0.112$\pm$0.015 & 111 & 57049.7055  \\
\wcen-V72     & 0.3845045 & 13.545$\pm$0.004 & 13.303$\pm$0.008 & 13.318$\pm$0.005 &  0.181$\pm$0.013 &  0.130$\pm$0.019 &  0.102$\pm$0.010 & 111 & 57049.8543  \\
\wcen-V35     & 0.3868332 & 13.530$\pm$0.007 & 13.283$\pm$0.008 & 13.290$\pm$0.008 &  0.175$\pm$0.013 &  0.134$\pm$0.015 &  0.097$\pm$0.012 & 101 & 57049.7183  \\
\wcen-V81     & 0.3893851 & 13.535$\pm$0.004 & 13.300$\pm$0.008 & 13.285$\pm$0.006 &  0.177$\pm$0.013 &  0.099$\pm$0.014 &  0.110$\pm$0.009 & 111 & 57050.0175  \\
\wcen-V70     & 0.3908107 & 13.517$\pm$0.005 & 13.275$\pm$0.006 & 13.268$\pm$0.005 &  0.179$\pm$0.028 &  0.125$\pm$0.017 &  0.102$\pm$0.011 & 111 & 57050.0154  \\
\wcen-V136    & 0.3919260 & 13.491$\pm$0.009 & 13.245$\pm$0.008 & 13.248$\pm$0.010 &  0.138$\pm$0.020 &  0.163$\pm$0.021 &  0.081$\pm$0.010 & 110 & 57049.7801  \\
\wcen-V39     & 0.3933860 & 13.570$\pm$0.005 & 13.328$\pm$0.007 & 13.320$\pm$0.007 &  0.179$\pm$0.015 &  0.112$\pm$0.011 &  0.117$\pm$0.010 & 111 & 57050.1432  \\
\wcen-V87     & 0.3959409 & 13.529$\pm$0.006 & 13.282$\pm$0.012 & 13.249$\pm$0.011 &  0.149$\pm$0.013 &  0.081$\pm$0.017 &  0.087$\pm$0.015 & 101 & 57049.7780  \\
\wcen-V22     & 0.3961653 & 13.536$\pm$0.005 & 13.292$\pm$0.008 & 13.289$\pm$0.006 &  0.174$\pm$0.014 &  0.120$\pm$0.012 &  0.115$\pm$0.012 & 111 & 57050.0072  \\
\wcen-V30     & 0.4039713 & 13.521$\pm$0.006 & 13.263$\pm$0.009 & 13.246$\pm$0.008 &  0.151$\pm$0.009 &  0.073$\pm$0.011 &  0.105$\pm$0.013 & 111 & 57049.9279  \\
\wcen-V157    & 0.4058791 & 13.510$\pm$0.017 & 13.312$\pm$0.008 & 13.182$\pm$0.011 &  0.184$\pm$0.026 &  0.090$\pm$0.012 &  0.130$\pm$0.015 & 111 & 57049.8212  \\
\wcen-V66     & 0.4072308 & 13.481$\pm$0.006 & 13.239$\pm$0.006 & 13.230$\pm$0.006 &  0.163$\pm$0.015 &  0.113$\pm$0.014 &  0.097$\pm$0.009 & 111 & 57049.7452  \\
\wcen-V155    & 0.4139333 & 13.487$\pm$0.008 & 13.241$\pm$0.010 & 13.203$\pm$0.009 &  0.172$\pm$0.014 &  0.096$\pm$0.016 &  0.062$\pm$0.011 & 100 & 57049.8221  \\
\wcen-NV354   & 0.4194950 & 13.460$\pm$0.007 & 13.202$\pm$0.009 & 13.194$\pm$0.009 &  0.158$\pm$0.015 &  0.128$\pm$0.016 &  0.079$\pm$0.013 & 111 & 51277.1019  \\
\wcen-V117    & 0.4216425 & 13.482$\pm$0.005 & 13.237$\pm$0.009 & 13.209$\pm$0.007 &  0.197$\pm$0.022 &  0.129$\pm$0.016 &  0.079$\pm$0.013 & 111 & 57049.6234  \\
\wcen-V75     & 0.4222067 & 13.424$\pm$0.004 & 13.159$\pm$0.006 & 13.164$\pm$0.004 &  0.169$\pm$0.015 &  0.128$\pm$0.015 &  0.119$\pm$0.010 & 111 & 57049.9286  \\
\wcen-V147    & 0.4225113 & 13.393$\pm$0.006 & 13.180$\pm$0.007 & 13.152$\pm$0.007 &  0.182$\pm$0.021 &  0.131$\pm$0.016 &  0.110$\pm$0.013 & 111 & 57049.5814  \\
\wcen-V77     & 0.4259824 & 13.448$\pm$0.003 & 13.190$\pm$0.006 & 13.184$\pm$0.006 &  0.171$\pm$0.022 &  0.099$\pm$0.012 &  0.099$\pm$0.012 & 111 & 57049.9350  \\
\wcen-V24     & 0.4622215 & 13.396$\pm$0.005 & 13.125$\pm$0.008 & 13.123$\pm$0.006 &  0.162$\pm$0.015 &  0.120$\pm$0.013 &  0.093$\pm$0.013 & 111 & 57049.7840  \\
\wcen-V123    & 0.4749551 & 13.408$\pm$0.003 & 13.152$\pm$0.004 & 13.149$\pm$0.006 &  0.150$\pm$0.013 &  0.095$\pm$0.009 &  0.092$\pm$0.010 & 111 & 57050.0919  \\

\multicolumn{10}{c}{---RRab1; total used for template: 29 ($J$), 11 ($H$), 16 ($K_s$)---}                                                               \\
AV Peg         & 0.3903912 &  9.573$\pm$0.010 &       \ldots     &  9.318$\pm$0.010 &  0.404$\pm$0.035 &       \ldots     &  0.292$\pm$0.030 & 101 & 47123.7076   \\
V445 Oph       & 0.3970227 &  9.606$\pm$0.005 &  9.335$\pm$0.005 &  9.222$\pm$0.005 &  0.334$\pm$0.024 &  0.283$\pm$0.019 &  0.263$\pm$0.019 & 111 & 46981.3385   \\
RR Gem         & 0.3972830 & 10.474$\pm$0.020 &       \ldots     & 10.211$\pm$0.010 &  0.394$\pm$0.051 &       \ldots     &  0.287$\pm$0.029 & 101 & 47198.7870   \\
AR Per         & 0.4255984 &  8.989$\pm$0.010 &       \ldots     &  8.626$\pm$0.010 &  0.367$\pm$0.034 &       \ldots     &  0.278$\pm$0.025 & 101 & 47123.6655   \\
SW And         & 0.4422660 &  8.757$\pm$0.020 &  8.581$\pm$0.020 &  8.506$\pm$0.010 &  0.424$\pm$0.026 &  0.314$\pm$0.024 &  0.289$\pm$0.011 & 111 & 47065.7327   \\
M4-V12         & 0.4461098 & 11.726$\pm$0.040 & 11.362$\pm$0.200 & 11.253$\pm$0.033 &  0.475$\pm$0.075 &       \ldots     &       \ldots     & 100 & 55412.7833   \\
RR Leo         & 0.4523765 &  9.911$\pm$0.011 &       \ldots     &  9.655$\pm$0.010 &  0.554$\pm$0.046 &       \ldots     &  0.340$\pm$0.029 & 101 & 47227.5782   \\
M4-V19         & 0.4678111 & 11.603$\pm$0.067 & 11.226$\pm$0.200 & 11.181$\pm$0.046 &  0.462$\pm$0.076 &       \ldots     &       \ldots     & 100 & 55412.3131   \\
M4-V21         & 0.4720074 & 11.591$\pm$0.028 & 11.219$\pm$0.200 & 11.137$\pm$0.013 &  0.413$\pm$0.094 &       \ldots     &       \ldots     & 100 & 55412.6133   \\
DX Del         & 0.4726174 &  8.965$\pm$0.021 &  8.813$\pm$0.020 &  8.710$\pm$0.020 &  0.388$\pm$0.035 &  0.272$\pm$0.024 &  0.256$\pm$0.023 & 111 & 47300.6712   \\
\wcen-V112     & 0.4743560 & 13.760$\pm$0.016 & 13.460$\pm$0.013 & 13.482$\pm$0.016 &  0.623$\pm$0.079 &  0.404$\pm$0.040 &  0.296$\pm$0.018 & 111 & 50985.5829   \\
UU Vir         & 0.4755542 &  9.763$\pm$0.016 &  9.669$\pm$0.015 &  9.485$\pm$0.005 &  0.464$\pm$0.060 &       \ldots     &  0.352$\pm$0.023 & 101 & 46884.0244   \\
M4-V18         & 0.4787920 & 11.632$\pm$0.037 & 11.399$\pm$0.200 & 11.077$\pm$0.044 &  0.426$\pm$0.059 &       \ldots     &       \ldots     & 100 & 55412.8648   \\
BB Pup         & 0.4805437 & 11.168$\pm$0.020 & 10.966$\pm$0.020 & 10.883$\pm$0.020 &  0.437$\pm$0.045 &  0.334$\pm$0.022 &  0.323$\pm$0.029 & 111 & 47193.3909   \\
M4-V10         & 0.4907175 & 11.515$\pm$0.024 & 11.351$\pm$0.200 & 11.199$\pm$0.063 &  0.329$\pm$0.030 &       \ldots     &       \ldots     & 100 & 55412.2533   \\
M4-V11         & 0.4932087 & 11.640$\pm$0.019 & 11.219$\pm$0.200 & 11.220$\pm$0.034 &  0.362$\pm$0.037 &       \ldots     &       \ldots     & 100 & 55412.9542   \\
M4-V7          & 0.4987872 & 11.592$\pm$0.037 & 11.165$\pm$0.200 & 11.138$\pm$0.018 &  0.294$\pm$0.039 &       \ldots     &       \ldots     & 100 & 55412.2209   \\
\wcen-V74      & 0.5032142 & 13.604$\pm$0.005 & 13.379$\pm$0.014 & 13.354$\pm$0.007 &  0.544$\pm$0.048 &  0.313$\pm$0.035 &  0.338$\pm$0.038 & 111 & 55711.7447   \\
M4-V8          & 0.5082236 & 11.573$\pm$0.029 & 11.396$\pm$0.200 & 11.152$\pm$0.098 &  0.361$\pm$0.055 &       \ldots     &       \ldots     & 100 & 55412.5025   \\
\wcen-V23      & 0.5108703 & 13.717$\pm$0.005 & 13.445$\pm$0.010 & 13.396$\pm$0.008 &  0.441$\pm$0.054 &  0.282$\pm$0.033 &  0.357$\pm$0.029 & 111 & 49866.6429   \\
\wcen-V5       & 0.5152800 & 13.668$\pm$0.012 & 13.488$\pm$0.019 & 13.409$\pm$0.012 &  0.441$\pm$0.028 &  0.354$\pm$0.043 &  0.334$\pm$0.016 & 111 & 49865.6237   \\
\wcen-V59      & 0.5185514 & 13.624$\pm$0.017 & 13.397$\pm$0.013 & 13.387$\pm$0.013 &  0.385$\pm$0.066 &  0.312$\pm$0.055 &       \ldots     & 110 & 57049.8804   \\
\wcen-V8       & 0.5213259 & 13.577$\pm$0.004 & 13.341$\pm$0.009 & 13.305$\pm$0.005 &  0.449$\pm$0.056 &  0.344$\pm$0.018 &  0.295$\pm$0.017 & 111 & 49824.5018   \\
M4-V2          & 0.5356819 & 11.573$\pm$0.026 & 11.102$\pm$0.200 & 11.068$\pm$0.030 &  0.374$\pm$0.033 &       \ldots     &  0.275$\pm$0.027 & 101 & 55412.1842   \\
M4-V26         & 0.5412174 & 11.508$\pm$0.047 & 11.239$\pm$0.200 & 11.020$\pm$0.029 &  0.464$\pm$0.058 &       \ldots     &       \ldots     & 100 & 55412.4631   \\
M4-V36         & 0.5413092 & 11.566$\pm$0.042 & 11.213$\pm$0.200 & 11.084$\pm$0.028 &  0.281$\pm$0.039 &       \ldots     &       \ldots     & 100 & 55412.8657   \\
M4-V16         & 0.5425483 & 11.490$\pm$0.041 & 11.141$\pm$0.200 & 10.990$\pm$0.036 &  0.322$\pm$0.034 &       \ldots     &       \ldots     & 100 & 55412.2979   \\
M4-V24         & 0.5467833 & 11.467$\pm$0.026 & 11.131$\pm$0.200 & 11.123$\pm$0.097 &  0.373$\pm$0.055 &       \ldots     &       \ldots     & 100 & 55412.4631   \\
\wcen-V120     & 0.5485474 & 13.615$\pm$0.005 & 13.332$\pm$0.009 & 13.328$\pm$0.007 &  0.488$\pm$0.061 &  0.321$\pm$0.039 &  0.293$\pm$0.032 & 111 & 51218.8794   \\
\multicolumn{10}{c}{---RRab2; total used for template: 46 ($J$), 35 ($H$), 45 ($K_s$)---}                                                             \\
\wcen-V100     & 0.5527477 & 13.631$\pm$0.006 & 13.341$\pm$0.013 & 13.329$\pm$0.008 &  0.502$\pm$0.062 &  0.294$\pm$0.027 &  0.299$\pm$0.021 & 111 & 50975.6290  \\
RR Cet         & 0.5529680 &  8.786$\pm$0.010 &       \ldots     &  8.513$\pm$0.010 &  0.456$\pm$0.037 &       \ldots     &  0.284$\pm$0.021 & 101 & 47123.7458  \\
TU UMa         & 0.5576992 &  8.898$\pm$0.020 &  8.704$\pm$0.020 &  8.636$\pm$0.020 &  0.441$\pm$0.028 &  0.314$\pm$0.025 &  0.291$\pm$0.024 & 111 & 47228.0206  \\
\wcen-V67      & 0.5644486 & 13.593$\pm$0.004 & 13.323$\pm$0.006 & 13.310$\pm$0.005 &  0.506$\pm$0.048 &  0.326$\pm$0.029 &  0.298$\pm$0.019 & 111 & 57049.4659  \\
\wcen-V44      & 0.5675378 & 13.610$\pm$0.005 & 13.326$\pm$0.006 & 13.304$\pm$0.007 &  0.456$\pm$0.044 &  0.306$\pm$0.030 &  0.300$\pm$0.020 & 111 & 50971.6089  \\
\wcen-V106     & 0.5699029 & 13.465$\pm$0.011 & 13.230$\pm$0.020 & 13.170$\pm$0.010 &  0.513$\pm$0.048 &  0.301$\pm$0.052 &       \ldots     & 100 & 51305.4665  \\
RV Oct         & 0.5711625 &  9.832$\pm$0.020 &  9.587$\pm$0.020 &  9.492$\pm$0.020 &  0.510$\pm$0.051 &  0.338$\pm$0.034 &  0.318$\pm$0.033 & 111 & 47690.0642  \\
M4-V9          & 0.5718945 & 11.453$\pm$0.035 & 11.018$\pm$0.200 & 10.965$\pm$0.038 &  0.539$\pm$0.031 &       \ldots     &       \ldots     & 100 & 55412.7595  \\
\wcen-V113     & 0.5733764 & 13.516$\pm$0.009 & 13.278$\pm$0.010 & 13.215$\pm$0.008 &  0.515$\pm$0.066 &  0.332$\pm$0.042 &  0.333$\pm$0.022 & 111 & 50978.5866  \\
\wcen-V51      & 0.5741424 & 13.466$\pm$0.006 & 13.179$\pm$0.009 & 13.161$\pm$0.008 &  0.414$\pm$0.040 &  0.312$\pm$0.021 &  0.326$\pm$0.020 & 111 & 51276.8553  \\
WY Ant         & 0.5743302 &  9.892$\pm$0.020 &  9.668$\pm$0.020 &  9.598$\pm$0.020 &  0.398$\pm$0.038 &  0.270$\pm$0.027 &  0.285$\pm$0.028 & 111 & 47193.8534  \\
\wcen-V73      & 0.5752037 & 13.489$\pm$0.004 & 13.236$\pm$0.007 & 13.210$\pm$0.009 &  0.449$\pm$0.048 &  0.309$\pm$0.027 &  0.271$\pm$0.025 & 111 & 57049.4408  \\
\wcen-V55      & 0.5816639 & 13.614$\pm$0.004 & 13.283$\pm$0.108 & 13.291$\pm$0.005 &  0.410$\pm$0.036 &       \ldots     &  0.250$\pm$0.015 & 101 & 57049.9012  \\
RX Eri         & 0.5872189 &  8.653$\pm$0.016 &       \ldots     &  8.356$\pm$0.011 &  0.387$\pm$0.062 &       \ldots     &  0.284$\pm$0.030 & 001 & 47226.4723  \\
\wcen-V25      & 0.5885155 & 13.458$\pm$0.009 & 13.194$\pm$0.008 & 13.157$\pm$0.012 &  0.437$\pm$0.054 &  0.364$\pm$0.045 &  0.293$\pm$0.029 & 111 & 50921.8161  \\
TT Lyn         & 0.5974357 &  8.894$\pm$0.010 &       \ldots     &  8.609$\pm$0.010 &  0.334$\pm$0.034 &       \ldots     &  0.248$\pm$0.024 & 101 & 44563.7163  \\
\wcen-V33      & 0.6023333 & 13.409$\pm$0.005 & 13.146$\pm$0.010 & 13.141$\pm$0.006 &  0.468$\pm$0.045 &  0.315$\pm$0.026 &  0.288$\pm$0.021 & 111 & 57050.1249  \\
\wcen-V90      & 0.6034053 & 13.420$\pm$0.016 & 13.194$\pm$0.012 & 13.102$\pm$0.010 &  0.545$\pm$0.069 &  0.370$\pm$0.030 &  0.274$\pm$0.017 & 111 & 50973.5512  \\
\wcen-V49      & 0.6046495 & 13.474$\pm$0.003 & 13.197$\pm$0.005 & 13.163$\pm$0.005 &  0.422$\pm$0.040 &  0.273$\pm$0.027 &  0.282$\pm$0.016 & 111 & 57049.5968  \\
UU Cet         & 0.6060382 & 11.098$\pm$0.010 & 10.888$\pm$0.010 & 10.807$\pm$0.010 &  0.305$\pm$0.024 &  0.247$\pm$0.018 &  0.253$\pm$0.021 & 111 & 48194.1587  \\
\wcen-V79      & 0.6082869 & 13.461$\pm$0.004 & 13.205$\pm$0.011 & 13.152$\pm$0.007 &  0.469$\pm$0.041 &  0.325$\pm$0.026 &  0.306$\pm$0.021 & 111 & 49922.5029  \\
\wcen-V118     & 0.6116195 & 13.224$\pm$0.013 & 12.949$\pm$0.013 & 12.879$\pm$0.012 &  0.414$\pm$0.042 &  0.255$\pm$0.043 &  0.280$\pm$0.021 & 101 & 50972.4694  \\
M4-V27         & 0.6120183 & 11.414$\pm$0.041 & 10.952$\pm$0.200 & 10.898$\pm$0.008 &  0.390$\pm$0.041 &       \ldots     &       \ldots     & 100 & 55412.7165  \\
\wcen-V20      & 0.6155878 & 13.409$\pm$0.006 & 13.114$\pm$0.007 & 13.100$\pm$0.007 &  0.465$\pm$0.039 &  0.261$\pm$0.021 &  0.311$\pm$0.019 & 111 & 57049.9046  \\ 
\wcen-V62      & 0.6197964 & 13.398$\pm$0.008 & 13.134$\pm$0.006 & 13.078$\pm$0.007 &  0.457$\pm$0.058 &  0.369$\pm$0.040 &  0.288$\pm$0.024 & 111 & 57049.8559  \\
M4-V5          & 0.6224011 & 11.405$\pm$0.022 & 11.172$\pm$0.200 & 10.943$\pm$0.010 &  0.189$\pm$0.025 &       \ldots     &       \ldots     & 100 & 55412.5487  \\
\wcen-V96      & 0.6245217 & 13.354$\pm$0.014 & 13.094$\pm$0.008 & 13.033$\pm$0.008 &  0.365$\pm$0.028 &  0.241$\pm$0.023 &  0.309$\pm$0.023 & 111 & 57049.3019  \\
SS Leo         & 0.6263448 & 10.184$\pm$0.005 &  9.979$\pm$0.005 &  9.904$\pm$0.005 &  0.541$\pm$0.034 &  0.322$\pm$0.018 &  0.332$\pm$0.018 & 111 & 46868.5395  \\
M4-V35         & 0.6270237 & 11.411$\pm$0.038 & 11.065$\pm$0.024 & 10.905$\pm$0.030 &  0.168$\pm$0.023 &  0.248$\pm$0.038 &  0.246$\pm$0.027 & 001 & 55412.5110  \\
\wcen-V4       & 0.6273185 & 13.378$\pm$0.008 & 13.106$\pm$0.008 & 13.077$\pm$0.009 &  0.494$\pm$0.054 &  0.333$\pm$0.031 &  0.316$\pm$0.020 & 111 & 57049.5110  \\
\wcen-V115     & 0.6304695 & 13.401$\pm$0.005 & 13.116$\pm$0.011 & 13.098$\pm$0.007 &  0.440$\pm$0.035 &  0.195$\pm$0.015 &  0.258$\pm$0.014 & 101 & 57049.3648  \\
\wcen-V146     & 0.6330968 & 13.426$\pm$0.022 & 13.219$\pm$0.018 & 13.101$\pm$0.017 &  0.462$\pm$0.055 &  0.283$\pm$0.044 &  0.295$\pm$0.029 & 101 & 54705.3835  \\
\wcen-V40      & 0.6340978 & 13.407$\pm$0.006 & 13.129$\pm$0.008 & 13.095$\pm$0.009 &  0.489$\pm$0.061 &  0.313$\pm$0.031 &  0.319$\pm$0.029 & 111 & 49863.7202  \\
\wcen-V122     & 0.6349212 & 13.390$\pm$0.006 & 13.086$\pm$0.008 & 13.081$\pm$0.006 &  0.442$\pm$0.055 &  0.196$\pm$0.018 &  0.295$\pm$0.020 & 111 & 57050.2454  \\
W Tuc          & 0.6422299 & 10.579$\pm$0.010 & 10.394$\pm$0.010 & 10.308$\pm$0.010 &  0.517$\pm$0.038 &  0.353$\pm$0.026 &  0.329$\pm$0.026 & 111 & 47493.8857  \\
\wcen-V86      & 0.6478414 & 13.379$\pm$0.006 & 13.075$\pm$0.006 & 13.051$\pm$0.008 &  0.430$\pm$0.053 &  0.386$\pm$0.048 &  0.290$\pm$0.024 & 111 & 50978.5945  \\
X Ari          & 0.6511784 &  8.289$\pm$0.006 &  8.021$\pm$0.005 &  7.930$\pm$0.005 &  0.422$\pm$0.033 &  0.311$\pm$0.034 &  0.287$\pm$0.024 & 111 & 45639.5108  \\
\wcen-V69      & 0.6532209 & 13.399$\pm$0.004 & 13.114$\pm$0.006 & 13.099$\pm$0.004 &  0.376$\pm$0.033 &  0.273$\pm$0.022 &  0.272$\pm$0.023 & 111 & 57050.3003  \\
\wcen-V132     & 0.6556445 & 13.344$\pm$0.013 & 13.061$\pm$0.008 & 12.987$\pm$0.011 &  0.480$\pm$0.040 &  0.304$\pm$0.026 &  0.260$\pm$0.019 & 111 & 54705.1500  \\
SU Dra         & 0.6603893 &  8.882$\pm$0.100 &       \ldots     &  8.627$\pm$0.100 &  0.440$\pm$0.104 &       \ldots     &  0.288$\pm$0.102 & 101 & 47227.8921  \\
\wcen-V41      & 0.6629338 & 13.374$\pm$0.007 & 13.082$\pm$0.007 & 13.043$\pm$0.009 &  0.381$\pm$0.027 &  0.339$\pm$0.028 &  0.328$\pm$0.019 & 111 & 57049.8439  \\
\wcen-V13      & 0.6690484 & 13.315$\pm$0.008 & 13.039$\pm$0.008 & 13.011$\pm$0.008 &  0.405$\pm$0.026 &  0.330$\pm$0.028 &  0.307$\pm$0.015 & 111 & 57048.9950  \\
\wcen-V114     & 0.6753083 & 13.355$\pm$0.011 & 13.123$\pm$0.013 & 13.031$\pm$0.009 &  0.373$\pm$0.028 &  0.272$\pm$0.029 &  0.269$\pm$0.018 & 111 & 50984.5544  \\
\wcen-V139     & 0.6768713 & 13.129$\pm$0.014 & 12.794$\pm$0.011 & 12.753$\pm$0.007 &  0.355$\pm$0.045 &  0.155$\pm$0.017 &  0.183$\pm$0.015 & 001 & 50972.5424  \\
\wcen-V149     & 0.6827238 & 13.327$\pm$0.007 & 13.046$\pm$0.010 & 13.032$\pm$0.008 &  0.420$\pm$0.053 &  0.283$\pm$0.026 &  0.352$\pm$0.030 & 111 & 57049.9876  \\
\wcen-V46      & 0.6869624 & 13.325$\pm$0.003 & 13.040$\pm$0.003 & 13.016$\pm$0.005 &  0.415$\pm$0.045 &  0.288$\pm$0.021 &  0.274$\pm$0.016 & 111 & 49821.6201  \\
\wcen-V102     & 0.6913961 & 13.329$\pm$0.004 & 13.030$\pm$0.006 & 12.993$\pm$0.006 &  0.453$\pm$0.044 &  0.209$\pm$0.018 &  0.290$\pm$0.017 & 111 & 50975.5249  \\
\wcen-V97      & 0.6918899 & 13.321$\pm$0.005 & 13.022$\pm$0.009 & 12.995$\pm$0.008 &  0.417$\pm$0.026 &  0.308$\pm$0.021 &  0.280$\pm$0.016 & 111 & 57049.3125  \\
\wcen-V141     & 0.6974361 & 13.275$\pm$0.012 & 12.977$\pm$0.012 & 12.923$\pm$0.011 &  0.317$\pm$0.023 &  0.170$\pm$0.018 &  0.233$\pm$0.017 & 111 & 50975.6468  \\
\multicolumn{10}{c}{---RRab3; total used for template: 16 ($J$), 14 ($H$), 13 ($K_s$)---}                                                             \\
\wcen-V7       & 0.7130342 & 13.289$\pm$0.005 & 13.018$\pm$0.006 & 12.976$\pm$0.006 &  0.428$\pm$0.041 &  0.305$\pm$0.025 &  0.312$\pm$0.028 & 111 & 49082.5766 \\
VY Ser         & 0.7140956 &  9.070$\pm$0.005 &  8.835$\pm$0.005 &  8.769$\pm$0.005 &  0.277$\pm$0.015 &  0.247$\pm$0.015 &  0.248$\pm$0.011 & 111 & 47655.8685 \\
\wcen-V34      & 0.7339550 & 13.256$\pm$0.007 & 12.967$\pm$0.005 & 12.931$\pm$0.007 &  0.325$\pm$0.029 &  0.248$\pm$0.020 &  0.281$\pm$0.019 & 111 & 52443.5106 \\
\wcen-V109     & 0.7440992 & 13.259$\pm$0.010 & 12.964$\pm$0.011 & 12.919$\pm$0.009 &  0.423$\pm$0.038 &  0.216$\pm$0.026 &  0.201$\pm$0.015 & 111 & 50984.5494 \\
\wcen-V111     & 0.7629011 & 13.234$\pm$0.012 & 12.984$\pm$0.014 & 12.861$\pm$0.010 &  0.359$\pm$0.030 &  0.300$\pm$0.028 &  0.217$\pm$0.018 & 111 & 57049.8840 \\
\wcen-V99      & 0.7661794 & 13.172$\pm$0.009 & 12.892$\pm$0.014 & 12.854$\pm$0.009 &  0.454$\pm$0.035 &  0.332$\pm$0.081 &  0.247$\pm$0.018 & 100 & 57050.0062 \\
\wcen-V54      & 0.7729093 & 13.232$\pm$0.004 & 12.917$\pm$0.005 & 12.884$\pm$0.004 &  0.308$\pm$0.033 &  0.281$\pm$0.035 &  0.253$\pm$0.032 & 111 & 53866.3923 \\
\wcen-V38      & 0.7790590 & 13.219$\pm$0.003 & 12.916$\pm$0.006 & 12.886$\pm$0.005 &  0.197$\pm$0.027 &  0.185$\pm$0.014 &  0.214$\pm$0.015 & 111 & 49869.7186 \\
\wcen-V26      & 0.7847215 & 13.222$\pm$0.005 & 12.927$\pm$0.007 & 12.875$\pm$0.006 &  0.317$\pm$0.027 &  0.276$\pm$0.024 &  0.230$\pm$0.020 & 110 & 57050.0872 \\
\wcen-V57      & 0.7944223 & 13.213$\pm$0.003 & 12.887$\pm$0.005 & 12.874$\pm$0.004 &  0.269$\pm$0.018 &  0.204$\pm$0.020 &  0.230$\pm$0.015 & 111 & 57049.3772 \\
\wcen-V268     & 0.8129334 & 13.186$\pm$0.008 & 12.864$\pm$0.005 & 12.831$\pm$0.011 &  0.260$\pm$0.023 &  0.227$\pm$0.025 &  0.189$\pm$0.022 & 111 & 51305.5583 \\
\wcen-V63      & 0.8259598 & 13.190$\pm$0.003 & 12.876$\pm$0.004 & 12.849$\pm$0.004 &  0.228$\pm$0.022 &  0.210$\pm$0.021 &  0.199$\pm$0.020 & 111 & 57050.0361 \\
\wcen-V128     & 0.8349918 & 13.131$\pm$0.005 & 12.833$\pm$0.008 & 12.764$\pm$0.091 &  0.300$\pm$0.023 &  0.236$\pm$0.021 &       \ldots     & 110 & 57050.0308 \\
\wcen-V144     & 0.8353219 & 13.140$\pm$0.013 & 12.866$\pm$0.007 & 12.768$\pm$0.008 &  0.263$\pm$0.029 &  0.222$\pm$0.020 &  0.278$\pm$0.022 & 101 & 54705.3353 \\
\wcen-V3       & 0.8412616 & 13.182$\pm$0.007 & 12.895$\pm$0.007 & 12.864$\pm$0.009 &  0.353$\pm$0.029 &  0.300$\pm$0.029 &  0.285$\pm$0.027 & 111 & 57049.7960 \\
\wcen-V104     & 0.8675259 & 13.217$\pm$0.011 & 12.884$\pm$0.010 & 12.856$\pm$0.011 &  0.174$\pm$0.015 &  0.199$\pm$0.017 &  0.168$\pm$0.014 & 111 & 57049.5137 \\
\end{longtable}
 \tablefoot{
 \tablefoottext{a}{Three-digit flag that indicates whether the variable was
 used for the light curve templates. The three digits correspond to the $J$, $H$ and $K_s$
 template, respectively. ``1'' indicates that the variable was used for the template, 
 ``0'' that it was not.}
 \tablefoottext{b}{Heliocentric Julian Day -- 2,400,000 days.}}
 \label{tab:sample}
\end{landscape}
}

\section{Near-Infrared light curve templates}\label{sect_template}
\subsection{Selection of the period bins}

\begin{figure*}[!htbp]
\centering
\includegraphics[width=12cm]{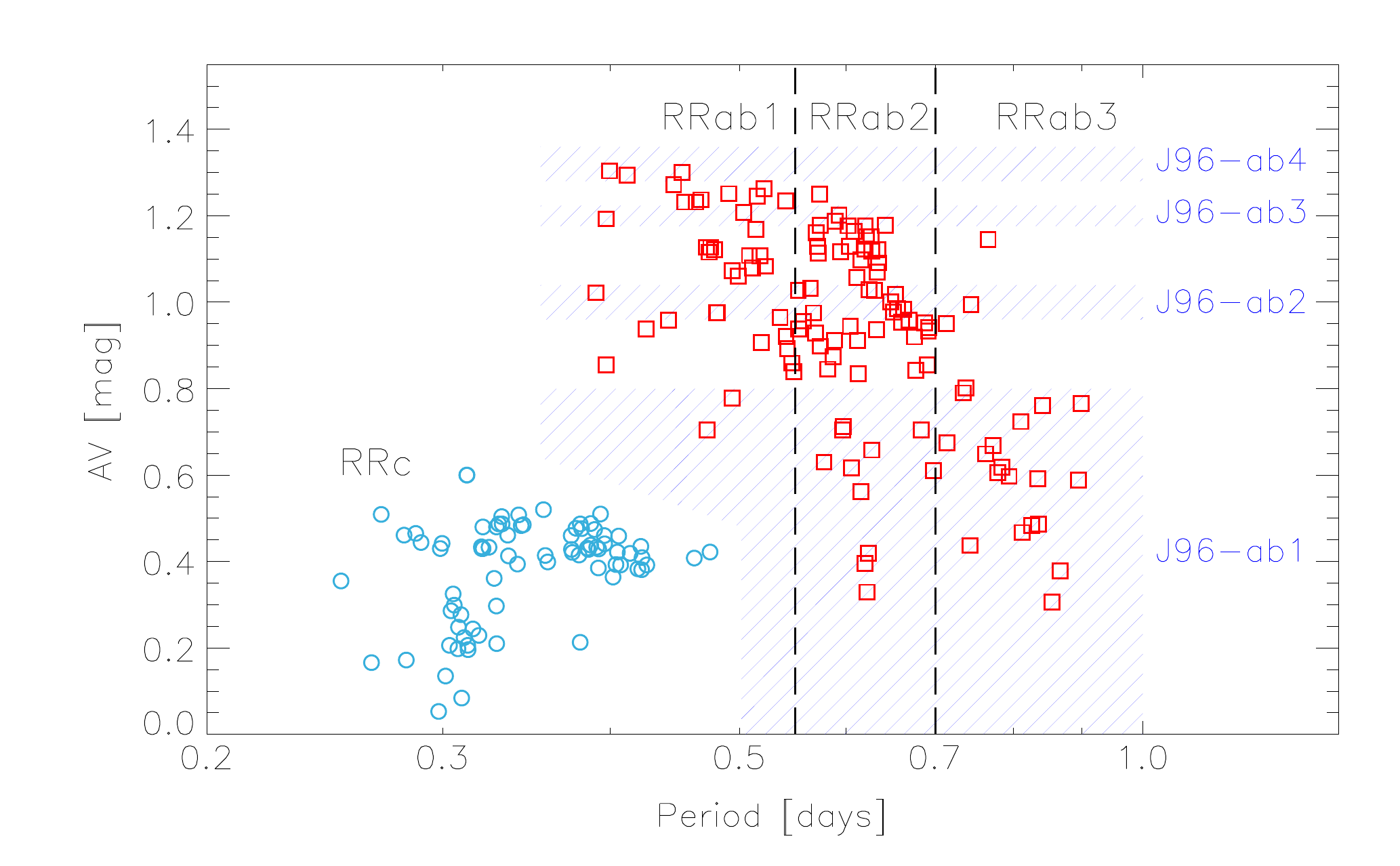}
\caption{Bailey diagram, $V$-band amplitude versus logarithmic period, 
for $\omega$ Cen RRLs. Blue circles mark RRc variables, while red squares 
mark RRab variables. The ranges in period for the RRab light-curve 
templates are indicated by vertical black dashed lines. 
The blue striped areas show the ranges in amplitude adopted for the 
light curve templates by J96. Note that they provided thresholds in 
the $B$ band, but here they have been rescaled by 1.25, i.e., the typical 
amplitude ratio ($AB/AV$) for RRab variables \citep{braga16}.}
\label{fig:bailey}
\end{figure*}

We defined the template bins according to the pulsation period of the variable. 
The reasons are manifold.
{\em i)} -- The period is a solid observable, since it can be firmly estimated also for 
variables showing multi-periodicity (Blazhko, mixed-mode). The same statement does 
not apply to the luminosity amplitude adopted by J96.  
{\em ii)} -- The period range covered by TDS variables (0.28-0.47 days for RRcs
and 0.39-0.87 days for RRabs) is much larger than the RRL sample 
adopted by J96 (0.25-0.34 days for RRc and 0.39-0.66 days for RRabs). 
{\em iii)} -- The optical luminosity amplitude is not a linear function of
the period \citep{cacciari05,kunder13}. Data plotted in the Bailey diagram 
(logarithmic period versus luminosity amplitudes, Fig.~\ref{fig:bailey}) 
clearly show that RRab and RRc variables with similar amplitudes can have 
significantly different pulsation periods. 
{\em iv)} -- The period is tightly correlated with the intrinsic parameters 
(stellar mass, luminosity, effective temperature) of the variable \citep{bono94b}.  

We have checked that, for RRc variables, one template bin
is enough because the shape of the 
light curve in the NIR bands is almost sinusoidal over the whole 
period range. On the other hand, the RRab variables were divided 
into three period bins for following reasons.
{\em i)} -- To improve the sampling along the light curve template we 
required at least ten variables per bin for each band, limiting 
the number of possible period bins.  
{\em ii)} -- The RRab variables display in the optical Bailey diagram 
a parabolic trend \citep{cacciari05} when moving from shorter to longer 
periods. Data plotted in Fig.~\ref{fig:bailey} show that the maximum is 
located at approximately 0.55 days. 
{\em iii)} -- We also decided to cut the sample at 0.7 days, because 
empirical evidence indicates that a transition---both in Blazhko properties 
and in the optical-to-NIR amplitude ratios---takes place across this boundary 
\citep{prudil2017,braga2018}. 

This means that the RRab variables were split into short (RRab1, P$\le$0.55 days), 
medium (RRab2, 0.55$<$P$<$0.70 days) and long (RRab3, P$\ge0.70$ days) 
period bins, while the RRc constitute a single period bin (0.28$<$P$<$0.47 days). 

It is worth mentioning that we could have extended the period range of the 
RRab3 template up to 0.9 days, by including \wcen-V91 and \wcen-V150.
However, both variables have light curves with a significantly different 
shape when compared to the other RRLs in RRab3 sub-sample. More statistics 
are required to establish whether RRLs  with periods longer than 
$\sim$0.87 days require a separated template bin. 

Finally we mention that the number of phase points per template bin is 
1,226 ($J$), 698 ($H$), 959 ($K_s$) for the RRc template; 
931 ($J$), 478 ($H$), 1,125 ($K_s$) for the RRab1 template; 
1,662 ($J$), 995 ($H$), 1,709 ($K_s$) for the RRab2 template; 
440 ($J$), 284 ($H$), 512 ($K_s$) for the RRab3 template.
The current data set is more than six times larger than the 
data set adopted by J96, and more than 2.5 times larger, if 
considering only the $K-s$-band data. 

\subsection{Normalization of the light curves}\label{sect_template}

The NIR light-curve templates we are developing provide the mean magnitude
 $\langle X \rangle$ of an RRL with an accuracy of the order of a few 
hundredths of a magnitude provided that the following data are available:\par 

{\it i)} the epoch ($t$) and the magnitude ($X_t$) of a phase point; 

{\it ii)} the period of the variable ($P$);

{\it iii)} the luminosity amplitude in either the $V$ or $B$ band 
($AV$ or $AB$);

{\it iv)} the epoch of the anchor point along the light curve. 
In this investigation, we adopt the epoch of the mean magnitude on the 
rising branch ($t_{ris}$). It was already demonstrated by \citep{inno15}
that for Classical Cepheids (CCs), $t_{ris}$ is a more precise anchor 
point than the more commonly-used epoch of the maximum light 
($t_{max}$).

We performed a number of simulations using optical and NIR light curves 
for which both $t_{ris}$ and $t_{max}$ were available and we found that the 
former is better defined when moving from the blue to the red edge of 
the RRL instability strip. The reasons why $t_{ris}$ is better defined than 
$t_{max}$ are twofold. {\it i)}  Large-amplitude RRab variables characterized by 
a ``sawtooth'' light curve show a cuspy maximum. This means that the phases 
across maximum light occur during a short time range, so an accurate 
estimate of the epoch of maximum light requires high time resolution. 
{\it ii)}  Some RRc variables display a well-defined dip just before maximum 
light \citep[U Com,][]{bono00c}. 
To properly identify and separate the two maxima, high time 
resolution is also required for these short-period variables.  

The mean NIR magnitude, $\langle X \rangle$, of a variable for which the 
aforementioned parameters are available can be estimated by using the following 
relation:  

\begin{equation}\label{eq_template}
\langle X \rangle = X_t - AX \cdot T({\phi}_t)
\end{equation}

where ${\phi}_t = \dfrac{t-t_{ris}}{P}$ is the difference in 
phase between the NIR phase point that was observed and the epoch, 
$t_{ris}$, of the anchor point, while $AX$ is the luminosity 
amplitude in the $X$ band. Note that the latter is typically 
unknown, but it can be estimated from the optical amplitude and 
empirical NIR-over-optical amplitude ratios \citep{braga2018}.
Note also that the light-curve templates must be normalized.

To generate the normalized light-curve templates, we adopted 
the magnitudes $m_{ijk}$ of the TDS variables, marked with an 
asterisk in Table~\ref{tab:allcv} and listed in 
Table~\ref{tab:sample}, where $i$ indicates the $i$-th phase
point of the empirical light curve, $j$ indicates the band (1 for $J$,
2 for $H$ and 3 for $K_s$), and $k$ indicates the 
$k$-th RRL in the TDS sample. We have transformed all the 
empirical $m_{ijk}$ measurements into normalized magnitudes 
$M_{ijk}$ by subtracting from each $k$-th RRL its mean magnitude 
in the $j$-th band (see Table~\ref{tab:sample} 
$\langle m_{jk} \rangle$) 
and by dividing for the $j$-th band amplitude (see Table~\ref{tab:sample} $A_{jk}$) 
according to the following relation: 
$M_{ijk} = \dfrac{m_{ijk} - \langle m_{jk} \rangle}{A_{jk}}$.
Fig.~\ref{fig:fourier} and \ref{fig:pegasus} show the final normalized light 
curves as a function of the pulsation phase for the TDS sample.

\section{Analytical fits to the light curve templates}\label{templatefit}

Once the normalized light curves for the three NIR bands and for the different 
period bins have been fixed we performed an analytical fit of the light curve
templates. We adopted two different fitting functions: Fourier series 
(Section~\ref{templatefit:four}) and periodic Gaussians (PEGASUS, 
Section~\ref{templatefit:pega}).

\subsection{Fourier fit}\label{templatefit:four}

\begin{figure*}[!htbp]
\centering
\includegraphics[width=8cm]{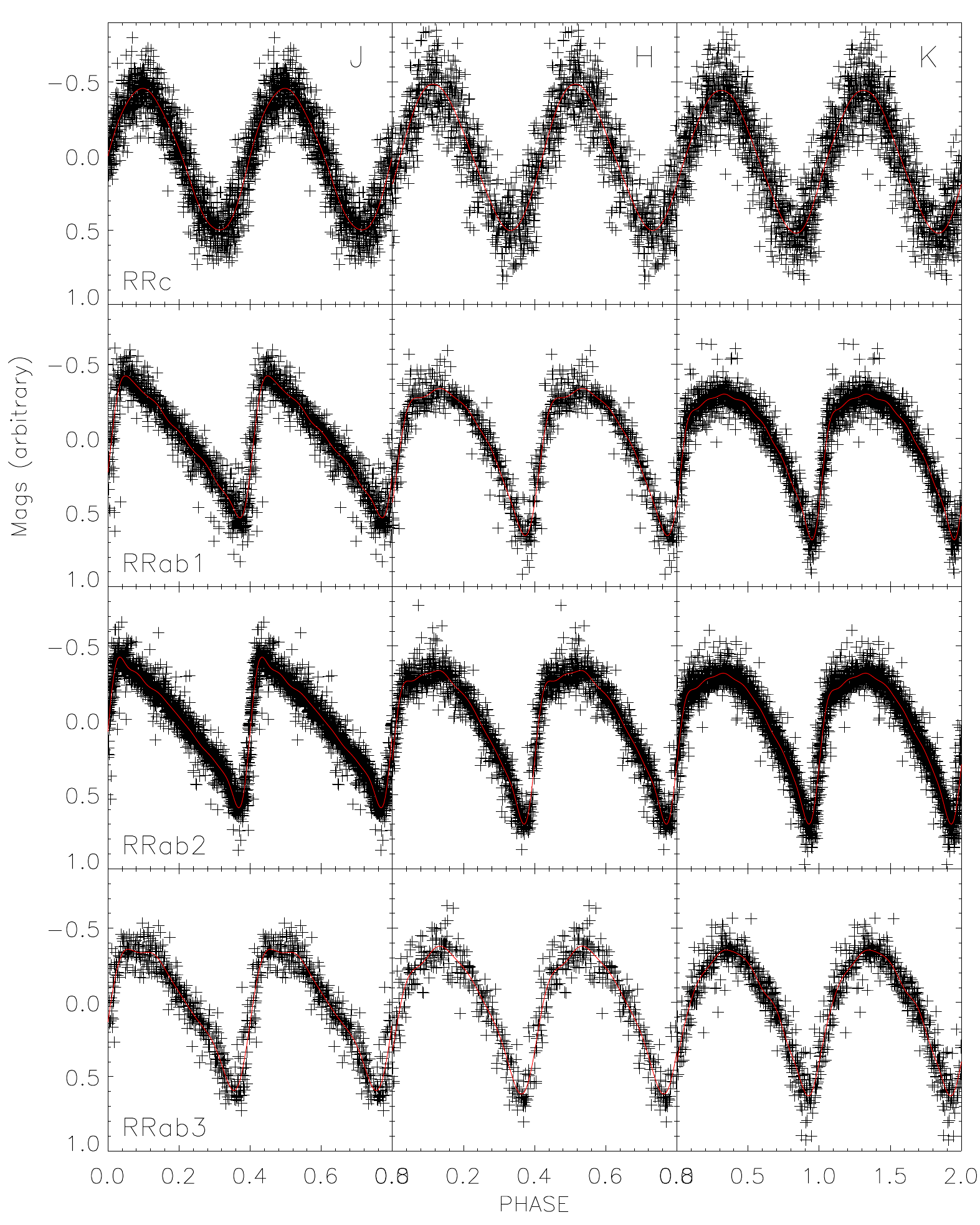}
\includegraphics[width=8cm]{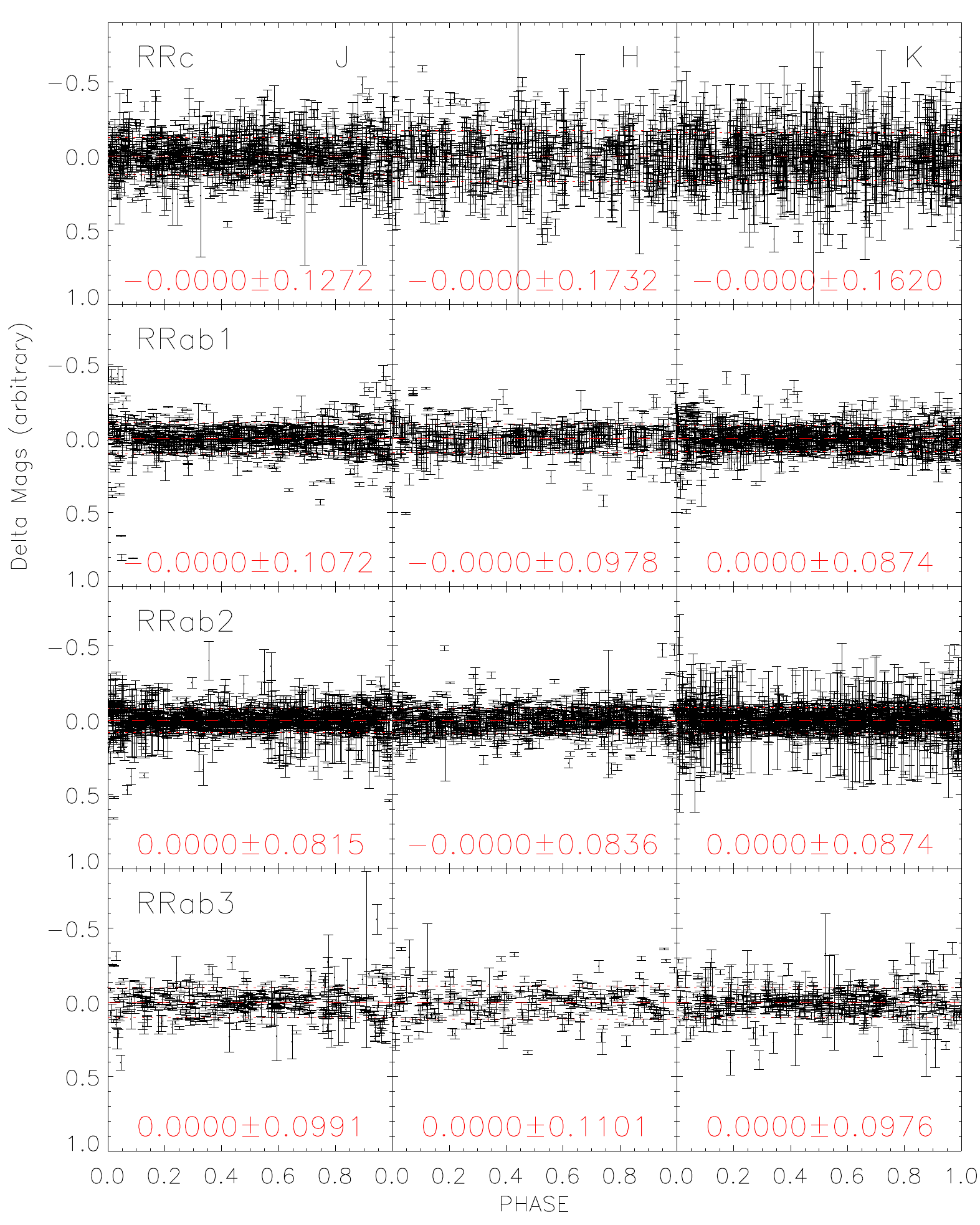}
\caption{Left panels: from left to right the different panels display 
the Fourier fits of the normalized  $JHK_s$ light curves. From top to 
bottom the panels show the four (RRc, RRab1, RRab2, RRab3) different 
ranges in period.  
Right panels: Same as the left, but for the residuals of the normalized 
light curves with the Fourier fits. The median and standard deviation of 
the median are labelled in red.}
\label{fig:fourier}
\end{figure*}

We have fit the normalized light curves with Fourier series of the $i$-th order

\begin{equation}\label{eq_fourier}
F(\phi) = A_0 + \Sigma_i A_i \cos{(2\pi i \phi - \phi_i)}
\end{equation}

with $i$ ranging from two to seven. The red lines plotted in the left panels of Fig.~\ref{fig:fourier} 
show the individual fits for the three different bands and for the four light-curve 
templates.

It is noteworthy that the agreement between the analytical fits and observations is,
within the errors, quite good over the entire pulsation cycle. In particular, 
the fits properly represent the data across the phases of minimum light in which the 
variation of the luminosity is more cuspy. Interestingly enough, we found that  
the residuals between the normalized light curves and the Fourier fits plotted in 
the right panels of the same figure are vanishing. They are typically smaller 
than the fourth decimal place. Moreover and even more importantly, the residuals 
do not show any phase dependence within the standard deviation (dashed red lines) 
of the analytical fits.  
In this context it is worth mentioning that the light-curve templates derived by 
J96 were obtained using 2$^{nd}$-order Fourier fits for the RRc variables and 6$^{th}$-order
Fourier fits for the RRab variables. We used different orders for almost
all the period bins, however, we adopted the 6$^{th}$ order
for the fit of the RRab3 $K_s$-band templates. This template 
includes roughly the same number of variables as the RRab1 template by J96 
($AB<$1.0 mag), however the coefficients of the fit are significantly 
different.

\subsection{PEGASUS fit}\label{templatefit:pega}

We also performed an independent fit of the normalized light curves using a series 
of periodic Gaussians, presented in \citet{inno15} with $i$ ranging from two to six. 

\begin{equation}\label{eq_pegasus}
P(\phi) = A_0 + \Sigma_i A_i \exp{\Big(-\sin{\Big(\dfrac{\pi (\phi - \phi_i)}{\sigma_i}\Big)^2}\Big)}
\end{equation}

\begin{figure*}[!htbp]
\centering
\includegraphics[width=8cm]{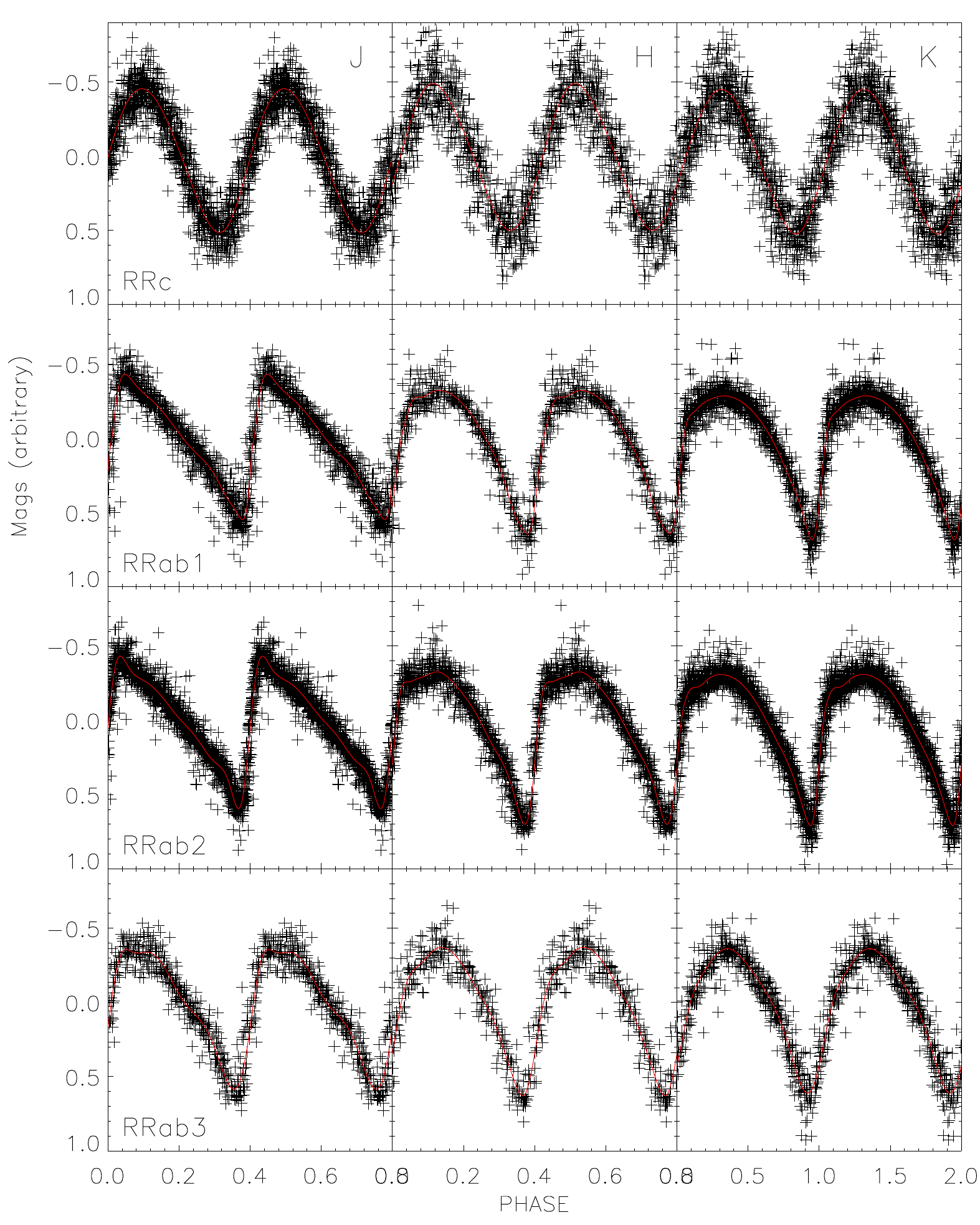}
\includegraphics[width=8cm]{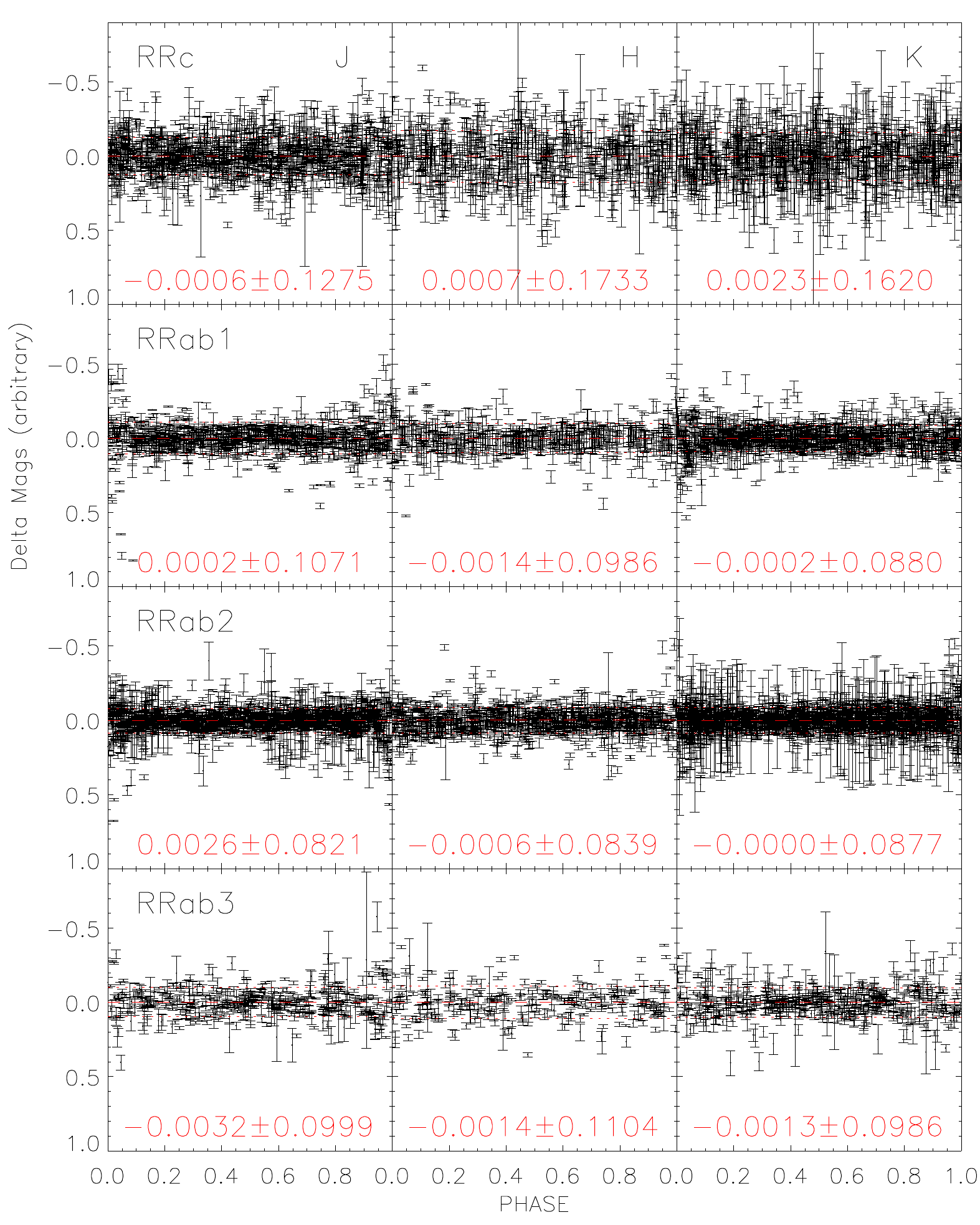}
\caption{Left panels: from left to right the different panels display 
the PEGASUS fits of the normalized  $JHK_s$ light curves. From top to 
bottom the panels show the four (RRc, RRab1, RRab2, RRab3) different 
ranges in period.  
Right panels: Same as the left, but for the residuals of the normalized 
light curves with the PEGASUS fits. The median and standard deviation of 
the median are labelled in red.}
\label{fig:pegasus}
\end{figure*}

Data plotted in the left panels of Fig.~\ref{fig:pegasus} show that 
PEGASUS fits follow the variation of the normalized light curves quite well over the 
entire pulsation cycle. This applies not only to the RRc, but also to the RRab light-curve 
templates. The main difference between the fits based either on PEGASUS or 
on Fourier series is that the former display a smoother variation over the 
entire pulsation cycle, while the latter show several small bumps/ripples. 
The NIR light curves with accurate photometry and very well sampled light curves 
do not display these bumps. This suggests that the bumps/ripples are spurious 
variations of the order of a few thousandths of a magnitude among the different 
variables included in a given period bin. 

The residuals between the normalized light curves and the PEGASUS fits are 
plotted in the right panels of the same figure. They are of the order of a few 
thousandths, i.e., slightly larger than the residuals of the Fourier fits. 
The difference is mainly due to the smoothness of the PEGASUS fits compared 
with the Fourier fits.

\subsubsection{Phases of minimum and maximum along the light curve template}\label{minmax}

Although there are solid reasons supporting the idea that 
$t_{ris}$ is easier to derive than the epoch of maximum light, 
$t_{max}$, and it provides a more precise epoch of reference, 
we are aware that all the recent surveys adopted $t_{max}$ as the reference epoch for 
RRLs and other variable stars. For this reason we also
provide the phases of both minimum and maximum 
(${\phi}_{max}$ and ${\phi}_{min}$) of the current light-curve 
templates (see Table~\ref{tab:minmax}). These pulsation phases---which can 
be considered typical---provide the opportunity to use the current templates 
to estimate the mean magnitude of variables for which only $t_{max}$ and/or 
$t_{min}$ is available in the literature. 

\begin{table*}[!htbp]
 \footnotesize
 \caption{Phases of minimum and maximum of the light curves templates.}
  \centering
 \begin{tabular}{llcccc}
 \hline\hline  
Template & band & $\phi_{min(F)}$ & $\phi_{max(F)}$ &  $\phi_{min(P)}$ & $\phi_{max(P)}$ \\ 
 \hline 
RRc   & $J$  & 0.785 & 0.243 & 0.783 & 0.238 \\
RRc   & $H$  & 0.828 & 0.282 & 0.824 & 0.283 \\
RRc   & $K_s$& 0.836 & 0.307 & 0.831 & 0.314 \\
RRab1 & $J$  & 0.929 & 0.123 & 0.945 & 0.119 \\
RRab1 & $H$  & 0.934 & 0.336 & 0.949 & 0.328 \\
RRab1 & $K_s$& 0.949 & 0.332 & 0.950 & 0.328 \\
RRab2 & $J$  & 0.920 & 0.084 & 0.918 & 0.089 \\
RRab2 & $H$  & 0.927 & 0.330 & 0.932 & 0.310 \\
RRab2 & $K_s$& 0.929 & 0.325 & 0.942 & 0.310 \\
RRab3 & $J$  & 0.889 & 0.139 & 0.890 & 0.125 \\
RRab3 & $H$  & 0.907 & 0.340 & 0.914 & 0.349 \\
RRab3 & $K_s$& 0.928 & 0.343 & 0.916 & 0.344 \\
\hline
 \end{tabular}
 \tablefoot{Phase 0.000 corresponds to the epoch of
 the mean magnitude on the rising branch ($t_{ris}$).}
\label{tab:minmax}
 \end{table*} 
 
\section{Validation of the light curve templates}\label{validation}

\subsection{Validation based on $\omega$ Cen RR Lyrae}\label{validation_wcen}

To validate the light-curve templates, we need optical 
and NIR light curves of RRLs from which we can derive 
accurate estimates of their photometric 
properties (mean magnitudes, amplitudes and 
$t_{ris}$). However, to perform an independent check we cannot 
use RRLs in the TDS (Table~\ref{tab:sample}).
Therefore we defined a Template Validation Sample (TVS) 
including four \wcen RRLs: \wcen-V20 (RRc), 
\wcen-V57 (RRab1), \wcen-V107 (RRab2) and \wcen-V124 (RRab3).
The selection of the four TVS RRLs was based on the following 
criteria: {\em i)} -- the TVS RRLs have well-sampled $J$-, $H$- 
and $K_s$-band light curves and cover the four-light curve templates 
we are developing; {\em ii)} -- the estimate of epoch of reference 
($t_{ris}$, $t_{max}$) is very accurate.

\begin{table*}[!htbp]
 \caption{Optical-NIR photometric properties of the $\omega$TVS RRLs.}
  \centering
 \begin{tabular}{l l c c}
 \hline\hline  
ID\tablefootmark{a} & Template & $P$ & $t_{ris}$\tablefootmark{b} \\
 & & days & HJD \\
 \hline 
 \wcen-V83  & RRc   & 0.3566102 & 57049.8333  \\
 \wcen-V107 & RRab1 & 0.5141038 & 49860.6035 \\ 
 \wcen-V125 & RRab2 & 0.5928780 & 49116.6901  \\
 \wcen-V15  & RRab3 & 0.8106543 & 54705.5137 \\ 
\hline
 \end{tabular}
 \tablefoot{\tablefoottext{a}{Heliocentric Julian Date -- 2,400,000 days.}}
\label{tab:wcen1}
\end{table*} 

We estimated the mean NIR ($\langle JHK_s \rangle_{best}$) 
magnitudes of the TVS RRLs by fitting the light curves in intensity 
and then converting the mean intensity to mean magnitude. 
To estimate the mean NIR magnitude ($\langle JHK_s \rangle_{templ}$)
with the light curve template we followed two different paths 
based either on single phase point (Section~\ref{wcen_1punto}) or on three 
independent phase points (Section~\ref{wcen_3punti}).  
The key idea is to estimate the accuracy of the light-curve templates
from the difference $\Delta\langle JHK_s \rangle$ between 
the measured ($\langle JHK_s \rangle_{best}$) and the estimated 
($\langle JHK_s \rangle_{templ}$) mean magnitudes. The mean NIR 
magnitudes will be estimated from the Fourier and PEGASUS fits 
for both the single-phase point and the triple-phase points method. 
To discriminate among them we add suffixes to the subscript of 
the mean magnitudes $\langle JHK_s \rangle_{templ[P/F][1/3]}$, 
where [$P/F$] indicates that we used either the PEGASUS or 
the Fourier fit, and [1/3] indicates that we used either the 
single-phase point or the triple-phase point approach.

Finally, to provide a more quantitative comparison with the light-curve 
template available in the literature we also fit the TVS RRLs 
with the J96 templates. 

\subsubsection{Light curve templates applied to a single phase point}\label{wcen_1punto}

We extracted 100 phase points ($\phi_i$,$JHK_{s(i)}$, where 
$i$ runs from 1 to 100) starting from an evenly-spaced grid of 
phases $\phi_i$=[0.00, 0.01, ... 0.99]. For each $\phi_i$, we generated a 
random magnitude $JHK_{s(i)} = JHK_{s}(fit(\phi_i)) + r\sigma$. The two 
components of this extracted light curve are 
{\it i)} $JHK_{s}(fit(\phi_i))$, which is the value of the fit of the light 
curve at the phase $\phi_i$, and 
{\it ii)}  $r\sigma$, which simulates random noise: $\sigma$ is the standard 
deviation of the phase points around the fit and $r$ is a random number 
extracted from a normal distribution.

\begin{figure*}[!htbp]
\centering
\includegraphics[trim={0 1.217cm 0 0},clip,width=8cm]{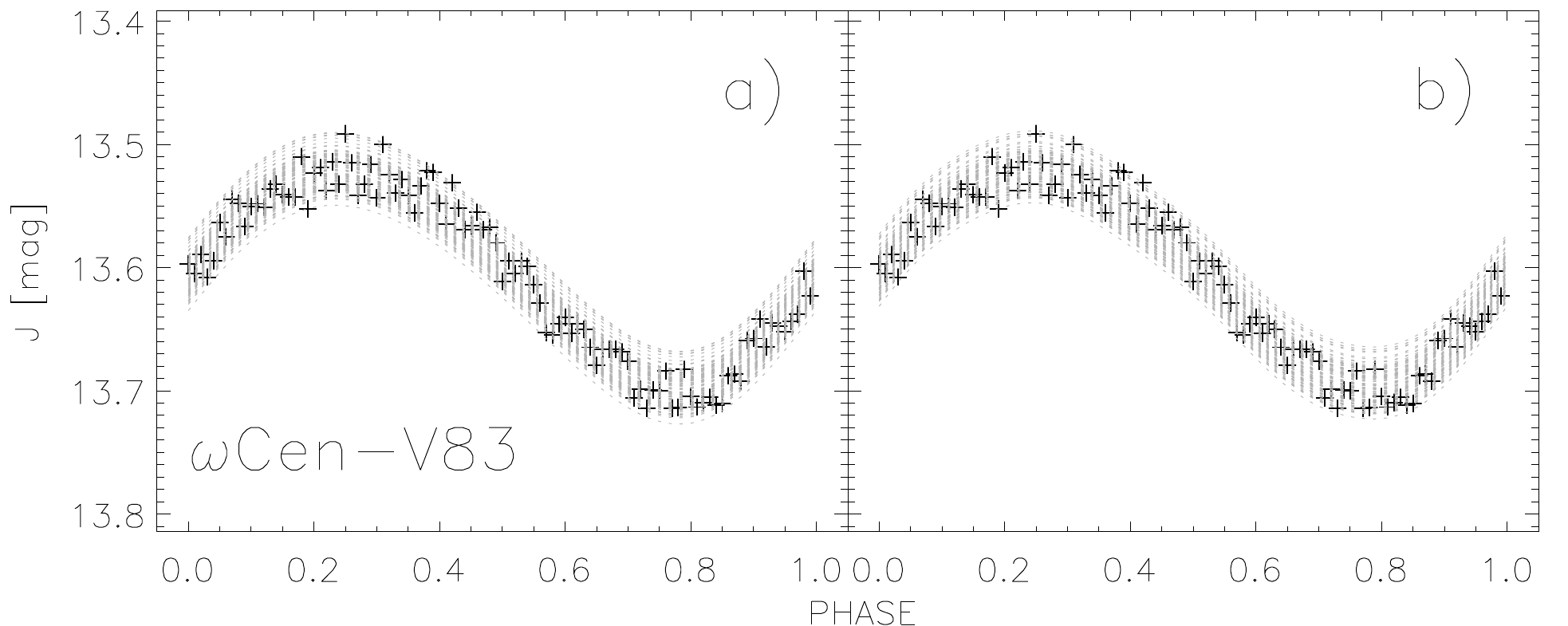}\\
\includegraphics[trim={0 1.217cm 0 0.136cm},clip,width=8cm]{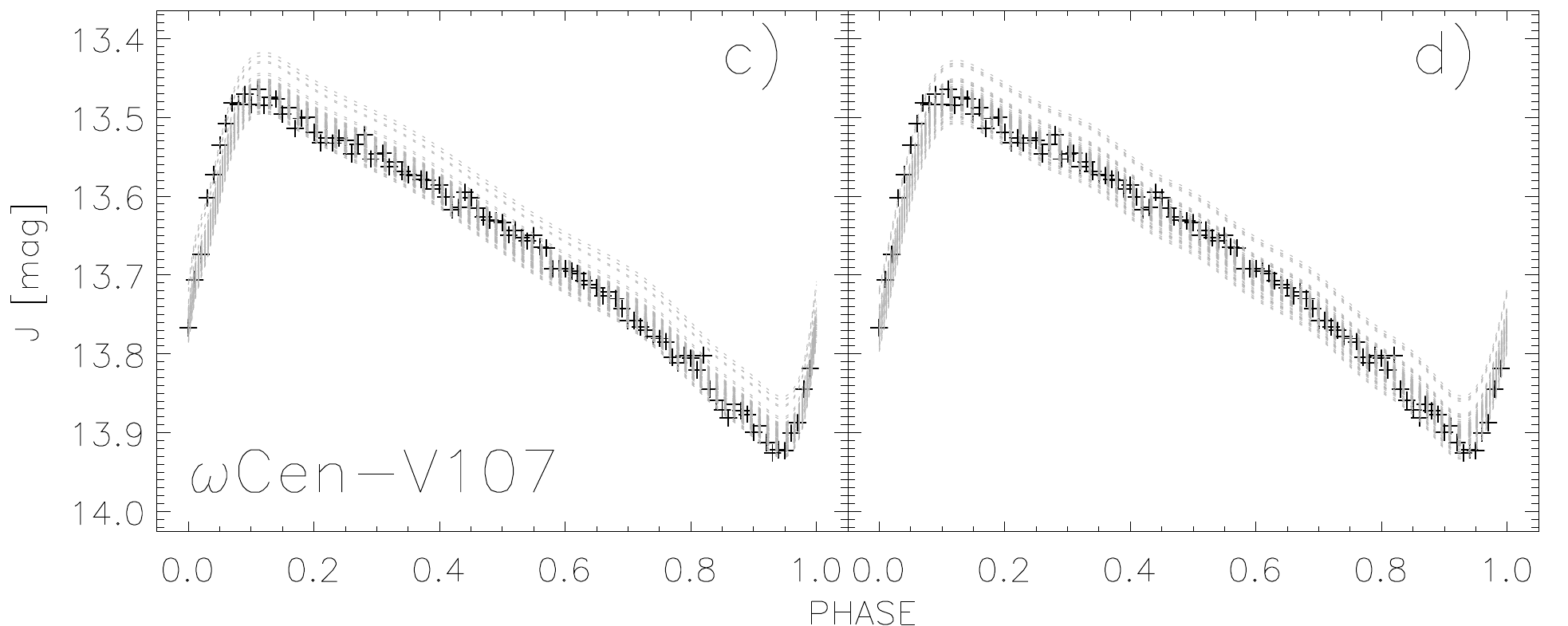}\\
\includegraphics[trim={0 1.217cm 0 0.136cm},clip,width=8cm]{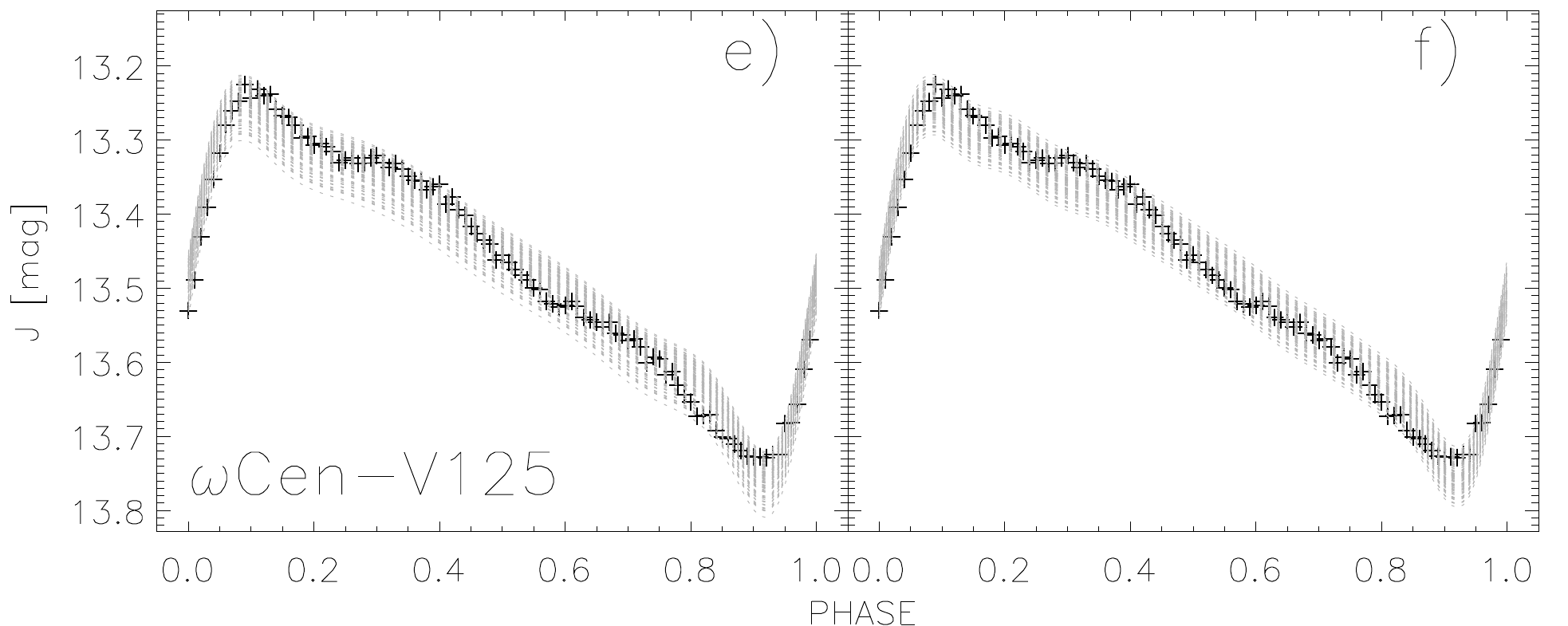}\\
\includegraphics[trim={0 0 0 0.136cm},clip,width=8cm]{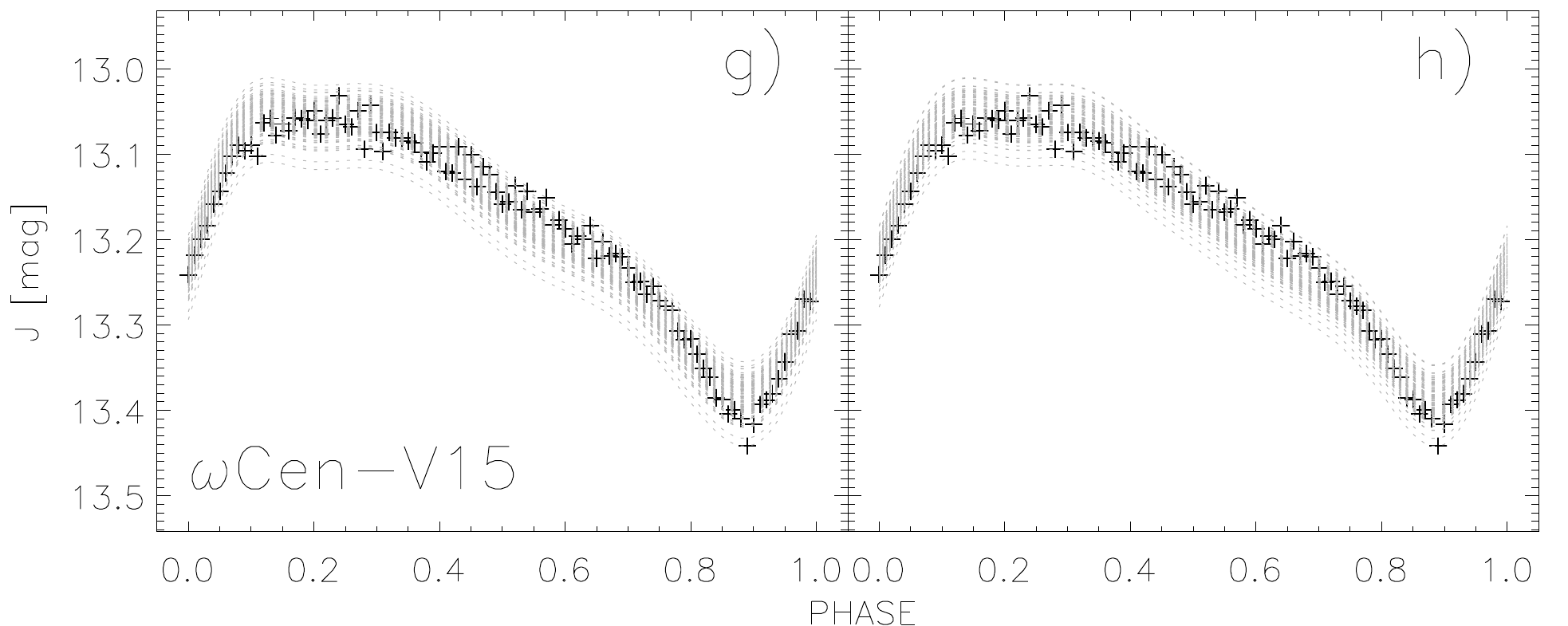}
\caption{a),b) panels: Black pluses show the randomly extracted $K_s$-band phase points over
the light curve of the RRc variable \wcen-V83. Gray dashed lines display the 
fit of the light curve template (Fourier, a); PEGASUS, b)) to the extracted 
phase points. The ID of the RRL is labeled.
d),e): Same as panels a) and b), but for the RRab variable \wcen-V107. 
The RRab1 light curve templates were adopted.
f),g): Same as panels a) and b), but for the RRab variable \wcen-V125. 
The RRab2 light curve templates were adopted.
h),i): Same as panels a) and b), but for the RRab variable \wcen-V15. 
The RRab3 light curve templates were adopted.}
\label{fig:checkj}
\end{figure*}

\begin{figure*}[!htbp]
\centering
\includegraphics[trim={0 1.217cm 0 0},clip,width=8cm]{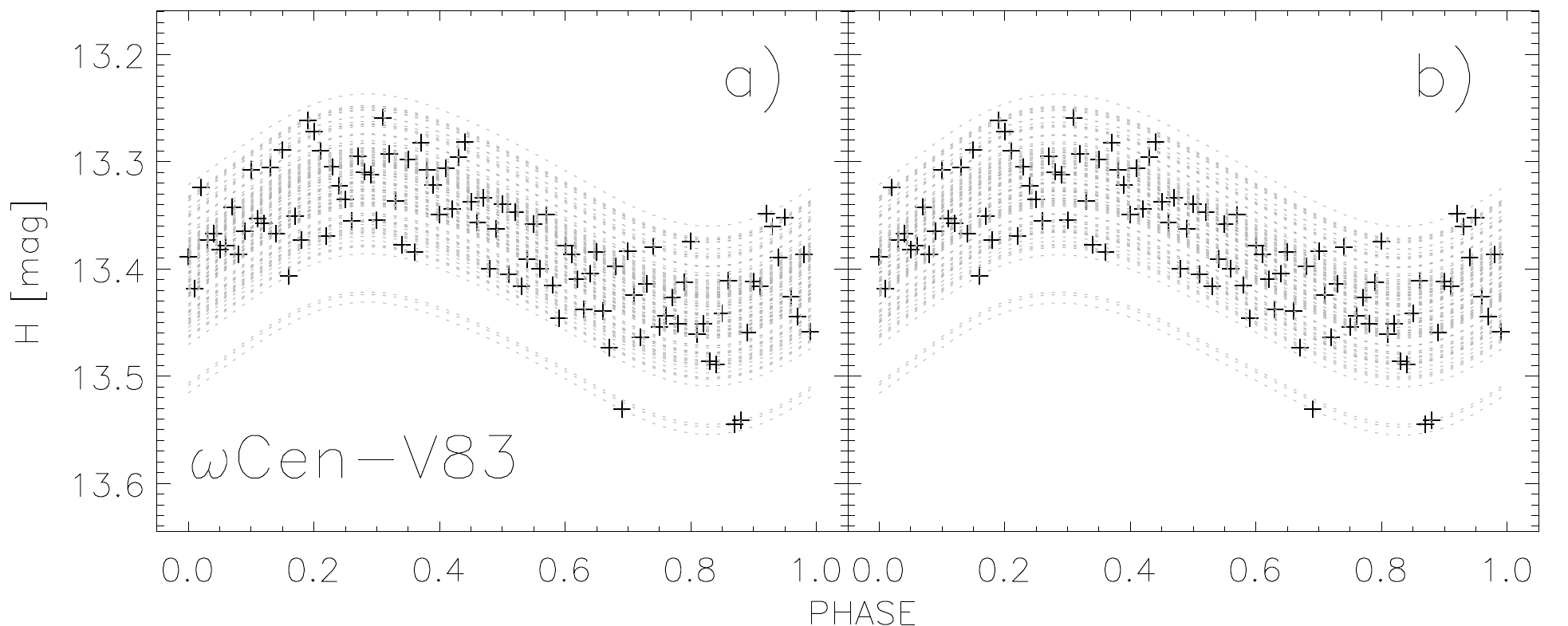}\\
\includegraphics[trim={0 1.217cm 0 0.136cm},clip,width=8cm]{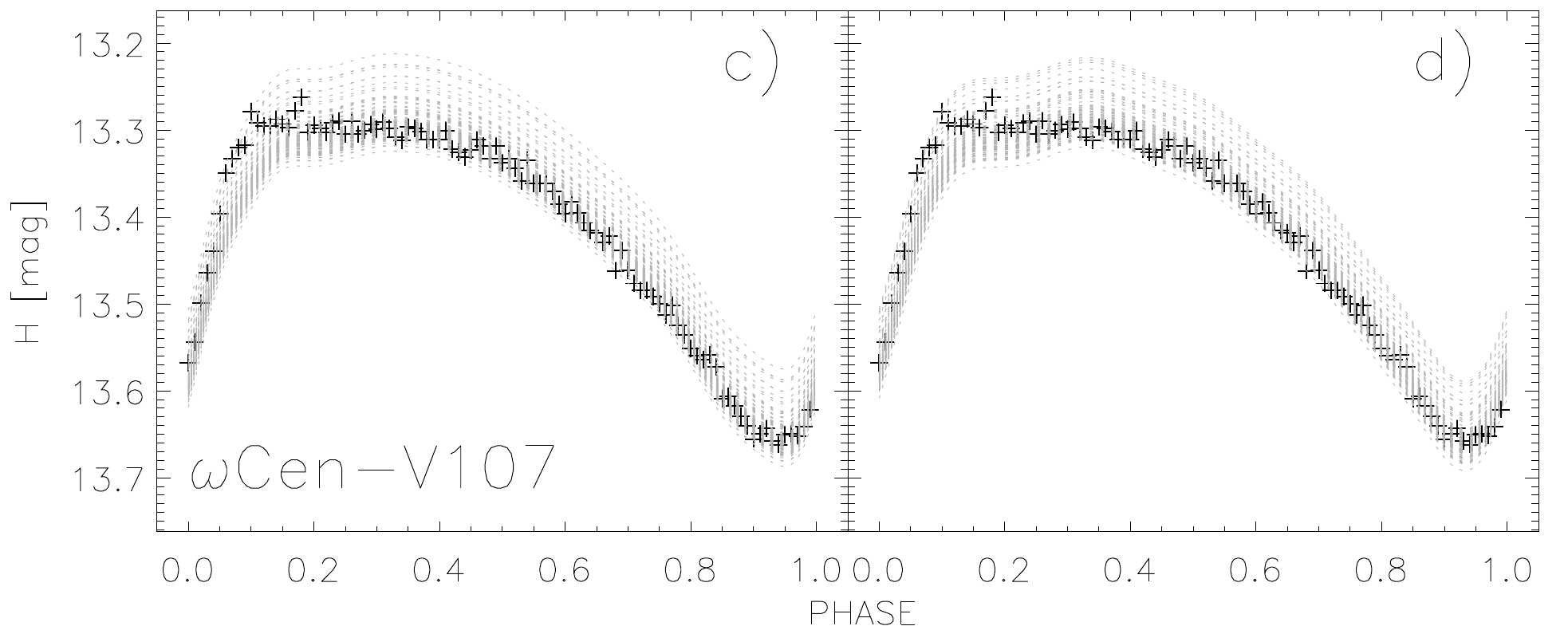}\\
\includegraphics[trim={0 1.217cm 0 0.136cm},clip,width=8cm]{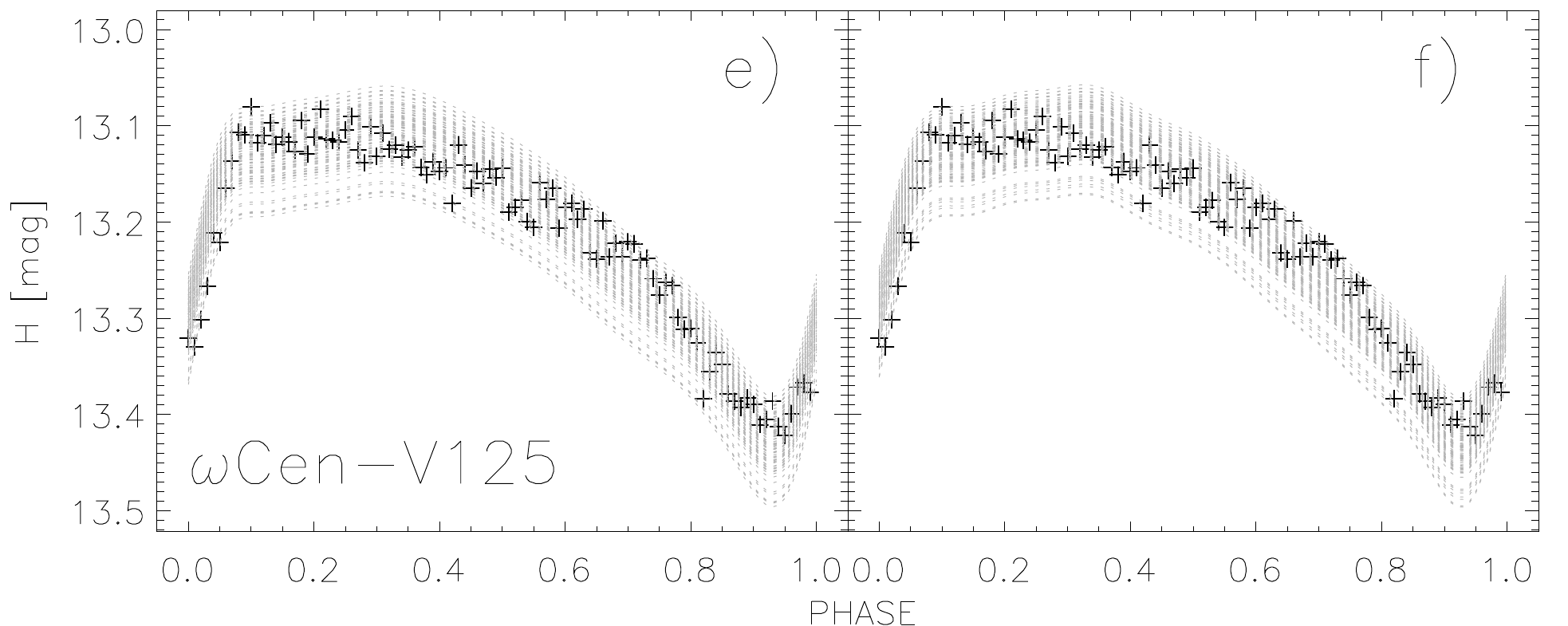}\\
\includegraphics[trim={0 0 0 0.136cm},clip,width=8cm]{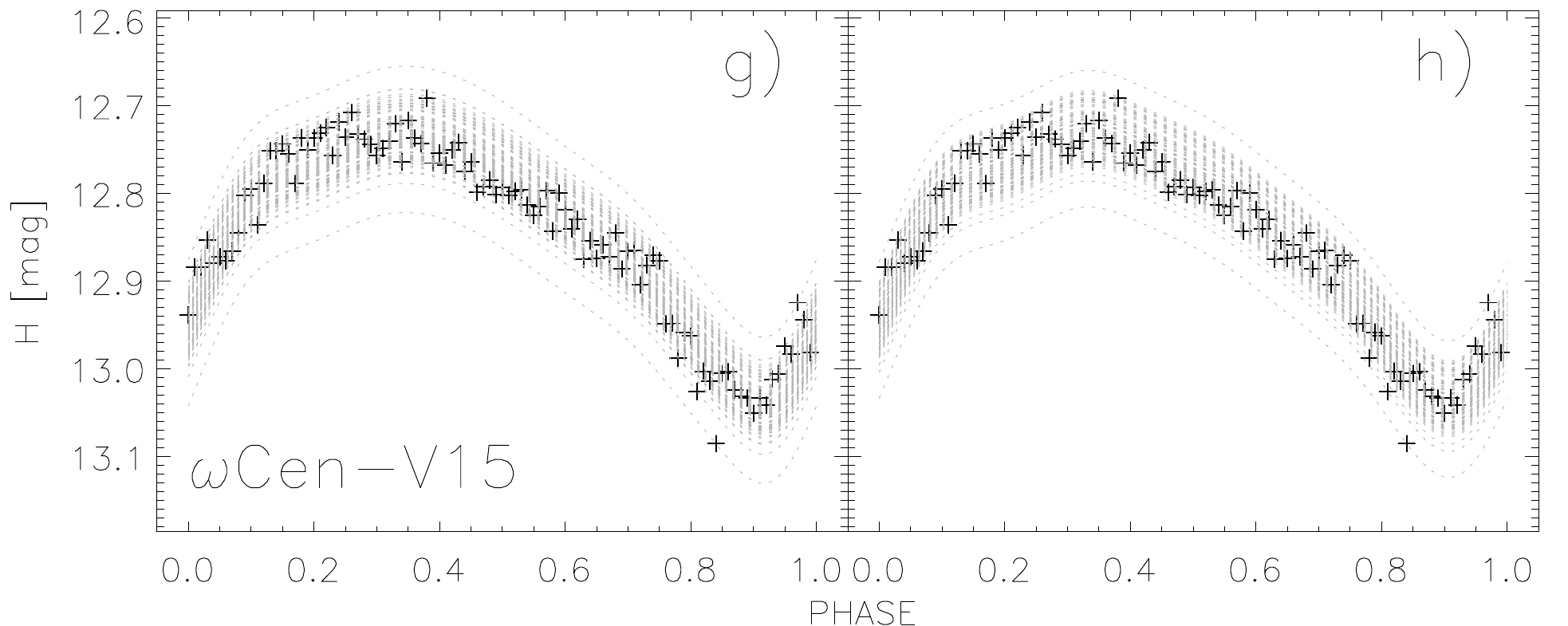}
\caption{Same as Fig.~\ref{fig:checkj}, but for the $H$ band light curve templates.}
\label{fig:checkh}
\end{figure*}

\begin{figure*}[!htbp]
\centering
\includegraphics[trim={0 1.28cm 0 0},clip,width=10cm]{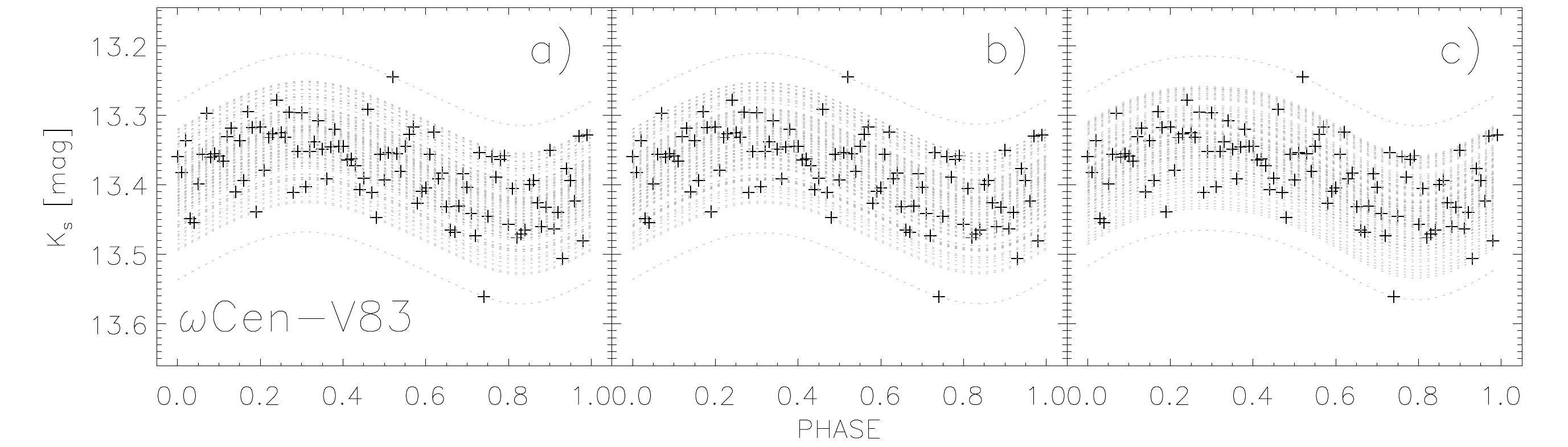}\\
\includegraphics[trim={0 1.28cm 0 0.15cm},clip,width=10cm]{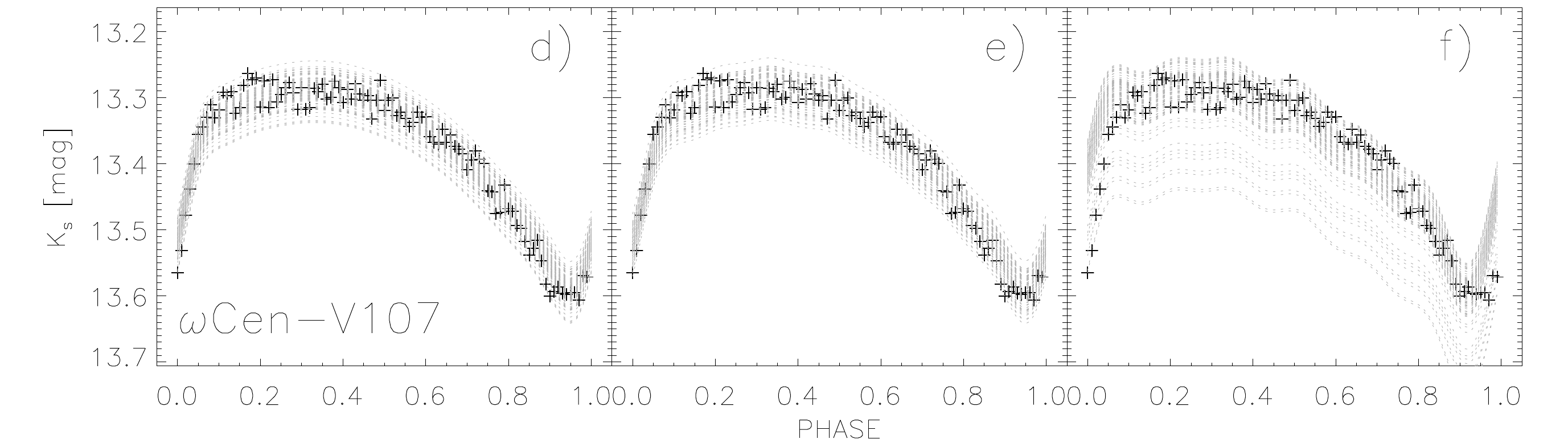}\\
\includegraphics[trim={0 1.28cm 0 0.15cm},clip,width=10cm]{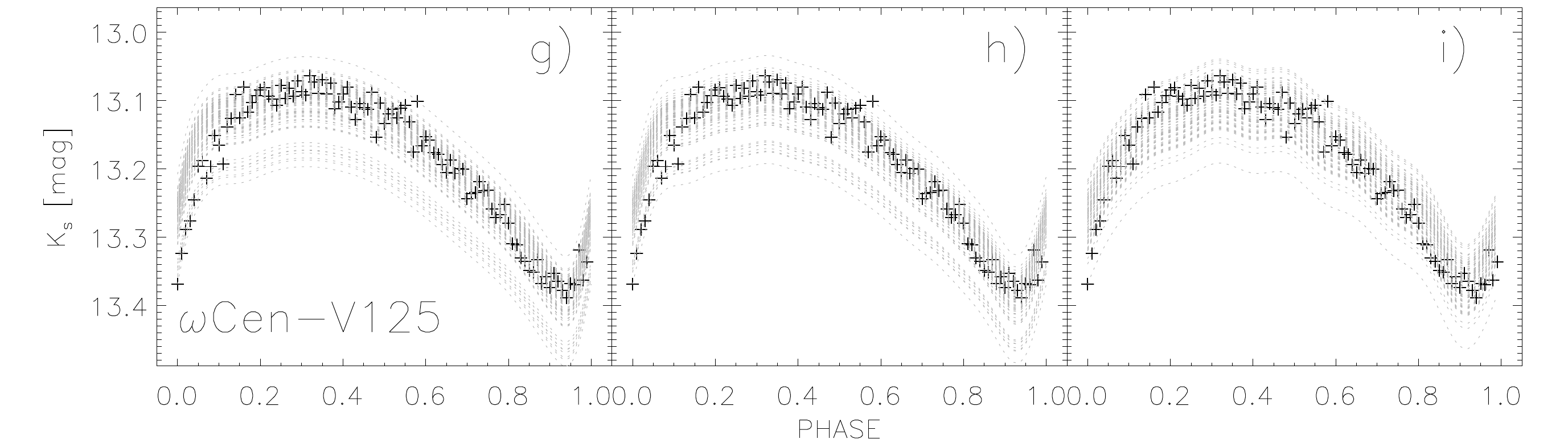}\\
\includegraphics[trim={0 0 0 0.15cm},clip,width=10cm]{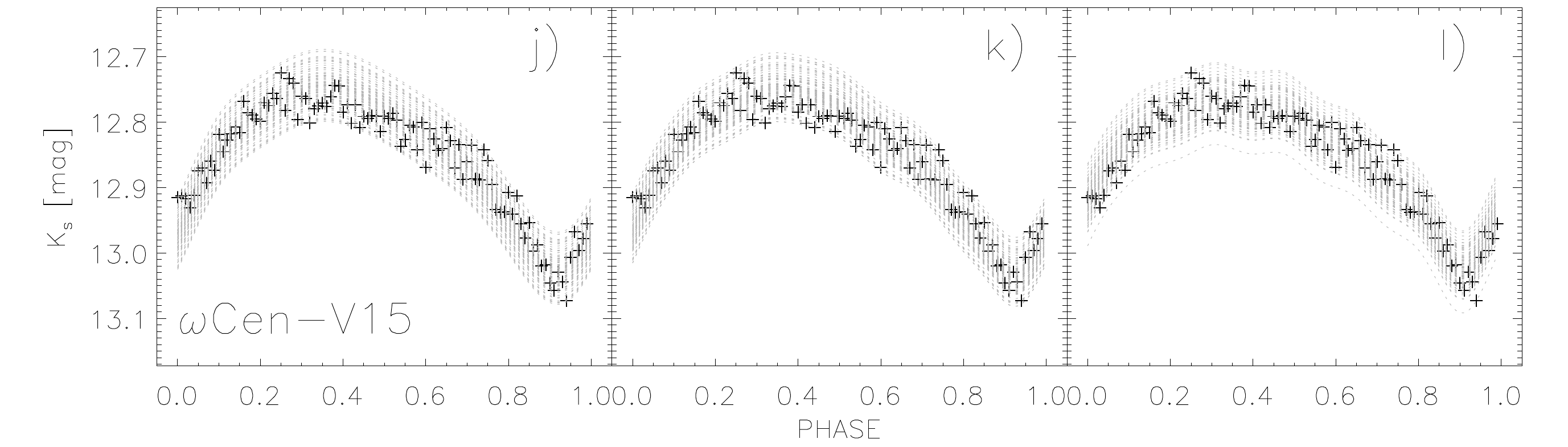}
\caption{Left (a),d),g),j)) and middle  (b),e),h),k)) panels are the same as Fig.~\ref{fig:checkj}, 
but for the $K_s$ band light curve templates. The right panels (c),f),i),l) display the fit based on 
the $K_s$ band the J96 light curve template.}
\label{fig:checkk}
\end{figure*}

We also derived, by applying the template with 
Equation~\ref{eq_template}, 100 estimates of 
$\langle JHK_s \rangle_{templ(i)}$, one for each 
extracted phase point. Subsequently, we estimated the 
median and the standard deviation of the median over the 100 
$\langle JHK_s \rangle_{templ(i)}$ extractions.  
Figures~\ref{fig:checkj}, \ref{fig:checkh} and \ref{fig:checkk}
display the extracted phase points and the fits based on the 
light-curve templates in the $J$, $H$ and $K_s$ bands.

The estimates of $\langle JHK_s \rangle_{templ}$---using 
the Fourier, PEGASUS and J96 templates---of the 
TVS RRLs are listed in Table~\ref{tab:oglevvv2}. The same 
Table also gives the difference in magnitude 
($\Delta\langle JHK_s \rangle$) among the different fits. 

It is worth noting (see Table~\ref{tab:wcen2}, columns 2, 3 and 4) that 
the mean of the residuals with respect to the measured magnitudes is 
at most 0.010 mag for all the templates.
In all cases, the standard deviations are larger than the 
residuals, meaning that the latter can be considered null
within the dispersion. The largest residuals are found in the $H$, 
band for the RRab1 template: the mean magnitudes estimated from the 
templates are $\sim$0.01 mag fainter than the measured mean magnitude. 
This happens because the fit of the $H$-band light curve has minor 
deviations from the light curve template, and the extracted single 
phase points follow these deviations. Note that in performing 
this test we are maximizing the uncertainty, since the error on the 
individual phase points is estimated as a Gaussian distribution with a 
$\sigma$ equal to the standard deviation of the analytical fit. Indeed, we
found that when using the individual measurements the residuals are 
systematically smaller.

The comparison between the new and the old $K_s$-band templates indicates 
that the former are on average better than the latter. Indeed, 
the residuals in the longest period bin (RRab3) of the new templates are 
one order of magnitude smaller than for the J96 template 
(--0.001~[Fourier]/0.000~[PEGASUS] mag vs --0.011 mag). Note, however, that 
the standard deviations are of the same order of magnitude of the difference
in offset between our templates and those of J96. Moreover, the 
standard deviation of the current RRab1 period bin is more than a factor of two
smaller than for the J96 template 
(0.016~[Fourier]/0.016~[PEGASUS] mag vs 0.038 mag). A glance at the data 
plotted in the right column of Fig.~\ref{fig:checkk}, and in particular in
the panels d, e) and f), clearly shows the difference.

\subsubsection{Light curve templates applied to three phase points}\label{wcen_3punti}

The application of the NIR light-curve templates to individual NIR measurements 
does require the knowledge of three parameters: {\it i)} the period, 
{\it ii)} the luminosity amplitude, and {\it iii)} the epoch of the anchor point ($t_{ris}$). 
The third parameter poses a severe limitation for RRLs because their periods 
range from a quarter of a day to less than one day. This means that either the pulsation 
period and the epoch of the anchor point have been estimated with very high
accuracy ($\sim$one part per million) or the separation between the time at 
which the optical and the NIR photometry were collected must be shorter 
than a few years.   

To overcome this limitation we decided to perform a number  of 
tests by assuming that three independent NIR measurements were available. 
The advantage of this approach is that the light curve template is used as 
a fitting function. The approach is quite simple and follows the following 
steps: {\it i)} an estimate of the NIR luminosity amplitude using the optical to NIR 
amplitude ratios available in the literature;   
{\it ii)} a least-squares fit of a light curve including at least three phase 
points, minimizing the $\chi^2$ of two parameters: 
a shift in phase ($\Delta\phi$) and a shift in magnitude ($\Delta mag$). 
The functions to be minimized are: 
 
\begin{equation}\label{eq_fourierfit}
F(\phi; \Delta\phi,\Delta mag) = \Delta mag + A_{NIR} \cdot (A_0^F + \Sigma_i A_i^F \cos{(2\pi i \phi - \phi_i - \Delta\phi))}
\end{equation}

and

\begin{equation}\label{eq_pegasusfit}
P(\phi; \Delta\phi,\Delta mag) = \Delta mag + A_{NIR} \cdot \Big(A_0^P + \Sigma_i A_i^P \exp{\Big(-\sin{\Big(\dfrac{\pi (\phi - \phi_i - \Delta\phi)}{\sigma_i^P}\Big)^2}\Big)\Big)}
\end{equation}

for the Fourier and PEGASUS templates, respectively.
To further investigate the difference between new and old light-curve 
templates, the same minimization was also performed using the J96 
templates: 

\begin{equation}\label{eq_j96}
J(\phi; \Delta\phi,\Delta mag) = \Delta mag + A_{NIR} \cdot (A_0^J + \Sigma_i A_i^J \cos{(2\pi i \phi - \phi_i - \Delta\phi))}
\end{equation}

To validate the templates with this approach, 
we generated 100 triplets of phase points ($\phi_{ij}$,$JHK_{s(ij)}$, where 
$i$ runs from 1 to 100 and $j$ from 1 to 3). The phases are randomly
extracted from a uniform distribution between 0 and 1. The extracted 
magnitudes, $JHK_{s(ij)}$, were treated following the approach 
discussed in Section~\ref{wcen_1punto}.

\begin{figure*}[!htbp]
\centering
\includegraphics[trim={0 0 2cm 0},clip,width=12cm]{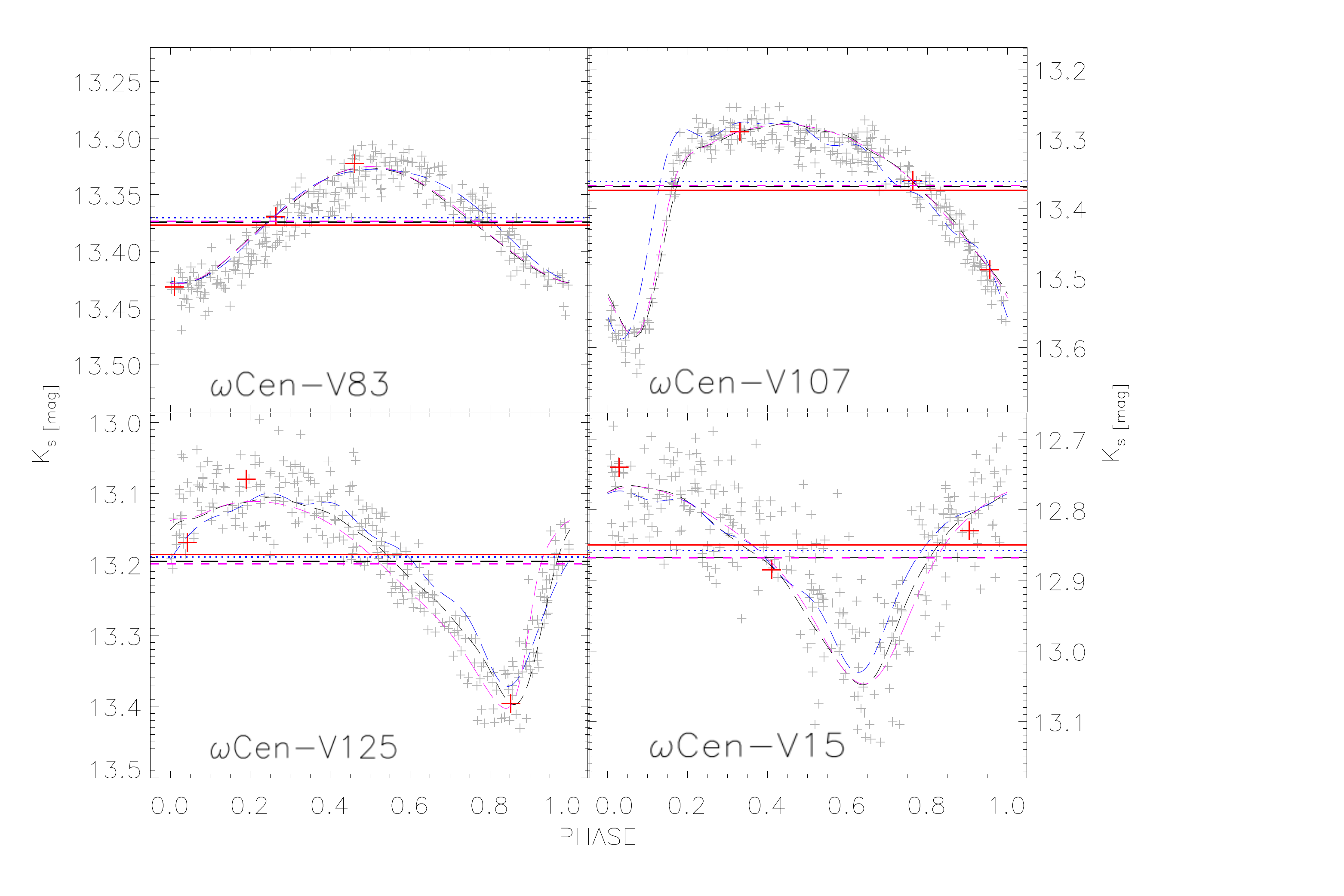}\\
\caption{Top-left: Fourier, PEGASUS and J96 template fits applied to 
RRc variable \wcen-V83. Gray crosses show the randomly extracted phase 
points. The red crosses display the three phase points of a single 
random extraction. 
The horizontal red line shows the mean magnitude of the variable
based on the intensity average fit of the empirical data. 
The black dashed curve and the horizontal black long-dashed line 
show the fit with the Fourier template and its mean 
magnitude. The magenta dashed curve and the horizontal magenta 
long-dashed line display the same, but for the PEGASUS fit.  
The blue dotted curve and the horizontal blue dotted line display 
the same, but for the J96 fit.
Top-right: Same as the top-left, but for the RRab variable \wcen-V107 (RRab1).
Bottom-left: Same as the top-left, but for  the RRab variable \wcen-V125 (RRab2).
Bottom-right: Same as the top-left, but for the RRab variable \wcen-V15 (RRab3). 
Note that, in these panels, the light curves are not phased using $t_{ris}$, 
but to an arbitrary epoch (HJD=2,350,000) to underline that the three phase 
points method is independent of the reference epoch.}
\label{fig:check3points}
\end{figure*}

Once the 100 three-phase point light curves were generated, we
performed the fits using Eq.\ref{eq_fourierfit},  
\ref{eq_pegasusfit} and \ref{eq_j96}. The individual $K_s$-band 
fits are displayed in Fig.~\ref{fig:check3points}.
We computed 100 estimates of the mean $\langle JHK_s \rangle_{templ(i)}$ 
magnitudes as the integral in intensity over the template fits. 
The final mean magnitude ($\langle JHK_s \rangle_{templ}$) 
and its uncertainty were determined as the median and the 
standard deviation of the median over the 100 random estimates of 
$\langle JHK_s \rangle_{templ(i)}$ (see Section~\ref{wcen_1punto}).
The Table~\ref{tab:wcen2} also shows the magnitude differences 
$\Delta\langle JHK_s \rangle$ between the template estimates of the 
mean magnitudes $\langle JHK_s \rangle_{templ(i)}$ and the best 
estimates of the mean magnitudes based on the fit of the light curve 
($\langle JHK_s \rangle_{best}$).

\begin{table*}[!htbp]
 \footnotesize
 \caption{NIR photometric properties of the $\omega$TVS RRLs.}
  \centering
 \begin{tabular}{l r r r r r r r}
 \hline\hline  
 & $\langle mag \rangle_{best}$ & $\langle mag \rangle_{templF1}$ & $\langle mag \rangle_{templP1}$  & $\langle mag \rangle_{templJ1}$ & $\langle mag \rangle_{templF3}$ & $\langle mag \rangle_{templP3}$ & $\langle mag \rangle_{templJ3}$ \\
 & mag & mag & mag & mag & mag & mag & mag \\
\hline 
\multicolumn{8}{c}{---\wcen-V83 (RRc)---}\\
 $J:$          & 13.602$\pm$0.004 & 13.602$\pm$0.014 & 13.602$\pm$0.014 &     \ldots        &13.602$\pm$0.025 & 13.603$\pm$0.025  &     \ldots       \\
 $\Delta J:$   &     \ldots       &  0.000$\pm$0.014 &  0.000$\pm$0.014 &     \ldots        & 0.000$\pm$0.025 &  0.001$\pm$0.025  &     \ldots       \\
 $H:$          & 13.377$\pm$0.006 & 13.375$\pm$0.037 & 13.374$\pm$0.037 &     \ldots        &13.378$\pm$0.028 & 13.378$\pm$0.028  &     \ldots       \\
 $\Delta H:$   &     \ldots       &--0.002$\pm$0.037 &--0.003$\pm$0.037 &     \ldots        & 0.001$\pm$0.028 &  0.001$\pm$0.028  &     \ldots       \\
 $K_s:$        & 13.377$\pm$0.005 & 13.381$\pm$0.038 & 13.381$\pm$0.038 &  13.376$\pm$0.038 &13.377$\pm$0.026 & 13.377$\pm$0.026  & 13.375$\pm$0.025 \\
 $\Delta K_s:$ &     \ldots       &  0.004$\pm$0.038 &  0.004$\pm$0.038 & --0.001$\pm$0.038 & 0.000$\pm$0.026 &  0.000$\pm$0.026  &--0.002$\pm$0.025 \\
\multicolumn{8}{c}{---\wcen-V107 (RRab1)---}\\
 $J:$          & 13.658$\pm$0.008 & 13.662$\pm$0.016 & 13.663$\pm$0.015 &     \ldots        & 13.656$\pm$0.083 & 13.656$\pm$0.083 &     \ldots       \\
 $\Delta J:$   &     \ldots       &  0.004$\pm$0.016 &  0.005$\pm$0.015 &     \ldots        &--0.002$\pm$0.083 &--0.002$\pm$0.083 &     \ldots       \\
 $H:$          & 13.407$\pm$0.006 & 13.417$\pm$0.025 & 13.417$\pm$0.026 &     \ldots        & 13.397$\pm$0.044 & 13.396$\pm$0.044 &     \ldots       \\
 $\Delta H:$   &     \ldots       &  0.010$\pm$0.025 &  0.010$\pm$0.026 &     \ldots        &--0.010$\pm$0.044 &--0.011$\pm$0.044 &     \ldots       \\
 $K_s:$        & 13.373$\pm$0.006 & 13.377$\pm$0.016 & 13.376$\pm$0.016 & 13.365$\pm$0.038  & 13.372$\pm$0.026 & 13.372$\pm$0.027 & 13.370$\pm$0.046 \\
 $\Delta K_s:$ &     \ldots       &  0.004$\pm$0.016 &  0.003$\pm$0.016 &--0.008$\pm$0.038  &--0.001$\pm$0.026 &--0.001$\pm$0.027 &--0.003$\pm$0.046 \\
\multicolumn{8}{c}{---\wcen-V125 (RRab2)---}\\
 $J:$          & 13.460$\pm$0.005 & 13.462$\pm$0.019 & 13.461$\pm$0.021 &     \ldots        & 13.451$\pm$0.088 & 13.453$\pm$0.084 &     \ldots       \\
 $\Delta J:$   &     \ldots       &  0.002$\pm$0.019 &  0.001$\pm$0.021 &     \ldots        &--0.009$\pm$0.088 &--0.007$\pm$0.084 &     \ldots       \\
 $H:$          & 13.206$\pm$0.009 & 13.207$\pm$0.027 & 13.208$\pm$0.027 &     \ldots        & 13.205$\pm$0.072 & 13.206$\pm$0.067 &     \ldots       \\
 $\Delta H:$   &     \ldots       &  0.001$\pm$0.027 &  0.002$\pm$0.027 &     \ldots        &--0.001$\pm$0.072 &  0.000$\pm$0.067 &     \ldots       \\
 $K_s:$        & 13.186$\pm$0.007 & 13.181$\pm$0.024 & 13.181$\pm$0.026 &  13.180$\pm$0.023 & 13.182$\pm$0.064 & 13.181$\pm$0.061 & 13.178$\pm$0.060 \\
 $\Delta K_s:$ &     \ldots       &--0.005$\pm$0.024 &--0.005$\pm$0.026 & --0.006$\pm$0.023 &--0.004$\pm$0.064 &--0.005$\pm$0.061 &--0.008$\pm$0.060 \\
\multicolumn{8}{c}{---\wcen-V15 (RRab3)---}\\
 $J:$          & 13.178$\pm$0.006 & 13.181$\pm$0.020 & 13.176$\pm$0.021 &     \ldots        & 13.169$\pm$0.071 & 13.168$\pm$0.075 &     \ldots       \\
 $\Delta J:$   &     \ldots       &  0.003$\pm$0.020 &--0.002$\pm$0.021 &     \ldots        &--0.009$\pm$0.071 &--0.010$\pm$0.075 &     \ldots       \\
 $H:$          & 12.843$\pm$0.008 & 12.849$\pm$0.026 & 12.847$\pm$0.027 &     \ldots        & 12.835$\pm$0.064 & 12.837$\pm$0.066 &     \ldots       \\
 $\Delta H:$   &     \ldots       &  0.006$\pm$0.026 &  0.004$\pm$0.027 &     \ldots        &--0.008$\pm$0.064 &--0.006$\pm$0.066 &     \ldots       \\
 $K_s:$        & 12.850$\pm$0.007 & 12.849$\pm$0.026 & 12.850$\pm$0.027 &  12.839$\pm$0.029 & 12.844$\pm$0.066 & 12.843$\pm$0.069 & 12.839$\pm$0.058 \\
 $\Delta K_s:$ &     \ldots       &--0.001$\pm$0.026 &  0.000$\pm$0.027 & --0.011$\pm$0.029 &--0.006$\pm$0.066 &--0.007$\pm$0.069 &--0.011$\pm$0.058 \\
 \end{tabular}
 \tablefoot{\tablefoottext{a}{The complete ID is OGLE-BLG-RRLYR-NNNNN, where ``NNNNN''
 is the ID appearing in the first column.}}
\label{tab:wcen2}
 \end{table*}
 
\begin{figure*}[!htbp]
\centering
\includegraphics[trim={0 0 2cm 0},clip,width=12cm]{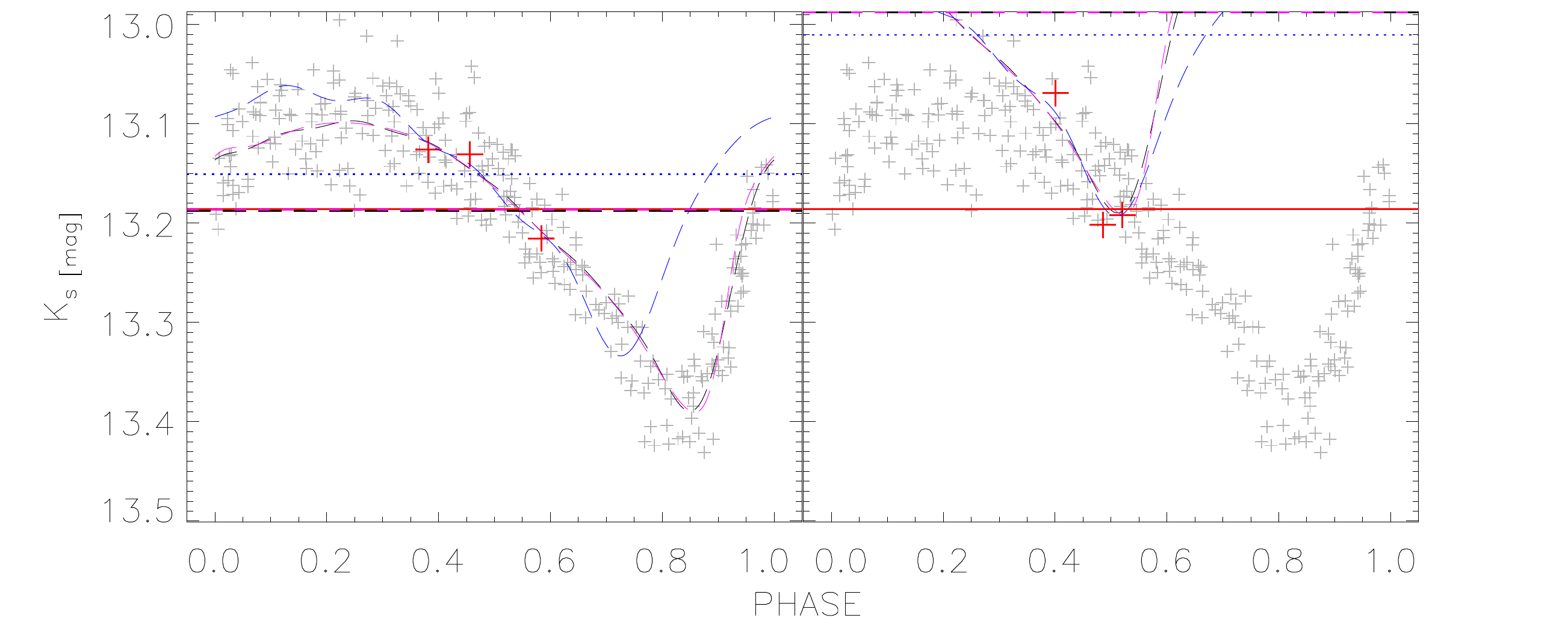}\\
\caption{Left: Same as Fig.~\ref{fig:check3points}, but for random extraction on the 
light curve of the RRab variable \wcen-V125 (RRab2). In this specific case the three 
randomly extracted phase points are close in phase. The J96 fit does not provide an 
accurate estimate of the mean magnitude. Right: same as the left, but for an extraction 
where the three randomly extracted phase points have a smaller difference in phase. In 
this case the light curve templates do not provide an accurate estimate of the $K_s$ 
band mean magnitude.}
\label{fig:check3pointsbad}
\end{figure*}

Data Plotted in Fig.~\ref{fig:check3points} show that the residuals are similar 
to the fits based on a single phase point. Indeed, the residuals are, within 
the standard deviations, zero. However, the standard deviations of the template 
fits based on three phase points are larger than those based on a single phase 
point. The difference is mainly caused by the fact that the three 
randomly-selected phase points span, in some of the extractions, a very small 
range ($\Delta \phi\le0.05$) in pulsation phase 
(see Fig.~\ref{fig:check3pointsbad}. This is also the reason why the residuals
are correlated with the difference in phase between the two closest points 
in phase ($\Delta \phi$). 

The current findings indicate that the light-curve templates used as 
fitting curves provide accurate mean magnitudes when {\it i)} the 
distance between the phase points is at least 0.1 pulsation cycles. Otherwise,
we suggest averaging the two close phase points. {\it ii)} 
the number of available phase points is modest, i.e., larger than two, but smaller 
than a dozen. Classical analytical fits (Fourier, Spline, PLOESS, PEGASUS...) 
become more accurate for a larger number of measurements. 

\subsection{Validation based on OGLE + VVV RR Lyrae}\label{oglevvv}

An independent path to validate the current light curve templates is offered by 
the two different long-term photometric surveys collecting time-series data in 
the optical (OGLE, \citep{udalski92}) and the NIR (VVV, \citep{minniti2010})  
of a significant fraction of the Galactic Bulge. The photometric catalogs 
provided by these surveys can be simultaneously used to validate the 
$K_s$-band templates. Note that we cannot validate the $J$- and $H$-band 
templates because the VVV survey only collected $K_s$-band time series.

The validation relies on the OGLE-IV catalog of 38,257 Bulge 
RRLs \citet{soszynski14}. Using a searching radius of 2\sec, 
we found 2,517 matches in the VVV point source catalog. We used 
a very small searching radius because this provides a faster 
selection of the good matches. Obviously, the completeness is 
modest, but the validation only requires a few variables per 
period bin. Among them we selected 22 RRLs and the criteria 
we adopted for the selection are the following: 
{\em i)} -- good coverage of the $K_s$-band light curve, and in fact
they all have at least 38 phase points (80\% of them have at least 
49 phase points);   
{\em ii)} -- good coverage of both the $V$- and the $I$-band light 
curve to provide accurate estimates of the luminosity amplitudes 
($AV$, $AI$) and of the epochs of the mean magnitudes on the 
rising branch ($t_{ris(V)}$, $t_{ris(I)}$). The phasing of optical 
and NIR light curves was performed using the pulsation period 
provided by OGLE. The distribution of these variables among the 
different period bins is the following: RRc (six), RRab1 (five), 
RRab2 (five) and RRab3 (six).  These variables were called the 
``Bulge Template Validation Sample'' (BTVS) and their pulsation 
properties are listed in Table~\ref{tab:oglevvv}).

\begin{table*}[!htbp]
 \footnotesize
 \caption{Optical properties of the Bulge RRLs adopted to validate the light crve template.}
  \centering
 \begin{tabular}{l l c c c c c c c}
 \hline\hline  
ID (OGLE)\tablefootmark{a} & ID (VVV) & $P$ & $\langle V \rangle$ & $AV$  & $t_{risV}$\tablefootmark{b} &  $\langle I \rangle$ & $AI$ & $t_{risI}$\tablefootmark{b} \\
 & & days & mag & mag & HJD & mag & mag & HJD \\
 \hline 
\multicolumn{9}{c}{---RRc---}\\
15624 & 515514387864 & 0.30180268 & 15.576$\pm$0.004 & 0.433$\pm$0.026 & 7974.3143 & 14.844$\pm$0.004 & 0.249$\pm$0.005 & 57937.4921 \\
34149 & 515618496995 & 0.33202391 & 16.075$\pm$0.005 & 0.468$\pm$0.050 & 7679.3197 & 15.345$\pm$0.005 & 0.254$\pm$0.008 & 57581.7032 \\
35612 & 515576843509 & 0.35095553 & 15.565$\pm$0.005 & 0.388$\pm$0.026 & 7681.0746 & 14.880$\pm$0.004 & 0.219$\pm$0.006 & 57610.1786 \\
11254 & 515548620097 & 0.38432825 & 16.514$\pm$0.005 & 0.447$\pm$0.018 & 7975.2994 & 15.249$\pm$0.004 & 0.253$\pm$0.005 & 57948.0109 \\
04844 & 515642803328 & 0.41666920 & 16.427$\pm$0.006 & 0.401$\pm$0.031 & 7975.1612 & 15.493$\pm$0.005 & 0.249$\pm$0.006 & 57962.2387 \\
34454 & 515597035098 & 0.46802773 & 15.579$\pm$0.004 & 0.442$\pm$0.022 & 7675.3195 & 14.845$\pm$0.004 & 0.252$\pm$0.005 & 57653.3189 \\
\multicolumn{9}{c}{---RRab1---}\\
13498 & 515504500749 & 0.40860801 & 16.796$\pm$0.007 & 1.032$\pm$0.043 & 7974.5872 & 15.151$\pm$0.005 & 0.614$\pm$0.010 & 57948.4284 \\
13432 & 515504357949 & 0.45645736 & 16.381$\pm$0.006 & 1.138$\pm$0.050 & 7974.0969 & 15.319$\pm$0.005 & 0.641$\pm$0.009 & 57952.1844 \\
02515 & 515633495772 & 0.47702220 & 16.644$\pm$0.007 & 0.737$\pm$0.049 & 7971.7347 & 14.960$\pm$0.004 & 0.314$\pm$0.009 & 57893.0166 \\
09543 & 515599952081 & 0.51556077 & 17.156$\pm$0.008 & 0.908$\pm$0.042 & 7974.2127 & 15.436$\pm$0.005 & 0.566$\pm$0.010 & 57974.2092 \\
14578 & 515597002824 & 0.54341333 & 16.319$\pm$0.006 & 0.924$\pm$0.063 & 7674.7014 & 15.289$\pm$0.004 & 0.582$\pm$0.015 & 57652.9607 \\
\multicolumn{9}{c}{---RRab2---}\\
14806 & 515567300731 & 0.56104602 & 16.127$\pm$0.012 & 1.203$\pm$0.093 & 7974.3310 & 15.281$\pm$0.004 & 0.771$\pm$0.018 & 57936.1768 \\
34618 & 515597253109 & 0.58514134 & 15.440$\pm$0.009 & 1.203$\pm$0.091 & 7675.1019 & 14.648$\pm$0.004 & 0.806$\pm$0.026 & 57652.8608 \\ 
11992 & 515526076762 & 0.61353035 & 16.418$\pm$0.007 & 0.838$\pm$0.032 & 7974.9419 & 15.178$\pm$0.004 & 0.544$\pm$0.005 & 57954.0771 \\ 
33059 & 515657828314 & 0.63856513 & 16.244$\pm$0.005 & 0.843$\pm$0.070 & 7672.5702 & 15.215$\pm$0.005 & 0.531$\pm$0.012 & 57649.5760 \\
08440 & 515539115406 & 0.67505760 & 16.897$\pm$0.006 & 0.508$\pm$0.022 & 7975.4227 & 15.033$\pm$0.004 & 0.222$\pm$0.004 & 57953.1386 \\
\multicolumn{9}{c}{---RRab3---}\\
13220 & 515535451732 & 0.70481317 & 16.662$\pm$0.005 & 0.392$\pm$0.017 & 7974.4909 & 15.297$\pm$0.005 & 0.256$\pm$0.005 & 57944.1746 \\
10755 & 515526242552 & 0.73419273 & 16.848$\pm$0.006 & 0.144$\pm$0.008 & 7975.4450 & 15.477$\pm$0.005 & 0.105$\pm$0.005 & 57954.1314 \\
35604 & 515551783166 & 0.77483099 & 15.670$\pm$0.004 & 0.665$\pm$0.052 & 7681.1624 & 14.760$\pm$0.004 & 0.419$\pm$0.009 & 57609.8699 \\
04325 & 515667731582 & 0.82203622 & 16.166$\pm$0.004 & 0.311$\pm$0.032 & 7665.9812 & 14.964$\pm$0.004 & 0.191$\pm$0.006 & 57568.9618 \\
14958 & 515597272287 & 0.84151947 & 16.538$\pm$0.006 & 0.782$\pm$0.070 & 7674.6534 & 15.266$\pm$0.004 & 0.483$\pm$0.011 & 57652.7685 \\
15775 & 515545653428 & 0.87622048 & 15.655$\pm$0.005 & 0.523$\pm$0.050 & 7680.5802 & 14.711$\pm$0.004 & 0.330$\pm$0.009 & 57609.5979 \\
\hline
 \end{tabular}
 \tablefoot{
 \tablefoottext{a}{The complete ID is OGLE-BLG-RRLYR-NNNNN, where ``NNNNN''
 is the ID appearing in the first column.}
  \tablefoottext{b}{Heliocentric Julian Date -- 2,400,000 days.}}
\label{tab:oglevvv}
 \end{table*} 
 
The validation with the BTVS RRLs follows the approach adopted for the 
$\omega$ Cen RRLs (see Section~\ref{validation_wcen}). The key idea is to 
compare the mean magnitude estimated by using the template  
($\langle K_s \rangle_{templ}$) with the mean magnitude evaluated by using 
the $K_s$ band measurements ($\langle K_s \rangle_{best}$). Note that for these 
objects  we will compare eight independent estimates of 
$\langle K_s \rangle_{templ}$, because we will apply Fourier and PEGASUS fits  
to the light curve parameters based on the $V$- and on the $I$-band data. 
Moreover, the validation will be applied to both single phase points and 
triple phase points. We will add suffixes to 
the subscript of $\langle K_s \rangle_{templ[P/F][V/I][1/3]}$. 
where [$P/F$] indicates that we used either the PEGASUS or 
the Fourier fit, [$V/I$] indicates that we used either the $V$- or
the $I$-band data, and [1/3] indicates that we used either the single
phase point or the triple phase points.

\begin{figure*}[!htbp]
\centering
\includegraphics[trim={0 1.28cm 0 0},clip,width=12cm]{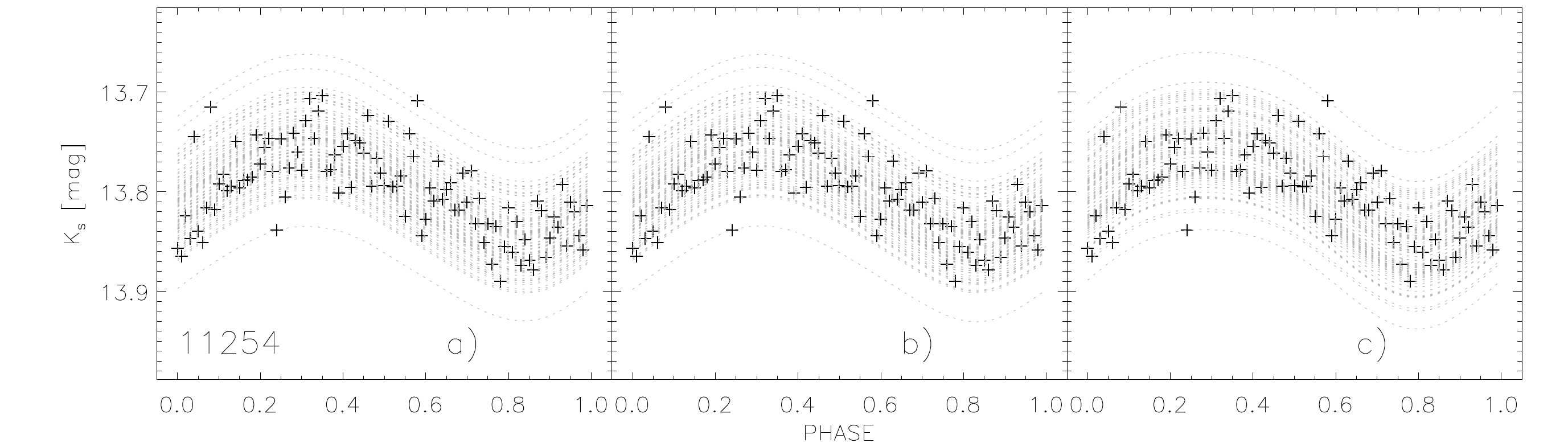}\\
\includegraphics[trim={0 1.28cm 0 0.15cm},clip,width=12cm]{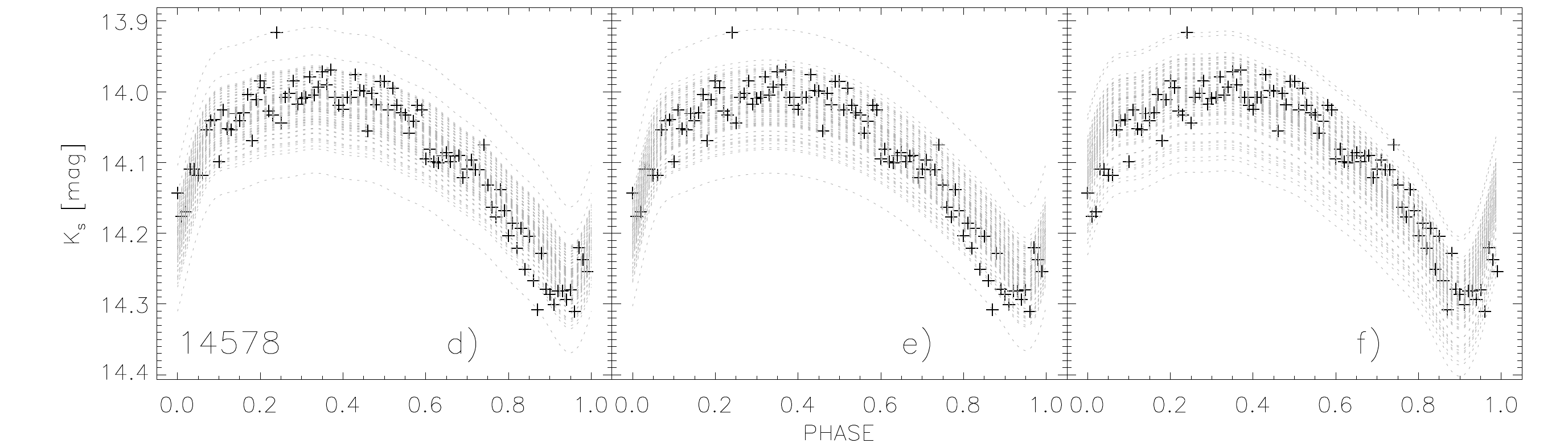}\\
\includegraphics[trim={0 1.28cm 0 0.15cm},clip,width=12cm]{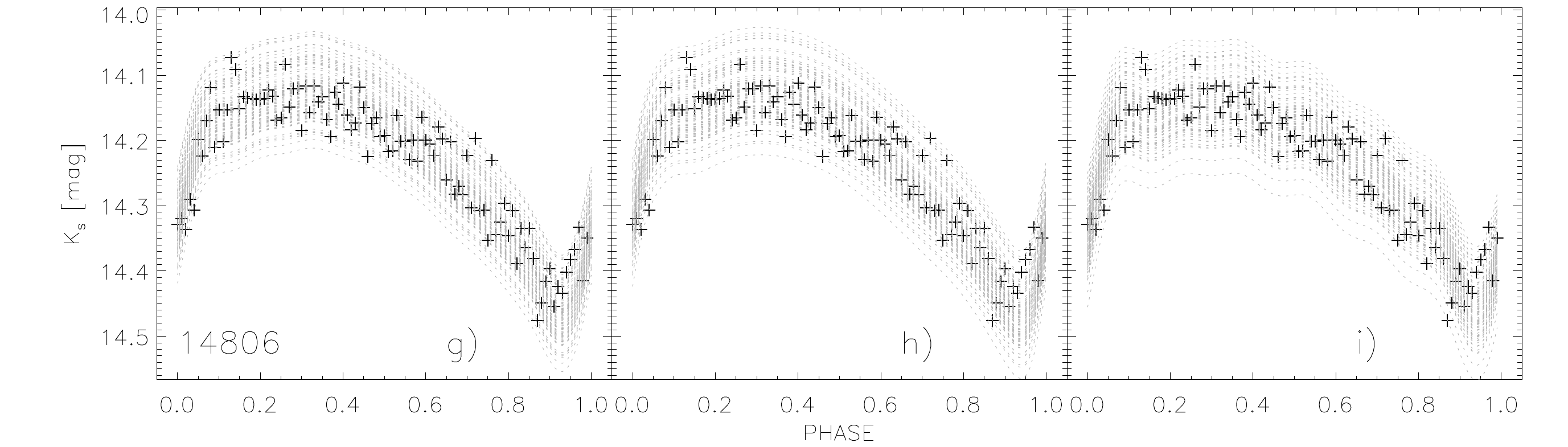}\\
\includegraphics[trim={0 0 0 0.15cm},clip,width=12cm]{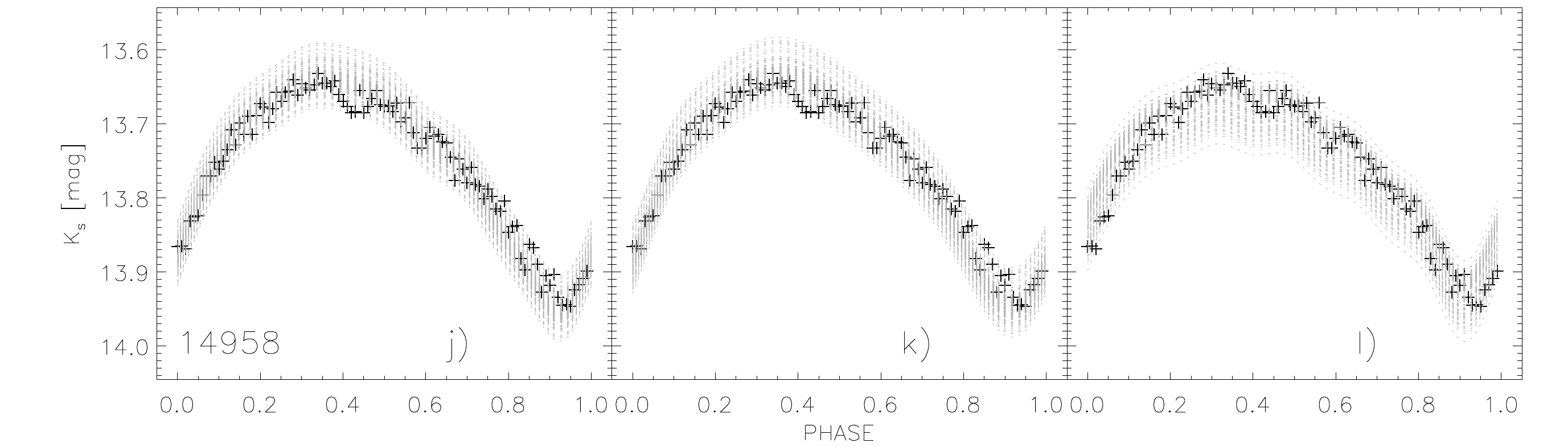}\\
\caption{a),b),c) panels: Black pluses represent the randomly extracted $K_s$-band phase points over
the light curve of the OGLE RRc variable 11254. Gray dashed lines display the template fits to the 
individual phase points. The Fourier (a), PEGASUS (b) and J96 (c) RRc light curve templates are 
also displayed. The ID of the RRL is labeled.
d),e),f): Same as panels a),b),c), but for the OGLE RRab variable 14578. The RRab1 light curve templates are displayed.
g),h),i): Same as panels a),b),c), but for the OGLE RRab variable 14806. The RRab2 light curve templates are displayed.
j),k),l): Same as panels a),b),c), but for the OGLE RRab variable 14958. The RRab3 light curve templates are displayed.}
\label{fig:ogle}
\end{figure*}

The two methods are identical to those described in 
Section~\ref{wcen_1punto} and \ref{wcen_3punti}. The only difference 
is that in this case we have more than one RRL per template bin.
Therefore, we also estimate the median difference 
$\langle\Delta\langle K_s \rangle \rangle$ for all the RRLs in the 
period bin. The results are listed in Table~\ref{tab:oglevvv2}.
Fig.~\ref{fig:ogle} displays the fits to four BTVS RRLs, one for 
each template bin.


\begin{sidewaystable*}
 \scriptsize
 \caption{NIR photometric properties of the BTVS RRLs.}
  \centering
 \begin{tabular}{l c c c c c c c c c c c}
 \hline\hline  
ID (OGLE)\tablefootmark{a} & $\langle K_s \rangle_{best}$ & $\langle K_s \rangle_{templFV1}$ & $\langle K_s \rangle_{templFI1}$  & $\langle K_s \rangle_{templPV1}$ & $\langle K_s \rangle_{templPI1}$  & $\langle K_s \rangle_{templJ1}$  & $\langle K_s \rangle_{templFV3}$ & $\langle K_s \rangle_{templFI3}$  & $\langle K_s \rangle_{templPV3}$ & $\langle K_s \rangle_{templPI3}$ & $\langle K_s \rangle_{templJ3}$  \\
 & mag & mag & mag & mag & mag & mag & mag & mag & mag & mag & mag \\
 \hline 
\multicolumn{10}{c}{---RRc---}\\
15624 & 13.991$\pm$0.021 & 13.988$\pm$0.039 & 13.996$\pm$0.034 & 13.986$\pm$0.038 & 13.991$\pm$0.035 & 13.993$\pm$0.036 & 13.989$\pm$0.027 & 13.991$\pm$0.029 & 13.990$\pm$0.025 & 13.994$\pm$0.029 & 13.991$\pm$0.029 \\
34149 & 14.450$\pm$0.018 & 14.447$\pm$0.018 & 14.450$\pm$0.016 & 14.446$\pm$0.017 & 14.449$\pm$0.015 & 14.445$\pm$0.022 & 14.452$\pm$0.014 & 14.450$\pm$0.011 & 14.450$\pm$0.013 & 14.450$\pm$0.011 & 14.450$\pm$0.011 \\
35612 & 13.906$\pm$0.012 & 13.907$\pm$0.026 & 13.905$\pm$0.023 & 13.902$\pm$0.024 & 13.901$\pm$0.021 & 13.900$\pm$0.027 & 13.906$\pm$0.019 & 13.905$\pm$0.015 & 13.907$\pm$0.015 & 13.906$\pm$0.018 & 13.905$\pm$0.015 \\
11254 & 13.797$\pm$0.027 & 13.794$\pm$0.028 & 13.798$\pm$0.025 & 13.802$\pm$0.029 & 13.795$\pm$0.025 & 13.796$\pm$0.030 & 13.797$\pm$0.021 & 13.796$\pm$0.019 & 13.796$\pm$0.020 & 13.797$\pm$0.019 & 13.796$\pm$0.019 \\
04844 & 14.221$\pm$0.024 & 14.222$\pm$0.069 & 14.226$\pm$0.066 & 14.218$\pm$0.077 & 14.222$\pm$0.073 & 14.221$\pm$0.070 & 14.214$\pm$0.044 & 14.223$\pm$0.038 & 14.211$\pm$0.039 & 14.220$\pm$0.042 & 14.223$\pm$0.038 \\
34454 & 13.841$\pm$0.011 & 13.841$\pm$0.022 & 13.840$\pm$0.020 & 13.840$\pm$0.023 & 13.835$\pm$0.024 & 13.836$\pm$0.025 & 13.836$\pm$0.016 & 13.837$\pm$0.014 & 13.836$\pm$0.017 & 13.840$\pm$0.016 & 13.837$\pm$0.014 \\
 & $\langle \Delta \langle K_s \rangle\rangle$: & --0.001$\pm$0.002  &  0.000$\pm$0.003  &  --0.004$\pm$0.004   &--0.002$\pm$0.003  &--0.003$\pm$0.003  & --0.001$\pm$0.003  & --0.001$\pm$0.002 &--0.001$\pm$0.004  &--0.001$\pm$0.002 &--0.005$\pm$0.003 \\\\
\hline
\multicolumn{10}{c}{---RRab1---}\\
13498 & 13.803$\pm$0.024 & 13.814$\pm$0.035 & 13.811$\pm$0.043 & 13.803$\pm$0.039 & 13.801$\pm$0.040 & 13.806$\pm$0.036 & 13.802$\pm$0.033 & 13.792$\pm$0.041 & 13.794$\pm$0.032 & 13.798$\pm$0.043 & 13.792$\pm$0.041 \\
13432 & 13.488$\pm$0.018 & 13.488$\pm$0.084 & 13.486$\pm$0.072 & 13.489$\pm$0.078 & 13.492$\pm$0.076 & 13.491$\pm$0.075 & 13.501$\pm$0.056 & 13.501$\pm$0.056 & 13.527$\pm$0.058 & 13.485$\pm$0.052 & 13.501$\pm$0.056 \\
02515 & 12.622$\pm$0.007 & 12.639$\pm$0.053 & 12.631$\pm$0.030 & 12.638$\pm$0.052 & 12.631$\pm$0.031 & 12.636$\pm$0.058 & 12.644$\pm$0.033 & 12.636$\pm$0.019 & 12.645$\pm$0.030 & 12.635$\pm$0.021 & 12.636$\pm$0.019 \\
09543 & 13.297$\pm$0.013 & 13.299$\pm$0.078 & 13.307$\pm$0.074 & 13.302$\pm$0.074 & 13.286$\pm$0.069 & 13.298$\pm$0.075 & 13.279$\pm$0.068 & 13.301$\pm$0.053 & 13.284$\pm$0.055 & 13.283$\pm$0.052 & 13.301$\pm$0.053 \\
14578 & 14.086$\pm$0.014 & 14.084$\pm$0.029 & 14.087$\pm$0.032 & 14.090$\pm$0.030 & 14.078$\pm$0.032 & 14.079$\pm$0.038 & 14.077$\pm$0.022 & 14.084$\pm$0.033 & 14.076$\pm$0.037 & 14.082$\pm$0.026 & 14.084$\pm$0.033 \\
\hline
 & $\langle \Delta \langle K_s \rangle\rangle$: &   0.002$\pm$0.010  &  0.008$\pm$0.006  &    0.004$\pm$0.007   &--0.002$\pm$0.008  &  0.003$\pm$0.008  &   0.000$\pm$0.016  &   0.003$\pm$0.010 &--0.009$\pm$0.029  &--0.005$\pm$0.010 &--0.001$\pm$0.007 \\\\
\multicolumn{10}{c}{---RRab2---}\\
14806 & 14.226$\pm$0.020 & 14.229$\pm$0.037 & 14.228$\pm$0.043 & 14.230$\pm$0.038 & 14.231$\pm$0.044 & 14.227$\pm$0.046 & 14.223$\pm$0.028 & 14.223$\pm$0.046 & 14.228$\pm$0.048 & 14.220$\pm$0.051 & 14.223$\pm$0.046 \\
34618 & 13.635$\pm$0.009 & 13.637$\pm$0.026 & 13.636$\pm$0.025 & 13.636$\pm$0.028 & 13.640$\pm$0.029 & 13.629$\pm$0.034 & 13.634$\pm$0.025 & 13.635$\pm$0.025 & 13.633$\pm$0.026 & 13.638$\pm$0.022 & 13.635$\pm$0.025 \\
11992 & 13.677$\pm$0.024 & 13.682$\pm$0.042 & 13.670$\pm$0.035 & 13.673$\pm$0.036 & 13.675$\pm$0.039 & 13.670$\pm$0.037 & 13.666$\pm$0.040 & 13.668$\pm$0.028 & 13.672$\pm$0.044 & 13.662$\pm$0.048 & 13.668$\pm$0.028 \\
33059 & 13.945$\pm$0.014 & 13.940$\pm$0.038 & 13.945$\pm$0.037 & 13.937$\pm$0.039 & 13.948$\pm$0.036 & 13.935$\pm$0.032 & 13.940$\pm$0.028 & 13.940$\pm$0.030 & 13.946$\pm$0.042 & 13.936$\pm$0.032 & 13.940$\pm$0.030 \\
08440 & 12.802$\pm$0.017 & 12.806$\pm$0.054 & 12.806$\pm$0.050 & 12.800$\pm$0.050 & 12.796$\pm$0.050 & 12.796$\pm$0.060 & 12.803$\pm$0.030 & 12.798$\pm$0.029 & 12.800$\pm$0.039 & 12.795$\pm$0.027 & 12.798$\pm$0.029 \\
 & $\langle \Delta \langle K_s \rangle\rangle$: &   0.004$\pm$0.004  &  0.001$\pm$0.004  &  --0.003$\pm$0.005   &  0.004$\pm$0.006  &--0.006$\pm$0.004  & --0.003$\pm$0.005  & --0.004$\pm$0.004 &--0.002$\pm$0.003  &--0.007$\pm$0.007 &--0.005$\pm$0.007 \\\\
\hline
\multicolumn{10}{c}{---RRab3---}\\
13220 & 13.762$\pm$0.018 & 13.764$\pm$0.027 & 13.755$\pm$0.027 & 13.765$\pm$0.030 & 13.761$\pm$0.027 & 13.746$\pm$0.039 & 13.756$\pm$0.024 & 13.755$\pm$0.028 & 13.754$\pm$0.032 & 13.759$\pm$0.028 & 13.755$\pm$0.028 \\
10755 & 13.748$\pm$0.025 & 13.746$\pm$0.026 & 13.746$\pm$0.028 & 13.749$\pm$0.030 & 13.740$\pm$0.027 & 13.748$\pm$0.046 & 13.749$\pm$0.018 & 13.746$\pm$0.021 & 13.750$\pm$0.020 & 13.745$\pm$0.023 & 13.746$\pm$0.021 \\
35604 & 13.680$\pm$0.010 & 13.679$\pm$0.019 & 13.683$\pm$0.021 & 13.679$\pm$0.020 & 13.679$\pm$0.019 & 13.674$\pm$0.031 & 13.678$\pm$0.028 & 13.682$\pm$0.026 & 13.679$\pm$0.018 & 13.677$\pm$0.024 & 13.682$\pm$0.026 \\
04325 & 13.569$\pm$0.010 & 13.576$\pm$0.049 & 13.566$\pm$0.046 & 13.566$\pm$0.055 & 13.573$\pm$0.046 & 13.563$\pm$0.054 & 13.570$\pm$0.035 & 13.569$\pm$0.036 & 13.565$\pm$0.038 & 13.569$\pm$0.034 & 13.569$\pm$0.036 \\
14958 & 13.750$\pm$0.011 & 13.749$\pm$0.017 & 13.748$\pm$0.016 & 13.746$\pm$0.017 & 13.749$\pm$0.017 & 13.743$\pm$0.020 & 13.751$\pm$0.021 & 13.747$\pm$0.015 & 13.753$\pm$0.013 & 13.754$\pm$0.016 & 13.747$\pm$0.015 \\
15775 & 13.579$\pm$0.012 & 13.577$\pm$0.030 & 13.580$\pm$0.023 & 13.570$\pm$0.026 & 13.578$\pm$0.027 & 13.565$\pm$0.036 & 13.579$\pm$0.018 & 13.577$\pm$0.021 & 13.576$\pm$0.025 & 13.574$\pm$0.022 & 13.577$\pm$0.021 \\
 & $\langle \Delta \langle K_s \rangle\rangle$: & --0.001$\pm$0.004  &--0.002$\pm$0.004  &  --0.002$\pm$0.004   &--0.001$\pm$0.004  &--0.007$\pm$0.006  &   0.000$\pm$0.003  & --0.002$\pm$0.003 &--0.002$\pm$0.004  &--0.003$\pm$0.003 &--0.009$\pm$0.006 \\\\
\hline
 \end{tabular}
 \tablefoot{
 \tablefoottext{a}{The complete ID is OGLE-BLG-RRLYR-NNNNN, where ``NNNNN''
 is the ID appearing in the first column.}}
\label{tab:oglevvv2}
 \end{sidewaystable*}

In this context it is worth mentioning that two (OGLE ID: 34618, 11992) 
out of the 22 BTVS RRLs, both belonging to the RRab2 period bin, are 
Blazhko RRLs. The amplitude modulation is 0.2 mag in the $I$ band and 
0.3 mag in the $V$ band. The Blazhko modulation does not significantly 
affect the mean magnitude ($\Delta \langle K_s \rangle$) for two 
main reasons. {\it i)} The OGLE data are well sampled and we could estimate the 
average amplitude over the Blazhko cycle. {\it ii)} Blazhko variables with 
extreme amplitude modulation, i.e., 0.5 mag in $V$, and sampled only across 
the phases of the maximum will be affected by an error of the order of 0.4 mag 
in $V$ amplitude. The impact of this amplitude uncertainty on the mean magnitude 
estimated by using the template is minimal, indeed it is of the order of 
0.002 mag in the $J$ band and even smaller for the other bands.  
Note, however, that this limitation becomes severe for the J96 
RRab templates, because the different light-curve templates 
are based on the luminosity amplitude. The use of a wrong template 
causes a systematic error in the mean magnitude of the order of a 
few hundredths of a magnitude.

 \section{Application of the new light curve templates to Reticulum RRLs}\label{sect_reticulum}
 
Reticulum is an extragalactic Globular Cluster associated with
the halo of the Large Magellanic Cloud (LMC). It hosts a 
sizable sample of RRLs \citep[32 in total][]{walker92_reticulum}
and it is an interesting workbench, because the J96 light curve 
templates were adopted by \citet{dallora04} to derive the mean 
$K_s$-band magnitudes of 30 RRLs that were observed with SOFI 
at NTT.  
However, the mean $J$-band magnitudes were estimated as the 
mean of the measurements. The number of measurements was limited, 
typically 46 unbinned phase points, which means on average ten binned 
phase points (see below). This means that the classical analytical fits 
(spline, Fourier series) could be applied.  Moreover, the $J$-band light 
curve templates were not available. These are the reasons why the 
authors focused their cluster distance determinations only on the 
$K_s$ band PL relation.  The new light-curve templates will be used to 
provide new $J$- and $K_s$-band mean magnitudes, new NIR PL relations, 
and---in turn---new cluster distance determinations.

\subsection{Phasing of the data and application of the light curve templates}\label{reticulum_phasing}

\begin{figure*}[!htbp]
\centering
\includegraphics[trim={0.8cm 4.3cm 1.7cm 0},clip,width=6cm]{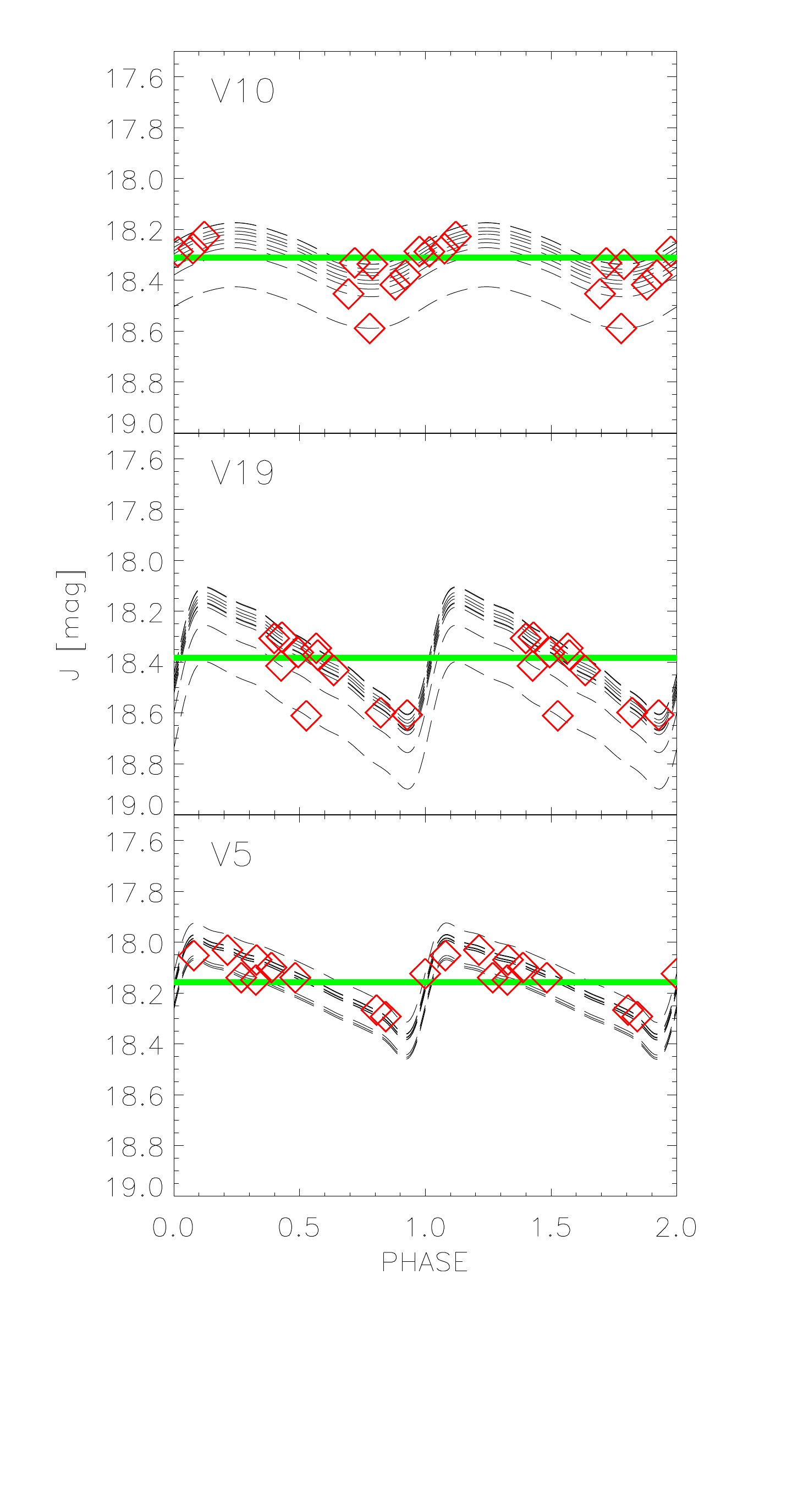}
\includegraphics[trim={0.8cm 4.3cm 1.7cm 0},clip,width=6cm]{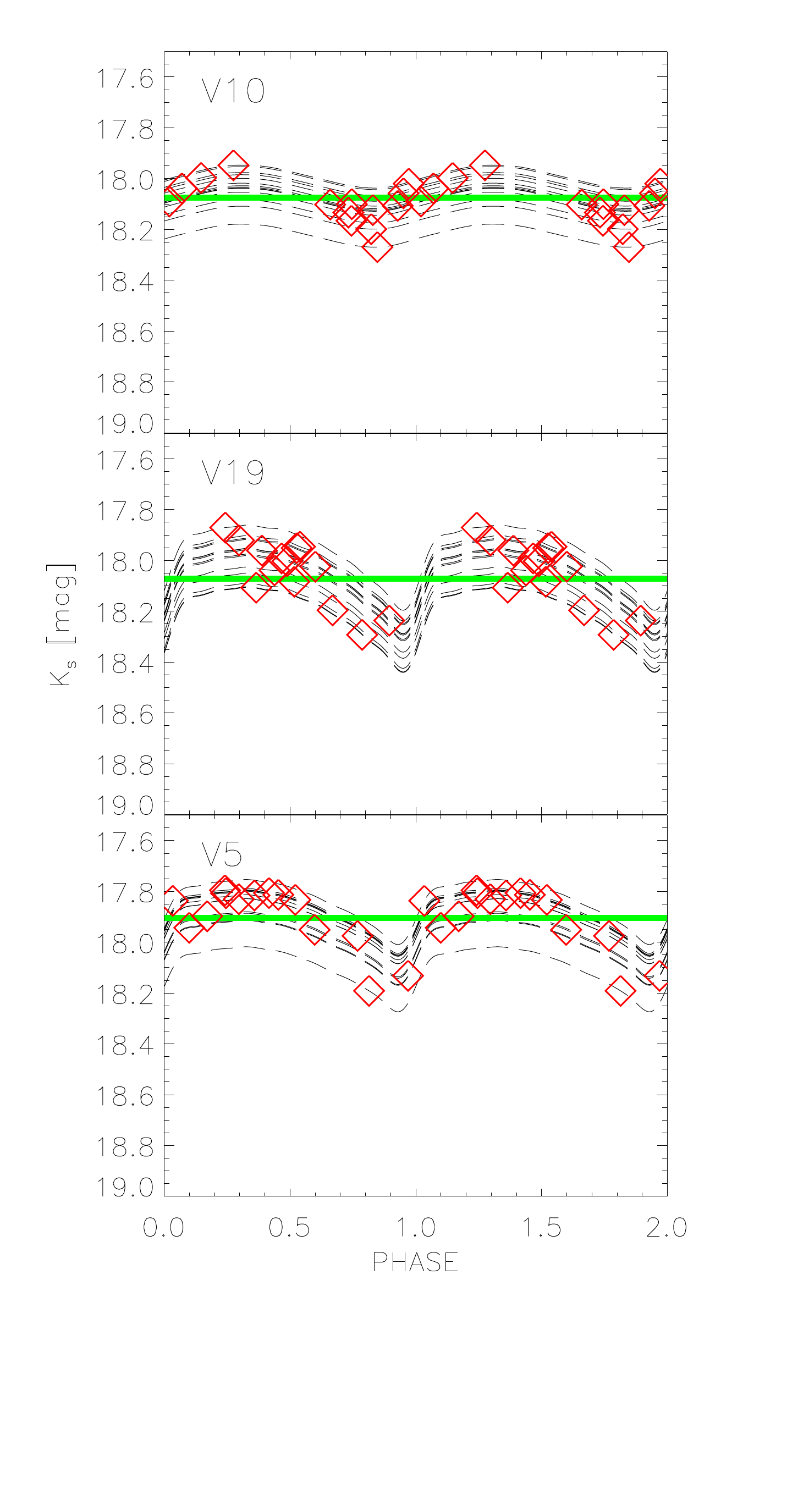}
\caption{Top: $J$ (left panel) and $K_s$ (right panel) band light curve for the 
RRc variable V10. The red diamonds display the binned 
phase points. The black dashed curves show the light curve template applied to 
the individual binned phase points. The thick green line displays the mean 
magnitude listed in Table~\ref{tab:reticulum}.
Middle: Same as the top, but for the RRab variable V19.
Bottom: Same as the top, but for the RRab variable V5.}
\label{fig:lcvsreticulum}
\end{figure*}

We plan to use the photometric data collected by \citet{dallora04}, but 
we will derive new NIR ($JK_s$) curves. In particular, we plan to take advantage 
of the new pulsation periods and epoch of the anchor point recently provided by 
\citep{kuehn2013a}. Moreover, the SOFI $JK_s$-band data were binned using the 
same approach adopted in \citet{braga2018}. The data collected in one dither 
pattern were binned into a single phase point using a time interval of 108 sec.  
The binned $J$- and $K_s$-band light curves have a number of phase points ranging 
from ten to fourteen. The $J$- and $K_s$-band light curves of three variables, 
V10 (RRc), V19 (RRab1) and V5 (RRab2), are displayed in Fig.~\ref{fig:lcvsreticulum} 
together with the template fits (black dashed lines) and the mean magnitude 
(green solid line).

We have folded the light curves with the periods published 
by \citet{kuehn2013a}. However, the decimal places provided 
in their Table 1 and 2 are limited and for seven RRLs (V3, V4, V11, V15, V24, V28, V32, 
using the new notation introduced by \citealt{kuehn2013a}), 
the folded light curves show significant phase drifts. Therefore, 
for these RRLs we estimated our own periods, based on their $V$-band 
light curves (see Table~\ref{tab:reticulum}).

\begin{table*}[!htbp]
 \caption{Optical and NIR photometric properties of Reticulum RRLs.}
  \centering
 \begin{tabular}{l l c c c c c}
 \hline\hline  
ID & Period & $\langle V \rangle$ & $AV$ & $t_{ris}$\tablefootmark{a} & $\langle J \rangle$ & $\langle K \rangle$ \\
 & days & mag & mag & days & mag & mag  \\
 \hline 
 V01 & 0.50993000                  & 19.030$\pm$0.018 &  1.15$\pm$0.05 & 55595.5036 &  \ldots        &      \ldots    \\    
 V02 & 0.61869000                  & 19.084$\pm$0.018 &  0.63$\pm$0.03 & 55595.6344 & 18.09$\pm$0.11 & 17.84$\pm$0.06 \\  
 V03 & 0.35354552\tablefootmark{b} & 19.053$\pm$0.018 &  0.42$\pm$0.02 & 55595.7200 & 18.24$\pm$0.04 & 18.03$\pm$0.08 \\  
 V04 & 0.35322097\tablefootmark{b} & 19.059$\pm$0.102 &  0.41$\pm$0.02 & 55595.6335 & 18.29$\pm$0.07 & 18.08$\pm$0.06 \\  
 V05 & 0.57185000                  & 19.042$\pm$0.018 &  0.90$\pm$0.04 & 55595.6783 & 18.15$\pm$0.05 & 17.90$\pm$0.07 \\  
 V06 & 0.59526000                  & 19.105$\pm$0.019 &  0.59$\pm$0.03 & 55595.9238 & 18.17$\pm$0.10 & 17.86$\pm$0.10 \\  
 V07 & 0.51044000                  & 19.011$\pm$0.019 &  1.14$\pm$0.04 & 55595.3900 & 18.21$\pm$0.06 & 18.00$\pm$0.11 \\  
 V08 & 0.64496000                  & 19.075$\pm$0.018 &  0.41$\pm$0.02 & 55595.6523 & 18.06$\pm$0.09 & 17.69$\pm$0.14 \\  
 V09 & 0.54496000                  & 19.007$\pm$0.018 &  0.80$\pm$0.04 & 55595.5549 & 18.22$\pm$0.05 & 17.94$\pm$0.07 \\  
 V10 & 0.35256000                  & 19.079$\pm$0.018 &  0.43$\pm$0.02 & 55595.2460 & 18.31$\pm$0.08 & 18.07$\pm$0.06 \\  
 V11 & 0.35539753\tablefootmark{b} & 19.072$\pm$0.020 &  0.44$\pm$0.03 & 55595.3895 & 18.34$\pm$0.07 & 18.11$\pm$0.07 \\  
 V12 & 0.29627000                  & 18.983$\pm$0.016 &  0.22$\pm$0.02 & 55595.5187 & 18.40$\pm$0.06 & 18.24$\pm$0.08 \\  
 V13 & 0.60958000                  & 19.093$\pm$0.019 &  0.72$\pm$0.04 & 55595.2470 & 18.13$\pm$0.08 & 17.81$\pm$0.07 \\  
 V14 & 0.58661000                  & 19.059$\pm$0.019 &  0.69$\pm$0.02 & 55595.6339 & 18.21$\pm$0.14 & 17.95$\pm$0.10 \\  
 V15 & 0.35427716\tablefootmark{b} & 19.092$\pm$0.019 &  0.42$\pm$0.03 & 55595.5856 & 18.31$\pm$0.07 & 18.11$\pm$0.10 \\  
 V16 & 0.52290000                  & 19.054$\pm$0.018 &  1.12$\pm$0.05 & 55595.7704 & 18.27$\pm$0.06 & 17.98$\pm$0.08 \\  
 V17 & 0.51241000                  & 19.041$\pm$0.019 &  1.14$\pm$0.10 & 55595.4844 & 18.25$\pm$0.22 & 18.09$\pm$0.11 \\  
 V18 & 0.56005000                  & 19.080$\pm$0.019 &  0.93$\pm$0.04 & 55595.4833 & 18.14$\pm$0.07 & 17.91$\pm$0.06 \\  
 V19 & 0.48485000                  & 19.056$\pm$0.019 &  1.22$\pm$0.04 & 55595.6953 & 18.38$\pm$0.09 & 18.07$\pm$0.08 \\  
 V20 & 0.56075000                  & 19.123$\pm$0.021 &  0.71$\pm$0.03 & 55595.6680 & 18.26$\pm$0.16 & 17.89$\pm$0.09 \\  
 V21 & 0.60700000                  & 19.094$\pm$0.019 &  0.70$\pm$0.03 & 55596.1339 & 18.19$\pm$0.17 & 17.76$\pm$0.12 \\  
 V22 & 0.51359000                  & 19.069$\pm$0.018 &  0.89$\pm$0.04 & 55595.6068 & 18.22$\pm$0.08 & 17.91$\pm$0.12 \\  
 V23 & 0.46863000                  & 19.162$\pm$0.021 &  0.95$\pm$0.04 & 55595.8827 & 18.33$\pm$0.18 & 18.10$\pm$0.09 \\  
 V24 & 0.34752424\tablefootmark{b} & 19.092$\pm$0.020 &  0.40$\pm$0.02 & 55595.8014 & 18.38$\pm$0.06 & 18.09$\pm$0.06 \\  
 V25 & 0.32991000                  & 19.048$\pm$0.018 &  0.50$\pm$0.02 & 55595.4580 & 18.39$\pm$0.06 & 18.21$\pm$0.14 \\  
 V26 & 0.65696000                  & 19.087$\pm$0.018 &  0.28$\pm$0.02 & 55595.3645 & 18.11$\pm$0.11 & 17.75$\pm$0.09 \\  
 V27 & 0.51382000                  & 19.062$\pm$0.020 &  1.22$\pm$0.07 & 55595.5439 & 18.16$\pm$0.16 & 17.95$\pm$0.09 \\  
 V28 & 0.31994112\tablefootmark{b} & 18.999$\pm$0.018 &  0.49$\pm$0.02 & 55595.3373 & 18.37$\pm$0.08 & 18.14$\pm$0.10 \\  
 V29 & 0.50815000                  & 19.063$\pm$0.018 &  1.14$\pm$0.04 & 55595.4709 & 18.36$\pm$0.11 & 18.06$\pm$0.07 \\  
 V30 & 0.53501000                  & 19.012$\pm$0.019 &  1.04$\pm$0.05 & 55595.7935 & 18.28$\pm$0.12 & 17.99$\pm$0.07 \\  
 V31 & 0.50516000                  & 19.087$\pm$0.018 &  1.07$\pm$0.05 & 55595.5749 & 18.31$\pm$0.08 & 18.03$\pm$0.07 \\  
 V32 & 0.35225470\tablefootmark{b} & 19.049$\pm$0.017 &  0.42$\pm$0.02 & 55595.7202 &  \ldots        &      \ldots  \\ 
 \hline
 \end{tabular}
 \tablefoot{\tablefoottext{a}{Heliocentric Julian Date -- 2,400,000 days.}
 \tablefoottext{b}{New pulsation periods from our own analysis.}}
\label{tab:reticulum}
 \end{table*}
 

Subsequently, we estimated $t_{ris}$ from the $V$-band light curves provided 
by \citet{kuehn2013a}. We fit the optical light curves using the PLOESS method 
described in \citet{braga2018}. We found that the difference between
our $V$-band mean magnitudes and those provided by  \citet{kuehn2013a} is 
negligible, with a mean of 0.003 mag, a standard deviation of 0.012 mag and 
a maximum difference of 0.035 mag. On the basis of the new periods and 
of the new epochs ($t_{ris}$), we folded the NIR light curves.

It is worth mentioning that Reticulum hosts six mixed-mode RRLs (RRd) 
and we have NIR data for five of them (except V32). We do not provide
templates for this type of variable, but since the dominant
mode is the first overtone, we decided to apply 
the RRc light curve template to these variables.
 
To apply the template, we need an estimate of the optical amplitudes
of the RRLs and of the NIR-to-optical amplitude ratios \citep{braga2018} to 
rescale the template function. We decided to adopt our own $V$-band 
amplitudes---estimated from the PLOESS fits derived in 
Section~\ref{reticulum_phasing}---because they differ from
those published by \citet{kuehn2013a}. The mean difference 
$\Delta AV = AV_{our}-AV_{K13}$ is --0.08 mag, with a standard deviation 
of 0.07 mag and a maximum difference of --0.32 mag. We obtained smaller 
luminosity amplitudes because, for Blazhko and RRd variables, we did not 
fit the brighter/fainter envelopes of the data (\citet{kuehn2013a}) 
since we are interested in the application of the template to determine 
their NIR mean magnitudes.

We then applied to each phase point of the NIR binned light curve 
both the PEGASUS and the Fourier light curve templates. This means 
that we estimated two mean magnitudes ($\langle J \rangle_i$, 
$\langle K_s \rangle_i$) per phase point, where $i$ indicates 
the $i$-th phase point. Interestingly enough, the Fourier and the 
PEGASUS templates provide, within the photometric uncertainty of the 
individual phase points, similar estimates of both $\langle J \rangle_i$ 
and $\langle K_s \rangle_i$.  The final values of $\langle J \rangle$ 
and $\langle K_s \rangle$ are the medians of all the $\langle J \rangle_i$ 
and $\langle K_s \rangle_i$. They are listed in Table~\ref{tab:reticulum}, 
together with their standard deviations.

\subsection{New empirical $J$ and $K_s$ PL relations and the distance to Reticulum}

We derived the PL relations in the $J$ and $K_s$ band after 
correcting the NIR mean magnitudes for reddening. Following the same 
arguments of \citet{muraveva18a}, we adopted the cluster reddening 
(E(\bmv)=0.03$\pm$0.02 mag) originally derived by \citep{walker92_reticulum}. 
We also adopted $R_V$=3.1 and the optical-to-NIR extinction ratios 
by \citet{cardelli89}. Note that in the current PL relations the periods 
of RRc and RRd variables were ``fundamentalized'', i.e., we adopted 
$\log P_F=\log P_{FO}$+0.128 \citep{kuehn2013a}. We obtained the 
following PL relations, where $J_0$ and $K_{s0}$ indicate the un-reddened 
magnitudes:

\begin{equation}\label{eq:plj}
J_0   = (17.78\pm0.05) - (1.58\pm0.17) \cdot \log{P}
\end{equation}

\begin{equation}\label{eq:plk}
K_{s0} = (17.29\pm0.04) - (2.40\pm0.15) \cdot \log{P}
\end{equation}

\begin{figure*}[!htbp]
\centering
\includegraphics[width=12cm]{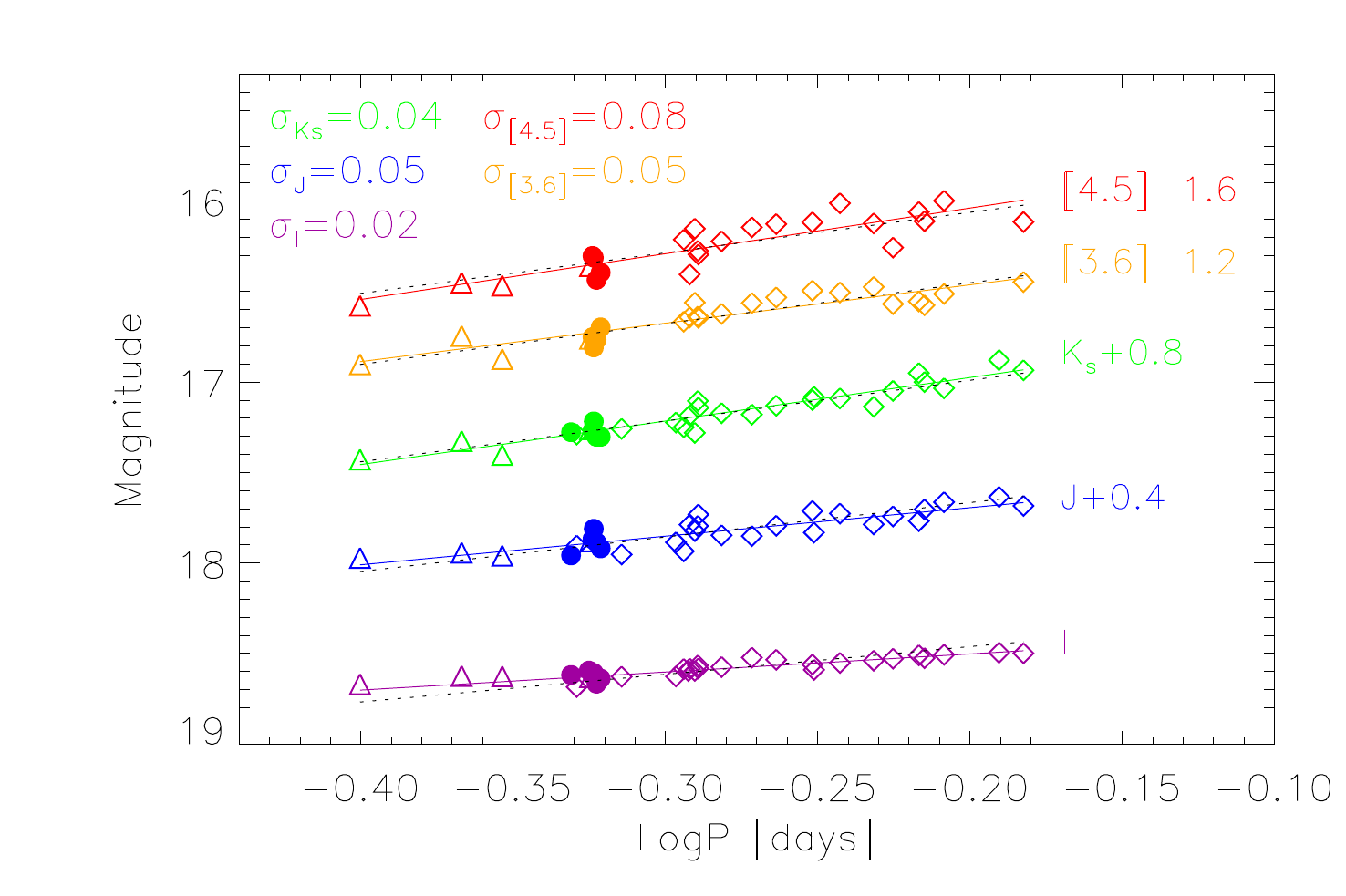}
\caption{$I$-, $J$-, $K_s$- [3.6]- and [4.5]-band PL relations of Reticulum RRLs. 
Diamonds display RRab variables, triangles RRc variables and 
circles the RRd variables. Purple, blue, green, orange and red symbols display the 
un-reddened mean magnitudes in the $I$, $J$, $K_s$ 
[3.6] and [4.5] bands, respectively. The solid lines
display the empirical PL relations (Equations~\ref{eq:plj}, \ref{eq:plk},
\ref{eq:pli1}, \ref{eq:pli2} and \ref{eq:pli}). 
The dashed black lines display the theoretical PLZ relations
by \citet{marconi15,neeley17}, at [Fe/H]=--1.70 (\citealt{suntzeff1992}, transformed  
into the \citealt{carretta09} metallicity scale) and 
artificially shifted in magnitude. The $I$ was only shifted for the current value 
of the true distance modulus. The standard deviation of the relations are 
labelled on the top-left corner.}
\label{fig:pljk}
\end{figure*}

The coefficients of the current empirical PL$K_s$ relation and their 
standard deviations are, within the errors, very similar to those 
obtained by \citet{dallora04}. The standard deviation 
of the PL$J$ relation is larger than in the PL$K_s$ relation 
(0.05 vs 0.04 mag), as suggested by theoretical predictions 
\citep[0.06 mag][]{marconi15}. Finally, we have estimated the true 
distance modulus ($\mu$) of Reticulum using the new NIR mean magnitudes
($J$,$K_s$) and the theoretical Global PLZ relations provided by 
\citet[][, Marconi et al (2018, in prep.)]{marconi15}. We have adopted
the spectroscopic iron abundance obtained by \citet{suntzeff1992} from 
Reticulum red giants, transformed into the \citep{carretta09} metallicity 
scale ([Fe/H]=--1.70). We found  
$\mu_J$=18.47$\pm$0.10$\pm$0.03 mag and $\mu_{Ks}$=18.49$\pm$0.09$\pm$0.05 mag, 
where the first is the standard error of the mean and the second the standard deviation. 
The latter was computed as the squared sum of the 
average uncertainty on the mean magnitudes only, since 
the uncertainty in the extinction and the propagation 
of the uncertainties in the calibrating PLZ coefficients 
vanish when square-summed.

The true distance modulus obtained by \citet{dallora04} from the same data, 
but using a different theoretical $K_s$-band PLZ \citep{bono03c} relation, was 
$\mu$=18.52$\pm$0.05 mag.  The three distance determinations agree quite well, and 
indeed the difference is within 1$\sigma$. 

The distance to Reticulum was estimated by \citet{kuehn2013a} using 
the visual mean magnitude-metallicity relation relation provided by 
\citet{catelan08}, a cluster metallicity of [Fe/H]=--1.66 
(\citealt{mackey04b}, in the \citealt{ZW84} scale) and a cluster 
reddening of E(\bmv)=0.016 mag \citep{schlegel98}. They found a 
true distance modulus of 18.40$\pm$0.20 mag. They also adopted 
the $I$-band PL relation provided by \citet{catelan04}, the same 
cluster reddening and the \citet{cardelli89} reddening law 
and they found a true distance modulus of 18.47$\pm$0.06 mag. 

The Reticulum true distance modulus was more recently estimated by 
\citet{muraveva18a} using Mid-Infrared (MIR) mean magnitudes, 
collected with IRAC at Spitzer, for 24 ([3.6]) and 23 ([4.5]) RRLs.  
They found true distance moduli of $\mu$=18.32$\pm$0.06 mag ([3.6]) 
and 18.34$\pm$0.08 mag ([4.5]) mag, adopting 
two empirical zero-points based on \gaia DR1 \citep{gaia_alldr} 
and \gaia  DR2 \citep{gaia_dr2,clementini2018} 
trigonometric parallaxes. They also adopted a third independent 
zero-point based on  \hst \citep{benedict11} trigonometric 
parallaxes for five field RRLs, and found that this calibration 
provides distances that are 0.10 mag larger than those based 
on the \gaia calibrations. Note that \citet{muraveva18a} adopted 
a different metal content ([Fe/H]=--1.66, \citealt{mackey04b}, 
in the \citealt{ZW84} scale), but the difference in cluster 
metallicity affects the distance only at the level of 0.01 mag.

The cluster distance found by \citet{muraveva18a} is smaller than the 
geometric distance to the LMC found by \citet{pietrzynski2013} 
($\mu_{LMC}$=18.493$\pm$0.008$\pm$0.047 mag) from late-type 
eclipsing binaries and by \citep{inno2016} ($\mu_{LMC}=18.48\pm0.10$ mag) 
from Classical Cepheids with optical/NIR ($VIJHK_s$; 
$\sim$4,000) and MIR ($w_1$, WISE photometric system; $\sim$2,600) 
measurements. Classical Cepheids are young (t$<$300 Myr), intermediate-mass 
stars and mainly trace the disk/bar of the galaxy. On the basis of their 
relative distances \citep{inno2016} found an LMC depth of the order of 
$\sim \pm$0.2 mag. This suggests that the intrinsic spread in distance 
along the line of sight is roughly the 10\% of its distance 
($\sim\pm$5 kpc).

To discuss in more detail the position of Reticulum compared with 
LMC barycenter we provide independent and homogeneous distance moduli 
based on both optical and MIR measurements available in the literature.  
This approach is further strengthened by the recent findings by 
\citet{muraveva18b} suggesting, based on a large sample of Gaia DR2 
trigonometric parallaxes \citep{gaia_dr2_arenout2018}, that 
the coefficients of the metallicity term predicted by pulsation 
models agree quite well with observations. We adopted the 
MIR mean magnitudes provided by \citet{muraveva18a} and the 
MIR theoretical PLZ relations provided by 
\citet[][see Fig.~\ref{fig:pljk}]{marconi15,neeley17}. 
We found the following empirical PL relations  

\begin{equation}\label{eq:pli1}
[3.6]_0   = (17.24\pm0.06) - (2.12\pm0.21) \cdot \log{P}
\end{equation}

\begin{equation}\label{eq:pli2}
[4.5]_0 = (17.13\pm0.08) - (2.52\pm0.29) \cdot \log{P}
\end{equation}

and true distance moduli of  $\mu_{[3.6]}=$18.30$\pm$0.06$\pm$0.05 
and of $\mu_{[4.5]}=$18.31$\pm$0.08$\pm$0.08 mag.
The new true distance moduli are in remarkable agreement with distances 
provided by \citet{muraveva18a}. The distances based on MIR mean magnitudes 
are systematically smaller then those based on NIR mean magnitudes, but the 
difference is of the order of 1$\sigma$. To further investigate 
the possible systematics affecting the current distance determinations we 
also estimated the true distance modulus from the $I$-band mean 
magnitudes provided by \citet{kuehn2013a}. 
We found the following empirical PL relation  

\begin{equation}\label{eq:pli}
I_0 = (18.30\pm0.03) - (1.00\pm0.10) \cdot \log{P}
\end{equation}

and  a true distance modulus of $\mu_I=$18.51$\pm$0.07$\pm$0.05 mag 
(see the purple line in Fig.~\ref{fig:pljk}).  Finally, we also adopted
the optical Period-Wesenheit (PW) relations for a threefold reason. 
{\it i)} These distance diagnostics are independent, by construction, of 
the reddening uncertainties. {\it ii)} Using some specific combinations
of filters, they are minimally affected by metal 
content \citep{marconi15}. {\it iii)} They mimic a period-luminosity-color relation 
\citep{madore82,marconi15,neeley17}. However, they rely on the 
assumption that the adopted reddening law is universal. 
We adopted the PW($V$,$B-I$) relation by \citet{marconi15} and the optical $VBI$ mean 
magnitudes provided by \citet{kuehn2013a} and we found 
$\mu=$18.52$\pm$0.03$\pm$0.03 mag. The mean of the homogeneous NIR 
(PLZ: $J$,$K_s$), MIR (PLZ: [3.6], [4.5]) and optical (PLZ: $I$; PW($V$,$B-I$))
distance determinations gives a mean cluster distance of 
$\mu=$18.47$\pm$0.02$\pm$0.06 mag\footnote{Note that the distances based 
on the $I$-band PL relation and on the PW($V$,$B-I$) relation are not 
independent. However, the inclusion of the former distance affects the 
mean cluster distance by less than 0.01 mag.}.

The current estimates support the evidence that Reticulum belongs to the 
LMC halo. In particular, the use of optical, NIR and MIR data suggests 
that it is located $\sim$1 kpc closer than the LMC barycenter, although 
it must be kept in mind that the systematics are of the same order of magnitude of this shift. 
Distance determinations based on MIR data and on \gaia trigonometric 
parallaxes suggest that Reticulum might be even closer  
($\sim$3 kpc, \citet{muraveva18a}). More accurate estimates require 
a novel approach as recently suggested by \citet{bono2018} to 
simultaneously estimate the cluster mean metallicity, reddening 
and distance.

\section{Summary and final remarks}\label{sect_conclusions}

In this work, we have provided the NIR ($JHK_s$) light curve templates of 
RRab and RRc variables. In the following, we summarize the most interesting  
results and discuss in more detail some relevant issues.

{\it Homogeneous photometry ---} We publish $JHK_s$ 
time series, in the 2MASS photometric system, of 254 
RRLs in the GGCs $\omega$ Cen, M4 and in the field of
the Milky Way. The latter sample was obtained from heterogeneous 
literature data in four different photometric systems 
(CIT, SAAO, UKIRT and ESO) which were homogenized. The overall
sample includes both photoelectric and CCD data, collected 
at telescopes in a wide range of diameter classes (1.3m to 8m).
We provide NIR ($JHK_s$) characterization (mean magnitudes, light amplitudes,  
epochs of the mean magnitude on the rising branch) for 94 RRab
and 51 RRc variables that were used to generate the light-curve templates. 

{\it Light-curve templates ---} We provide a total of 24 light-curve 
templates of RRLs: these are divided into Fourier and multi-Gaussian series 
(PEGASUS) fits of four period bins (one for the RRc and three for the RRab
variables) and three photometric bands ($J$, $H$ and $K_s$). The Fourier 
and PEGASUS series range from the fourth to the seventh order and from
the second to the sixth order, respectively. The Fourier templates show 
residuals with respect to the normalized cumulated light curves used to 
generate them that are smaller than those corresponding to the PEGASUS templates. However, 
the latter show fewer secondary, unphysical features (bumps and dips) and 
their residuals are still smaller than 0.005 normalized mag. We provide also the 
phases of minimum and maximum light for all the light-curve templates, 
in order to make it easier for future users to adopt the template even
when lacking the epoch of the mean magnitude on the rising branch, which
is less commonly reported than the epoch of maximum in large surveys.


{\it Template validation ---} We have validated our templates and compared
our $K_s$-band templates to those by J96. The tests were performed 
on both a subsample of four RRLs in $\omega$ Cen (one per template bin), 
that were not used to generate the templates, and on a set of 22 Galactic 
bulge RRLs for which we have $VI$ time series from OGLE and $K_s$-band 
time series from the VVV survey. We have checked that, 
within the dispersion, the mean magnitudes derived by applying the template
and the best estimate of the mean magnitude (i.e., the integral over the 
fit of the light curve, converted into intensities) are the same. The largest
offset is of 0.01 mag (with a standard deviation of 0.04 mag), for the 
$H$-band template of short-period RRab variables (RRab1 template bin), which
are also the ones with the largest amplitudes, meaning that they are more
prone to uncertainties. Compared to our $K_s$ templates, the J96 templates 
provide results which are similar, showing offsets either comparable 
or---sometimes---larger than ours.

{\it Reticulum ---} We have collected literature $JK_s$ time 
series for 30 over 32 RRLs in the LMC globular cluster 
Reticulum \citep{dallora04}. Using $BV$ time series for the same
RRLs \citep{kuehn2013a}, we derived the periods and $t_{ris}$
to apply our templates and estimated NIR mean magnitudes. We 
derived new empirical PL$JK_s$ relations, and in turn, new 
accurate and precise estimates of the distance to Reticulum. 
We found true distance moduli that agree quite well with each 
other ($\mu_J=$18.47$\pm$0.10$\pm$0.03 mag, 
$\mu_{Ks}=$18.49$\pm$0.09$\pm$0.05 mag) and with literature values. 
We adopted homogeneous calibrations for MIR ([3.6], [4.5]) and  
optical ($I$) PLZ relations and for the optical PW($V$,$B-I$) 
relation together with mean magnitudes provided by \citet{muraveva18a} 
and by \citet{kuehn2013a}. We found a mean cluster true distance 
modulus of $\mu=$18.47$\pm$0.02$\pm$0.06 mag. 
According to the most accurate and recent LMC distance determinations  
\citep{pietrzynski2013,inno2016}, the current $\mu$ estimate for 
Reticulum indicates that this cluster is $\sim$1 kpc closer to us 
than the LMC itself.

In the following we briefly outline some of the most relevant developments 
of the current project supporting the non-trivial effort for new NIR 
light-curve templates for RRLs. 

{\em Distance Scale} --- Future ground-based Extremely Large Telescopes 
(ELT, TMT, GMT) and space observing facilities (JWST, EUCLID, WFIRST) 
have been designed to reach their peak performance in the NIR regime. 
This means that a few NIR measurements of variables already identified 
and characterized in the NIR will allow us to fully exploit the RR Lyrae 
distance scale in Local Group and in Local Volume galaxies. Note that this 
opportunity fits within a context in which Gaia will provide exquisite 
calibration for both the zero point and the slope of the diagnostics we are 
currently using to estimate individual RR Lyrae distances 
\citep{gaia_alldr,gaia_brown2016,gaia_dr2_arenout2018}. Moreover, LSST 
will provide an unprecedented wealth of optical time series, and in turn a 
complete census of evolved variables in the nearby Universe \citep{oluseyi2012}. 
These are crucial prior conditions to reach a precision of the order of 1\% 
on individual RRL distances and an accuracy better than 3\% on the Hubble 
constant \citep[Carnegie RR Lyrae Program][]{beaton2016}. 

{\em Light curve characterization} --- Light-curve templates also provide the 
opportunity to improve the accuracy of the fit of the light curve when either 
a single or a few measurements are available. Note that this opportunity becomes 
even more relevant for NIR photometric surveys, like VVV+VVV-X
\citep{minniti2010}, that collect time-series 
data in the $K_s$-band and just a few measurements in the $J$ and 
$H$ bands. Accurate NIR mean magnitudes are, together with optical mean 
magnitudes, a fundamental ingredient for constraining the distance, the 
reddening and the metal content of field and cluster RR Lyrae using the 
recent algorithm (REDIME) suggested by \citet{bono2018}. 

{\em Envelope tomography} --- Our knowledge of linear 
and nonlinear phenomena taking place along the pulsation cycle of 
a variable star is still limited to a handful of objects. There is 
solid evidence that moving from the optical to the NIR regime 
luminosity changes are mainly dominated by variations of radius instead of 
temperature \citep{bono01,madore13}. 
However, we still lack accurate investigations 
of shock formation and propagation based on NIR spectroscopic 
diagnostics. The NIR light-curve templates provide the opportunity 
to trace the color ($V-K$) variation along the pulsation cycle, and 
in turn, the temperature variation. This information is crucial for 
estimating atmospheric parameters of spectra including a limited 
number of ionized/neutral heavy element lines 
\citep[][,Magurno et al. 2018, in preparation]{sollima06a}.   

It goes without saying that it is a real pleasure to develop a new 
tool to be used by the astronomical community, but it is 
even more appealing to use it on a broad range of stellar systems.

\begin{appendix}\label{sec:tris}
\section{Estimate of the phase of the anchor point ($t_{ris}$)}

To derive $t_{ris}$, we adopted the following approach. 
We selected, for each star, one filter for 
which either the optical ($B$, $V$) or the NIR ($J$) 
light curve is regular and well sampled.  
Then, we fit the light curve with either a PLOESS (all the 
MW RRLs and part of the $\omega$ Cen RRLs) or a spline fit 
(all the M4 RRLs and the remaining part of the $\omega$ Cen RRLs) 
and derived the mean magnitude from 
the intensity integral of the analytic fit. 

We then interpolated the phase at which the 
rising branch of the fit intersects the mean magnitude (${\phi}_0$). 
Finally, $t_{ris}$ could be obtained as 
$t_i-(({\phi}_i-{\phi}_0) \cdot P)-P$, where $t_i$ and 
${\phi}_i$ are the epoch and the phase of the $i$-th
phase point of the light curve, respectively. This specific 
phase point was called  the ``anchor point''. In principle, 
any of the phase points in the light curve could be an anchor point.
However, we have used an interactive procedure {\it 1)} to avoid
selecting an anchor point which deviates from the others due to 
either period changes or phase shifts with time; and
{\it 2)} to sort the phase points from the most recent to the 
oldest, to obtain a final estimate of $t_{ris}$ that is as recent 
as possible. The latter might seem a non-necessary requirement, 
but it is crucial to have reference epochs which are as close as 
possible in time to the NIR measurements over which the 
templates will be applied. 

\end{appendix}

\bibliographystyle{aa}

\end{document}